\documentclass{aa}  

\usepackage{graphicx}
\usepackage[colorlinks=true,linkcolor=blue,allcolors=blue]{hyperref}
\usepackage{txfonts}
\usepackage[usenames,dvipsnames]{xcolor} 
\usepackage{lscape}
\usepackage{longtable}
\usepackage{placeins}

\begin{document}

   \title{A study of interacting galaxies from the Arp-Madore Catalogue:}

    \subtitle{Triggering of star formation and nuclear activity}

   \author{Pedro H. Cezar
          \inst{1,2}
          \and
          Miriani G. Pastoriza\inst{3}
          \and Rogério Riffel \inst{1,3} \and Cristina Ramos Almeida \inst{1,2} \and Angela C. Krabbe \inst{4} \and Sandro B. Rembold \inst{5}
          }

   \institute{Instituto de Astrofísica de Canarias, Calle Vía Láctea, s/n, 38205 La Laguna, Tenerife, Spain\\ 
   \email{pedrocezar.astro@gmail.com}
         \and Departamento de Astrofísica, Universidad de La Laguna, 38206 La Laguna, Tenerife, Spain
         \and Departamento de Astronomia, Universidade Federal do Rio Grande do Sul, IF, CP 15051, Porto Alegre 91501-970, RS, Brazil \\
         \email{riffel@ufrgs.br}
        \and Departamento de Astronomia, Instituto de Astronomia, Geofísica e Ciências Atmosféricas da USP, Cidade Universitária, 05508-900 São Paulo, SP, Brazil 
        \and Departamento de Física, CCNE, Universidade Federal de Santa Maria, 97105-900 Santa Maria, RS, Brazil 
        }
        

\abstract{We present Gemini Multi-Object Spectrograph (GMOS) spectroscopic observations of 95 galaxies from the Arp \& Madore (1987) catalogue of peculiar galaxies. These galaxies have been selected because they appear to be in pairs and small groups. These observations have allowed us to confirm that 60 galaxies are indeed interacting systems. For the confirmed interacting sample, we have built a matched control sample of isolated galaxies. We present an analysis of the stellar populations and nuclear activity in the interacting galaxies and compare them with the isolated galaxies. We find a median light (mass) fraction of 55\% (10\%) in the interacting galaxies coming from stellar populations younger than 2 Gyr and 28\% (3\%) in the case of the isolated galaxies. More than half of the interacting galaxies are dominated by this young stellar population, while the isolated ones have most of their light coming from older stellar populations.   
We used a combination of diagnostic diagrams (BPTs and WHAN) to classify the main ionization mechanisms of the gas. 
The interacting galaxies in our sample consistently show a higher fraction of active galactic nuclei (AGN) relative to the control sample, which ranges between 1.6 and 4 depending on the combination of diagnostic diagrams employed to classify the galaxies and the number of galaxies considered. 
Our study provides further observational evidence that interactions drive star formation and nuclear activity in galaxies and can have a significant impact on galaxy evolution.}

               
    \titlerunning{A study of interacting galaxies of the Arp-Madore Catalogue}
    \authorrunning{Pedro H. Cezar et al.}

   \maketitle
%

\section{Introduction}\label{section:Intro}

It is well known that galaxy interactions and merger events play an important role in the evolution of galaxies and, in particular, their star formation histories. Interactions have been suggested as the drivers of violent star formation events, producing some of the most luminous objects in the Universe \citep{Sanders+96}. From early studies to the most recent ones, it has been found that star formation is enhanced in interactions and mergers as compared to isolated galaxies \citep{Donzelli+97,Woods+07,Knapen+15,Zaragoza-Cardiel+18,Pan+19,Azevedo+23}. However, just a few studies focused on analysing the stellar populations of these galaxies, most of them dedicated to single or a few pairs \citep{Krabbe+08,Krabbe+11,Krabbe+17,Alonso-Herrero+12,Dametto+14,Bessiere+14,Bessiere+17,Argudo+15,Cortijo-Ferrero+17,Cortijo-Ferrero+17c,Rosa+18}. As pointed out by \cite{Krabbe+17}, this analysis is crucial to understand not only the star-forming regions photoionizing the gas but also the full history of stellar populations born in the interacting galaxies. 

Galaxy interactions and mergers are also frequently associated with the triggering and feeding of Active Galactic Nuclei \citep[AGN; see][and references therein for a wide list of publications in the theme]{Heckman+14,Storchi+19}. This relation, however, is not exempt from controversy. Several simulations show that galaxy interactions trigger AGN activity \citep{Neistein+14,Gatti+15}. In particular, \cite{Neistein+14} propose in their work that minor mergers could be the main triggers of low to intermediate-luminosity AGN. However, recent theoretical work shows that only 15\% of the supermassive black hole (SMBH) growth is produced via galaxy mergers \citep{Steinborn+18,McAlpine+20}.

Observationally, several works found that in both radio-quiet and radio-loud AGN are more frequently hosted in galaxies showing signatures of morphological disturbance than non-active galaxies \citep{RamosAlmeida+11,RamosAlmeida+12,Ellison+19,Hernandez-Toledo+23,Rembold+24}, especially in the case of luminous quasars \citep{Bessiere+12,Pierce+22,Pierce+23,Araujo+23}. Another approach is to look for galaxies undergoing an interaction to verify the prevalence of AGN using optical excitation diagnostics or multiwavelength datasets. Once more, numerous works appear to indicate that interactions are capable of inducing and/or enhancing the nuclear activity \citep{Pastoriza+99,Koss+10,Ellison+11,Silverman+11,Satyapal+14,Fu+18,Jin+21,Steffen+23}. However, some works did not find a significant excess of AGN in interactions relative to field galaxies \citep{Cisternas+11,Bohm+13,Villforth+14,Karouzos+14,Argudo+16,Marian+19,Silva+21,Villforth+23}. These discrepancies could be due to selection effects,  including the AGN identification method  \citep{Satyapal+14,Ellison+19}, different methodologies, and datasets. \cite{Pierce+23} argue that in the absence of deep imaging, tidal features may be missed, and the fraction of morphological disturbance observed in AGN hosts could be underestimated. 
 The methodology used to identify AGN galaxies can also have an impact on the observed merger-AGN connection. Focusing on more advanced mergers, recent works targeting post-mergers \citep{Li+23} and galaxies presenting gas-stellar rotation misalignment \citep{Raimundo+23} show an excess of AGN relative to undisturbed galaxies.  

In this paper, we analyze the stellar populations and ionized gas excitation mechanisms of a sample of galaxy pairs or small group candidates from the \cite{ArpMadore+87} catalogue of peculiar galaxies. The sample of galaxy pairs includes interacting galaxies with smaller companions (Category 1, when one of the galaxies has less than half of the primary size), interacting galaxies with similar sizes (Category 2), groups (Categories 3 and 4), and disturbed galaxies with jets or tails (Categories 7 and 15). We first use the spectroscopic observations to confirm or discard the interacting nature of the pairs. The confirmed physical systems are then compared to a control sample of isolated galaxies built to match the interacting sample properties. In this way, we expect to be comparing similar galaxies but in different environments. With this comparison, we aim to address the impact of interactions and mergers on the stellar populations and nuclear activity of the galaxies. 

This paper is organized as follows: in  Sec.~\ref{section:dataandmeasuments} we present the observations, data reduction, and measurements used in this work. In Sec.~\ref{section:stellarpop} we show the results of the modelling of the stellar populations in the interacting and isolated galaxy samples. Sec.~\ref{section:reddening} presents the ionized gas and stellar populations extinctions. In Sec.~\ref{section:excitation} we show the results of the main excitation mechanisms responsible for ionizing the gas in the interacting and isolated samples. Sec.~\ref{section:starforming} focus on comparing the star formation strength of star-forming galaxies in the interacting and control samples. In Sec.~\ref{section:disc} we discuss the results and finally, Sec.~\ref{sec:conclusions} summarizes our findings. Throughout this paper, we adopt the Hubble constant as $H_0$ = 75\,km\,s$^{-1}$\,Mpc$^{-1}$ \citep{Spergel+07}. 

\section{Sample selection, observations, and analysis}
\label{section:dataandmeasuments}

Our parent interacting galaxy sample was selected from the \citet{ArpMadore+87} catalogue and it is composed of 43 candidates to pairs, small groups, and mergers. Most of them were selected from Category 1 (37/43 candidates), where one of the galaxies has less than half the size of its companion(s). Most of the main galaxies in Category 1 are spirals. The remaining six candidates are: AM 0338-320 (Category 2, pairs of galaxies of comparable sizes), AM 0103-502 (Category 3, triple systems), AM 0302-611 (Category 4, quartet systems), AM 2037-550 (Category 7, galaxies with jets), AM 2037-550, and AM 2220-661 (Category 15, galaxies with tails, loops of material, or debris). In total, these 43 galaxy systems comprise 100 individual galaxies. Most of these galaxies are infrared bright, and therefore, we expect a high incidence of star-forming and/or active galaxies.

\subsection{Observations and data reduction}\label{section:obs}

We present spectroscopic observations of 95 of the 100 galaxies included in our parent sample of 43 candidates to physical pairs, small groups, and mergers. To the best of our knowledge, these are the first spectroscopic observations for most of the sources. The sample is listed in Table \ref{tab:Sample} in Appendix~\ref{AP_A}, and further details are given in Section~\ref{section:samples}.

The data were obtained using the Gemini South Multi-Object Spectrograph (GMOS-S) at the
Gemini-South telescope, under the ``poor weather queue''
programme GS-2007A-Q-240, between October 2007 and January
2008. The instrument configuration was the 0.5\arcsec~slit with the
R150 grating binned to 2 $\times$ 4 pixels (spectral x spatial). Two grating
settings were used, centred at 8700 and 8800 {\AA}, in order to cover
the chip gaps. This setup provided full spectral coverage from
3500 {\AA} to beyond 8000 {\AA} (rest wavelength), at a spectral resolution
of 11 {\AA} at 6356 {\AA} - just enough to separate the [NII]+H$\alpha$ blend
while at the same time covering the whole optical spectrum for
redshifts up to $\sim$ 0.2. The slit was aligned to include the two galaxies in the pairs, and additional slit positions were used in the case of groups (see Fig.~\ref{fig:pairsimages}). The objects were nodded along the slit to improve sky subtraction. Exposure time was 1 h on source (2 $\times$ 900 s
per grating setting).

The data reduction followed standard procedures, using the
{\sc gemini.gmos} routine of the {\sc IRAF} package \citep{Tody+86,Tody+93}, except that no flat field
correction was applied to the spectroscopic data presented here. This
was due to the strong fringing effects present redwards of 750 nm,
which severely degraded the quality of the data towards the red.
For the large apertures extracted here, 
this has
no measurable effect on the data. To make a fair comparison among the galaxies, we have extracted all the spectra in a consistent aperture of 2.2\,kpc $\times$ 0.5\arcsec~(0.5\arcsec$\approx$ 380 pc at the mean distance of the sample).

The imaging data shown in Fig.~\ref{fig:pairsimages} correspond to
the acquisition images taken just before the spectroscopy to
centre the objects in the slit. The Sloan {\it r'} filter was used, and
reduction was standard, including bias and flat-field corrections. 


\subsection{The interacting and isolated galaxies samples}\label{section:samples}

\begin{figure*}
    \centering
    \includegraphics[width=0.65\columnwidth]{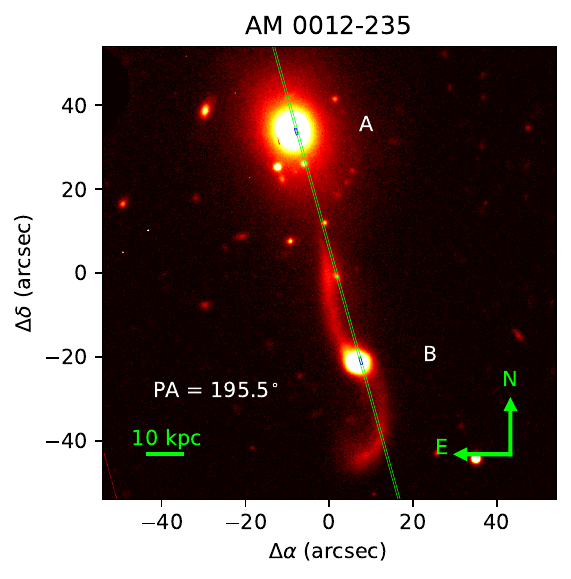}
    \includegraphics[width=0.65\columnwidth]{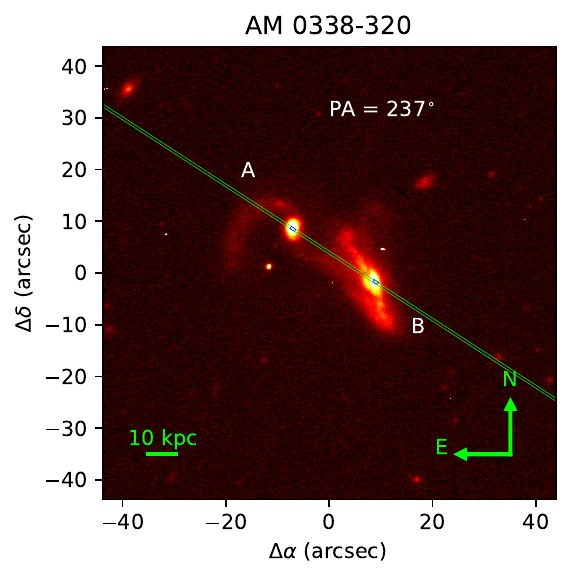}
    \includegraphics[width=0.65\columnwidth]{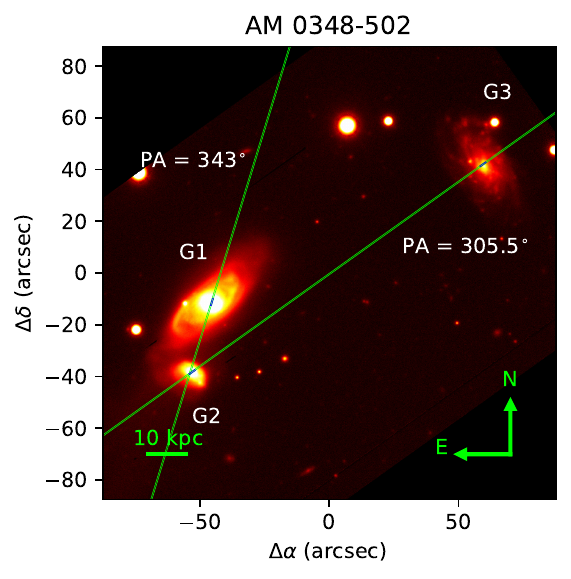}
    \includegraphics[width=0.65\columnwidth]{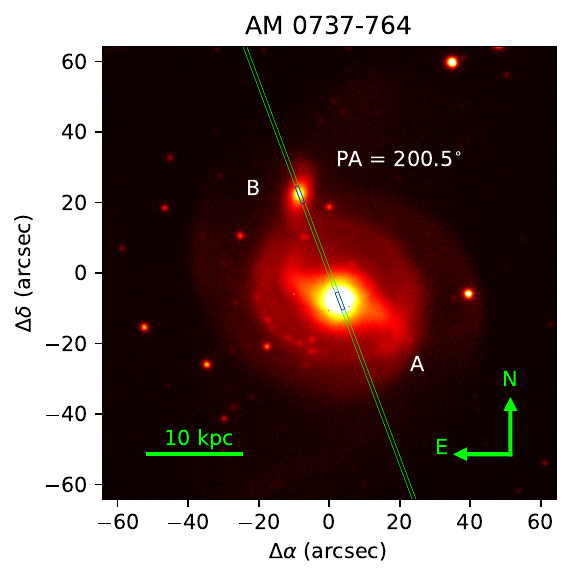}
    \includegraphics[width=0.65\columnwidth]{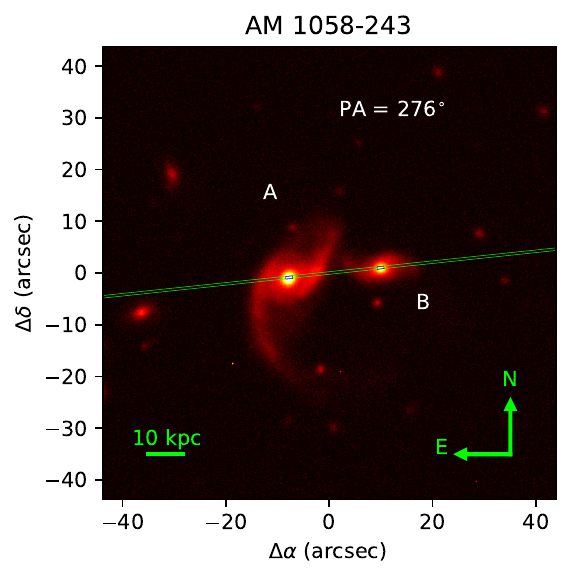}
    \includegraphics[width=0.65\columnwidth]{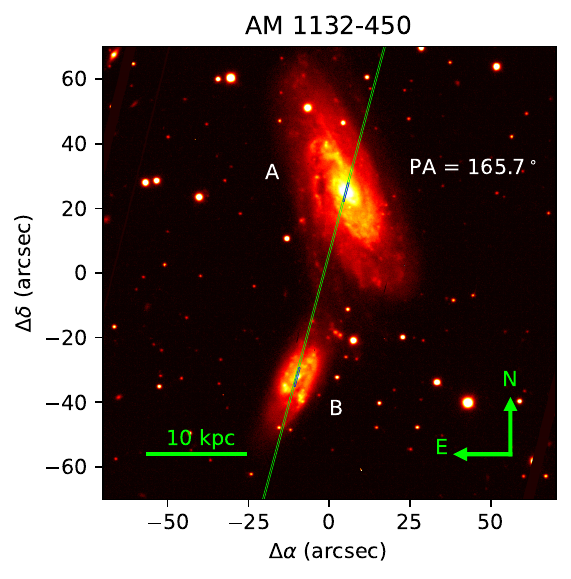}
    \includegraphics[width=0.65\columnwidth]{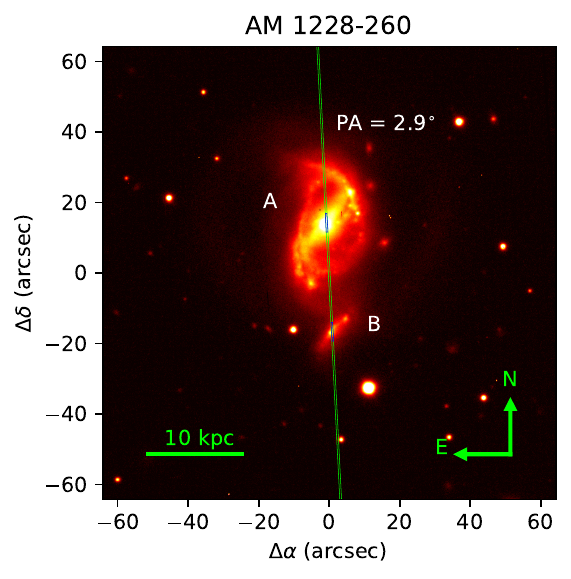}
    \includegraphics[width=0.65\columnwidth]{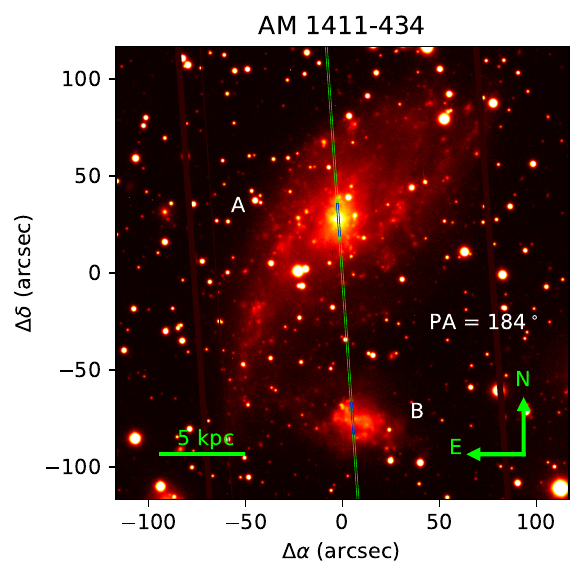}
    \includegraphics[width=0.65\columnwidth]{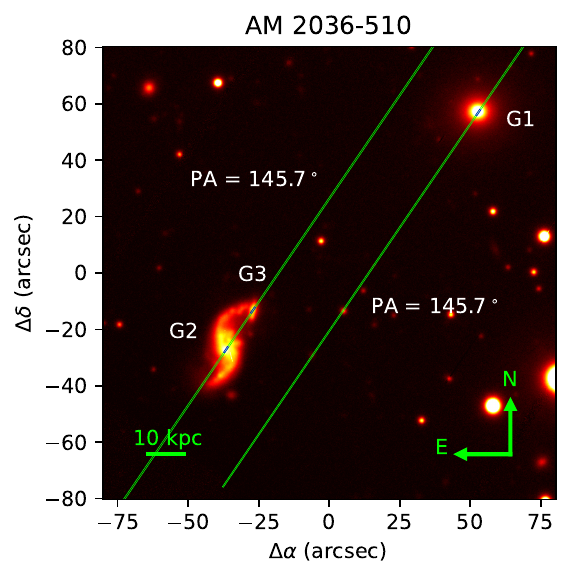}
    \includegraphics[width=0.65\columnwidth]{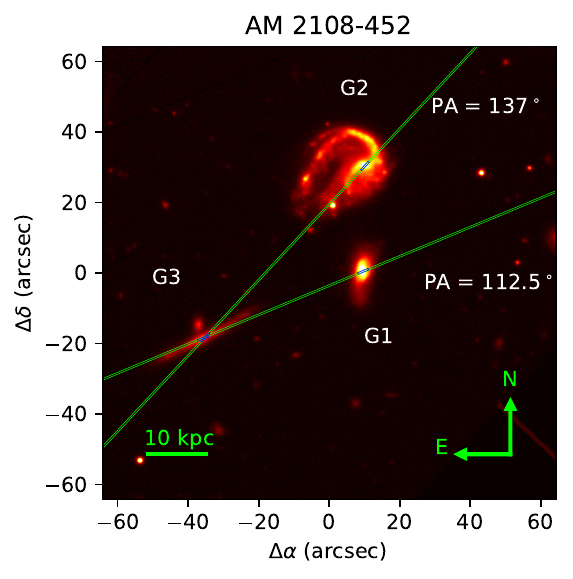}
    \includegraphics[width=0.65\columnwidth]{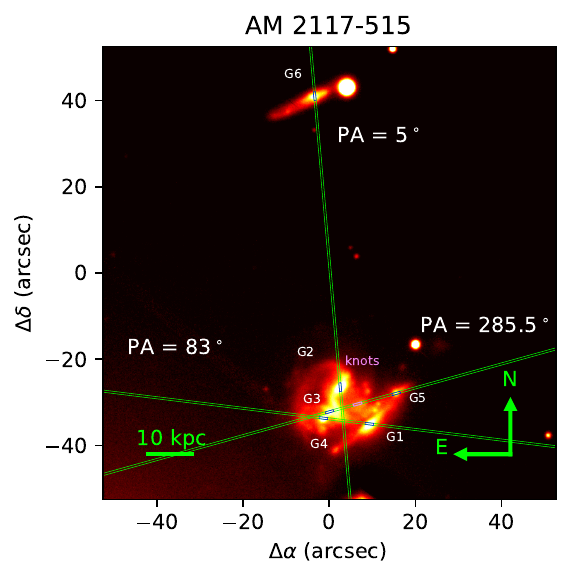}
    \includegraphics[width=0.65\columnwidth]{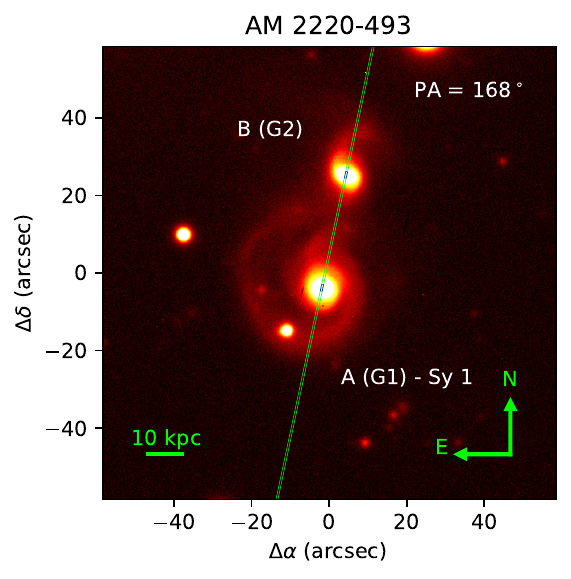}
    \caption{Examples of {\it r'} filter acquisition images obtained with GMOS-S of the interacting pairs and groups. The slits are represented in green, with the nuclear aperture used here shown in blue. The remaining images are shown in Fig.~\ref{fig:pairsimages2} in Appendix~\ref{AP_A}.}
    \label{fig:pairsimages}
\end{figure*}

\begin{figure*}
    \centering
    \includegraphics[width=0.97\columnwidth]{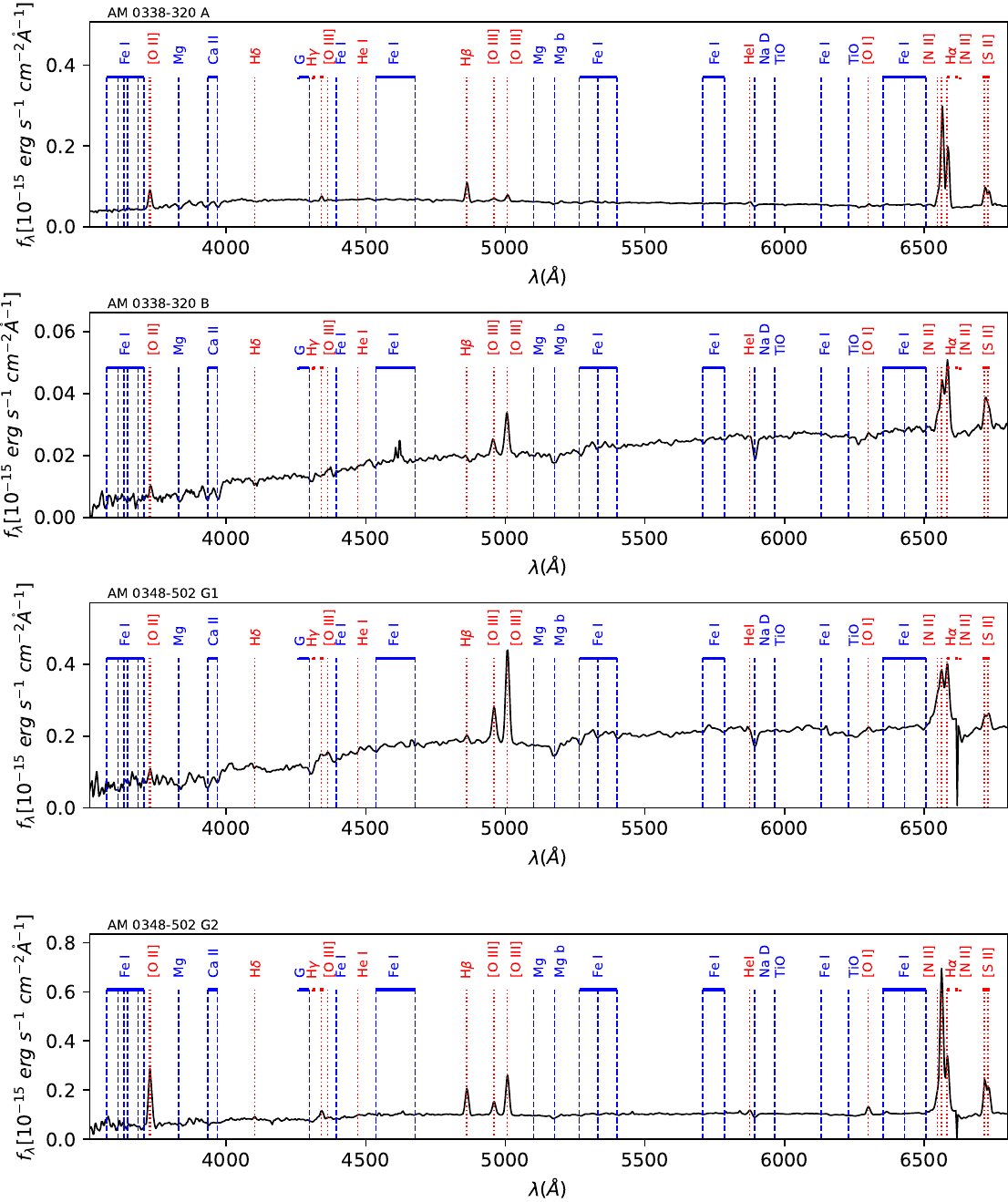}
    \includegraphics[width=0.97\columnwidth]{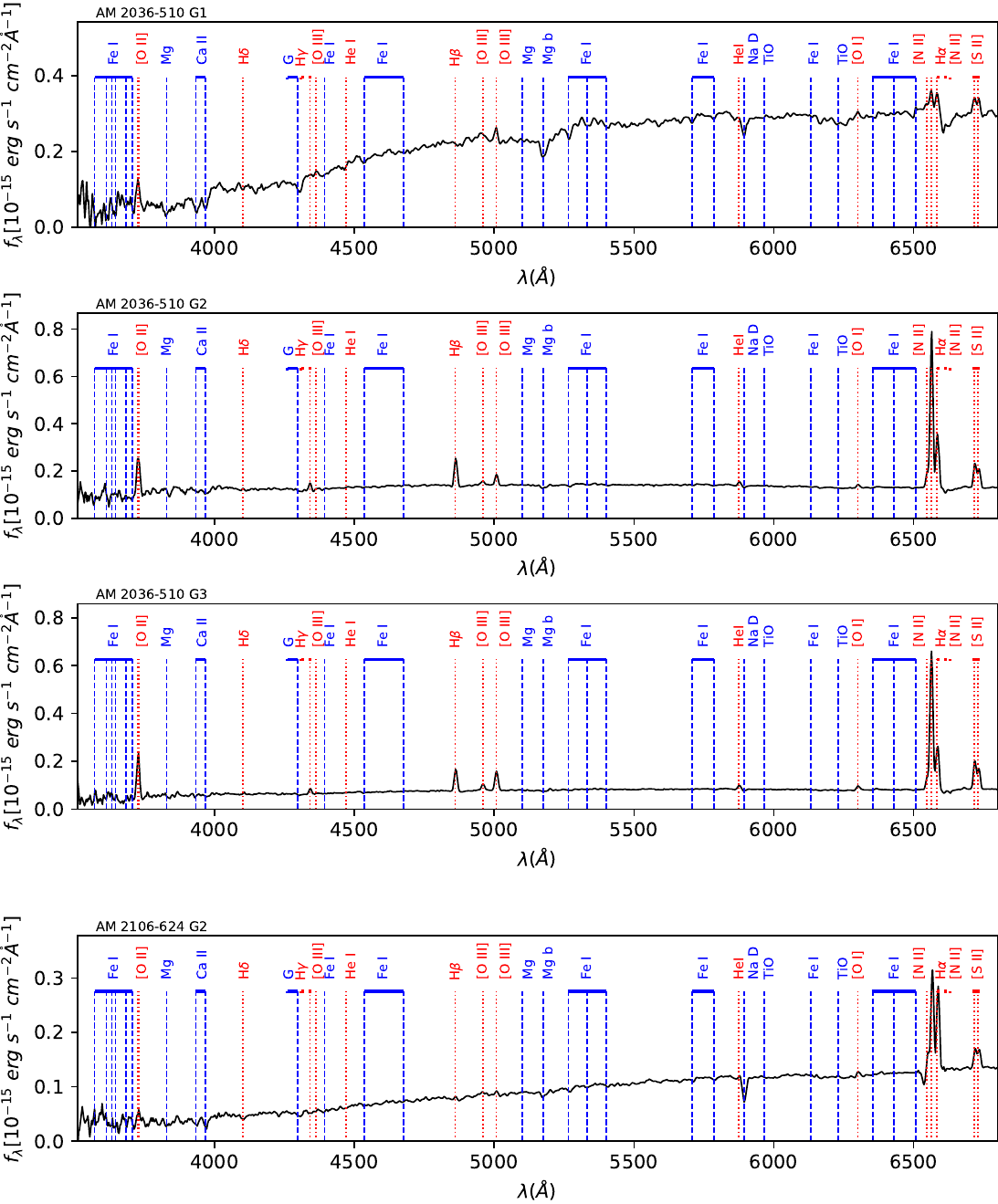}
    \includegraphics[width=0.97\columnwidth]{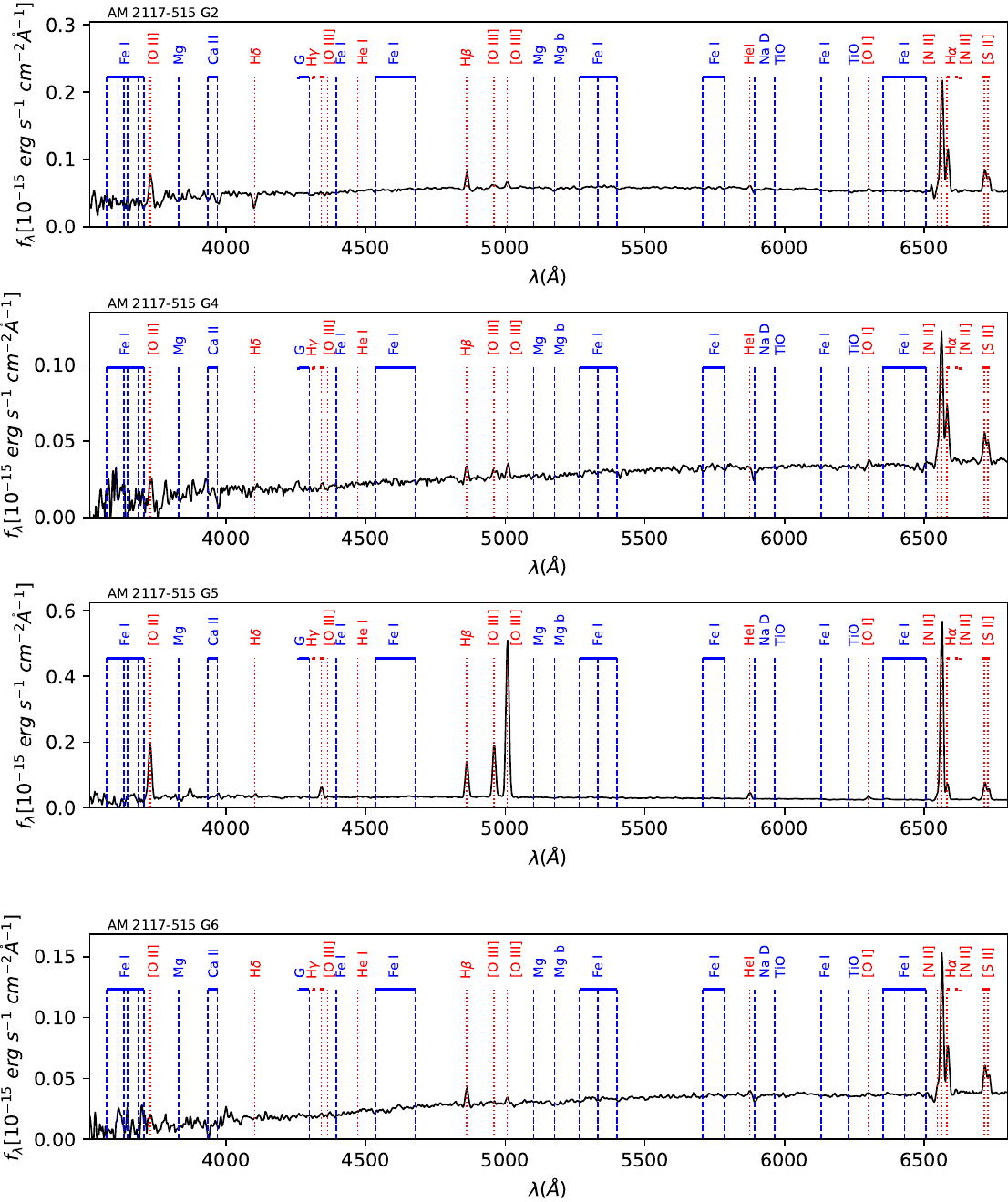} 
    \includegraphics[width=0.97\columnwidth]{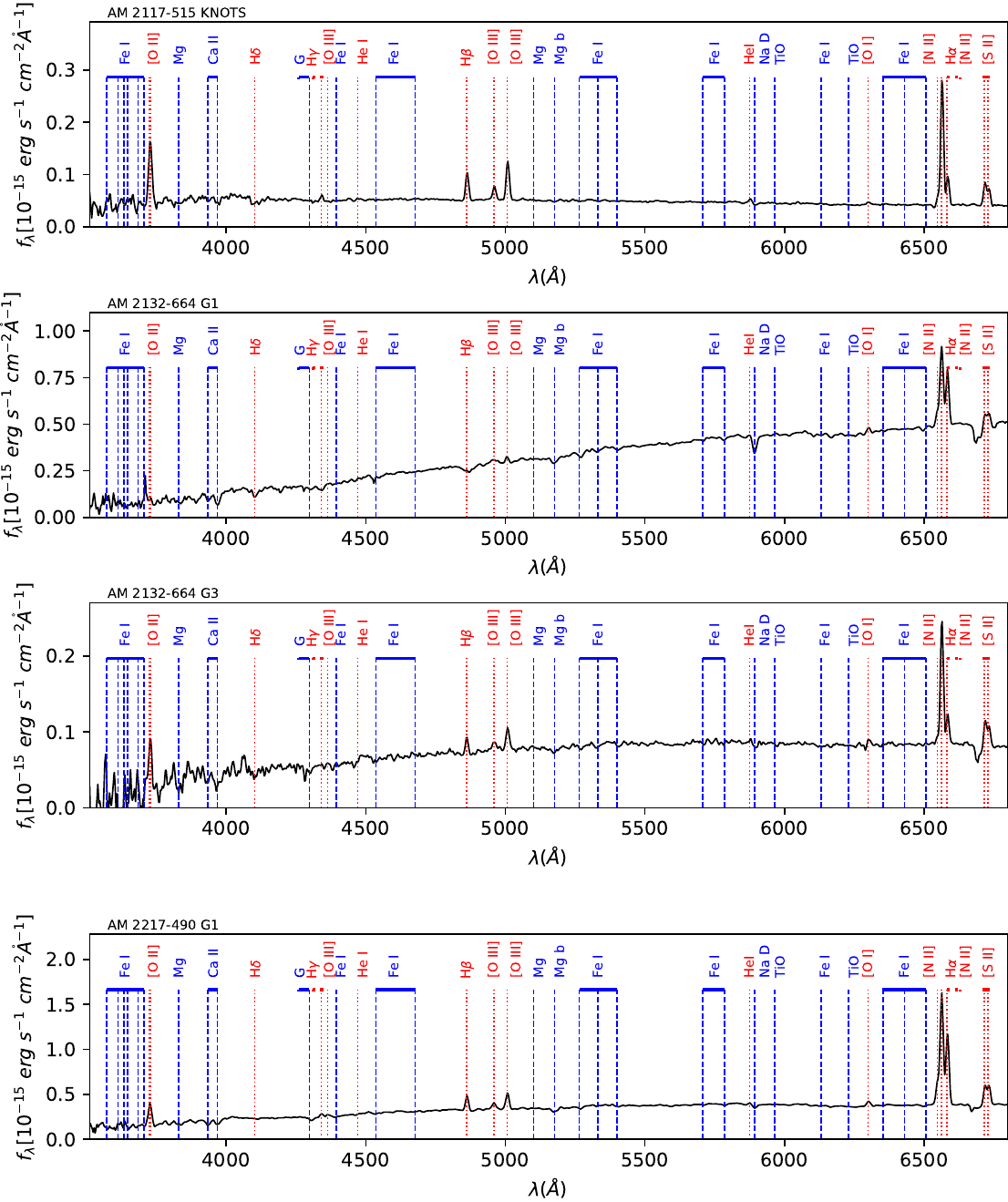}
    \caption{Examples of nuclear rest-frame spectra (2.2 kpc aperture) of the interacting pairs observed with GMOS-S. The main emission lines and absorption features are indicated as red and blue vertical lines, respectively. The remaining spectra are shown in Fig.~\ref{fig:pairsspectra2} in Appendix~\ref{AP_A}.}
    \label{fig:pairsspectra}
\end{figure*}

The redshift for the interacting galaxies was measured using the {\sc IRAF} tasks {\sc rvidlines} and  {\sc xcor}, for emission and absorption line spectra, respectively, and adopting $H_0$ = 75\,km\,s$^{-1}$\,Mpc$^{-1}$ \citep{Spergel+07}. As in previous works, in order to determine if the galaxies belong to a physical system, we constrain the difference between their radial velocities to be $\Delta v_{r}$ < 500 km\,s$^{-1}$ \citep{Patton+00,Lin+08,Ventou+17}. Sixty galaxies were found to be in truly interacting systems (19 pairs and 6 small groups), although we only have GMOS-S spectra for 59 of them. We do not have a GMOS-S spectrum of AM 2106-624 G1 because the slit was misplaced, but it was confirmed as a member of the pair with AM 2106-624 G2 thanks to a previous radial velocity estimate (see Table \ref{tab:radial_vel} in Appendix~\ref{AP_A}). Therefore, the following analysis was carried out using the 59 interacting galaxies with spectra, which we will refer to as the interacting (INT) sample hereafter. For them we measured redshifts in the interval 0.006 $\le$ {\it z} $\le$ 0.097. Figs.~\ref{fig:pairsimages} and~\ref{fig:pairsspectra} show examples of {\it r'} filter acquisition images and observed optical spectra of confirmed pairs and groups. 
The remaining 40 galaxies include 22 that are just close in projection, and 18 for which either the data did not allow us to constrain their redshifts or were not available. Table~\ref{tab:radial_vel} in Appendix~\ref{AP_A} presents the radial velocities for the interacting systems and Table~\ref{tab:radial_vel_nonpairs} in Appendix~\ref{AP_B} for the non-interacting galaxies, together with information about the sources which could not be either confirmed or discarded as interacting.

In order to find an adequate control sample of non-interacting galaxies (i.e., isolated sources), comparable in mass and redshift with those of our sample, we first used the broadband magnitudes (JHK) from the Two Micron All Sky Survey (2MASS) extended source catalogue \citep{Skrutskie+06} of the INT galaxies as a proxy for their stellar mass \citep{Davies+15}. We then searched for galaxies with optical spectra in the Sloan Digital Sky Survey (SDSS) III Data Release 9 \citep[DR9,][]{York+00,Eisenstein+11,Ahn+12} matched in redshift and 2MASS JHK magnitudes in the extended source catalogue to each of the INT galaxies. The candidates to control sample galaxies for each INT galaxy were then visually inspected using the SDSS images available, and kept only those having similar morphologies as the interacting galaxy they match with in infrared magnitude and redshift. In particular, we searched for galaxies with the same galaxy type as each of the INT galaxies (i.e., spirals, lenticulars, or ellipticals) and stellar structures (i.e., bars, rings, and/or spiral arms) and we discarded galaxies with clear signs of interaction and/or close companions of similar mass at distances < 100 kpc.

We note that we selected just one control sample galaxy for each pair member of the INT sample, which is the minimum required for comparison while still being possible to analyse the galaxies individually.
Fourteen of the interacting galaxies have either too disrupted optical morphologies or do not have infrared magnitudes available in the 2MASS database, thus no control sample galaxies were selected for them. We did not discard the INT galaxies without controls from the analysis because we checked that the results were consistent including them or not.

\begin{figure}
    \centering
    \includegraphics[width=\columnwidth]{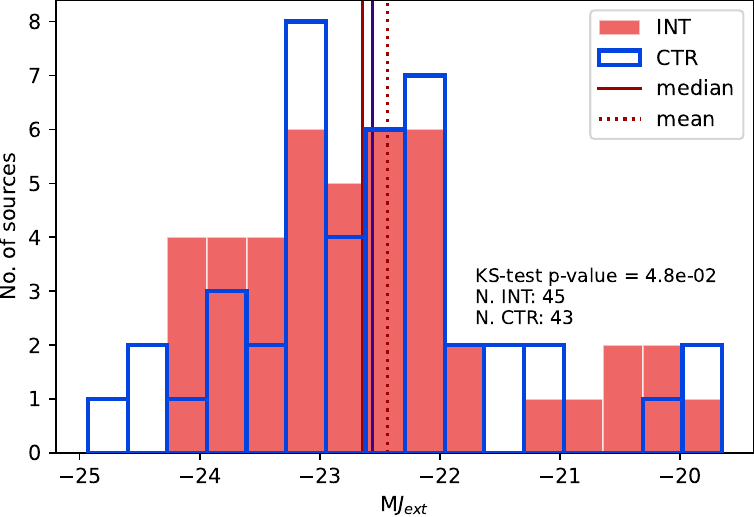}
    \includegraphics[width=\columnwidth]{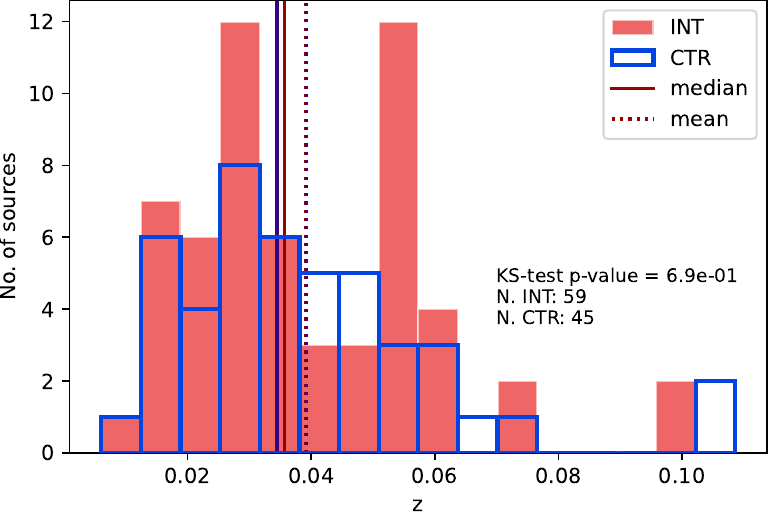}
    \caption{Distribution of absolute magnitudes (using the available extended $J_{ext}$ 2MASS magnitudes) and redshifts for INT (red filled histograms) and CTR (blue continuous line) samples. The plot also shows the samples median, mean values, and their Kolmogorov-Smirnov (KS) test p-values.}
    \label{fig:sampleshist}
\end{figure}

Our final control sample of isolated galaxies (CTR sample hereafter) consists of 45 objects, which match the INT sample in redshift, 2MASS J absolute magnitude, and morphology. Figure~\ref{fig:sampleshist} shows the redshift and absolute J magnitude distribution of the INT and CTR samples. The median, mean, and Kolmogorov-Smirnov (KS) test p-values confirm the similarity between them. Table~\ref{tab:radial_vel} in Appendix~\ref{AP_A} lists the INT galaxies, radial velocities, and the SDSS DR9 information of the corresponding CTR sample galaxies. The optical spectra of the CTR sample are from SDSS, which uses a 3\arcsec~fiber and covers the spectral range 3800–9200 \AA, with a spectral resolution (R) varying from R$\sim$1500 at 3800 \AA~to R$\sim$2500 at 9000 \AA. This corresponds to FWHM=2.78 \AA~at 7000 \AA~\citep{Alam+15, Abolfathi+18}.




\subsection{Stellar populations modeling}\label{section:stellarfitting}

To perform a stellar population analysis of the galaxies by means of full-spectral fitting of their spectra, we employed the {\sc starlight} code \citep{CidFernandes+05,CidFernandes+18}. This code combines the spectra of a set of $N_{\star}$ simple stellar population (SSP) template spectra denoted as $b_{j,\lambda}$. These templates are weighted in varying proportions to replicate the observed spectrum $O_\lambda$. The modeling process involves normalizing both the observed and modeled spectra, represented as $M_\lambda$, at a user-defined wavelength $\lambda_0$. The vector, $x_j$, represents the fractional contribution of the $j$th SSP to the light at the normalization wavelength $\lambda_0$. Additionally, {\sc starlight} incorporates a Gaussian distribution $G(v_\star, \sigma_\star)$ to address velocity shifts ($v_{\star}$) and velocity dispersion ($\sigma_\star$). The model spectra can be expressed as:

\begin{equation}
	M_\lambda = M_{\lambda_0} \left[ \sum_{n=1}^{N_\star} x_j\,b_{j,\lambda}\,r_\lambda \right] \otimes G(v_\star, \sigma_\star),
\end{equation}

where $M_{\lambda_0}$ represents the flux of the synthetic spectrum at the wavelength $\lambda_0$.  To account for reddening, it incorporates the term $r_\lambda = 10^{-0.4 (A_\lambda - A_{\lambda_0})}$.
In the process of fitting, the code searches for the best-fit parameters by minimizing $\chi^2$, defined as:

\begin{equation}
	\chi^2 = \sum_{\lambda_i}^{\lambda_f} [(O_\lambda - M_\lambda) \omega_\lambda]^2,
\end{equation}

where $\omega_\lambda$ are the weights of each pixel in the fitting. For more details see \citet{CidFernandes+05}.

We employ a "base of elements" known as the $GM$, as detailed in \citet{CidFernandes+13,CidFernandes+14}. This base is constructed using the {\sc Miles} \citep{Vazdekis+10} and \citet{GonzalezDelgado+05} models and has been updated with the {\sc Miles} V11 models \citep{Vazdekis+16}. Our selection includes 21 ages (t= 0.001, 0.006, 0.010, 0.014, 0.020, 0.032, 0.056, 0.1, 0.2, 0.316, 0.398, 0.501 0.631, 0.708, 0.794, 0.891, 1.0, 2.0, 5.01, 8.91, and 12.6 Gyr) and four metallicities (Z= 0.19, 0.40, 1.00, and 1.66 Z$_\odot$). See \citet{Riffel+21,Riffel+23} and references therein for more details on applications of this base. 

\begin{figure}
    \centering
    \includegraphics[width=\columnwidth]{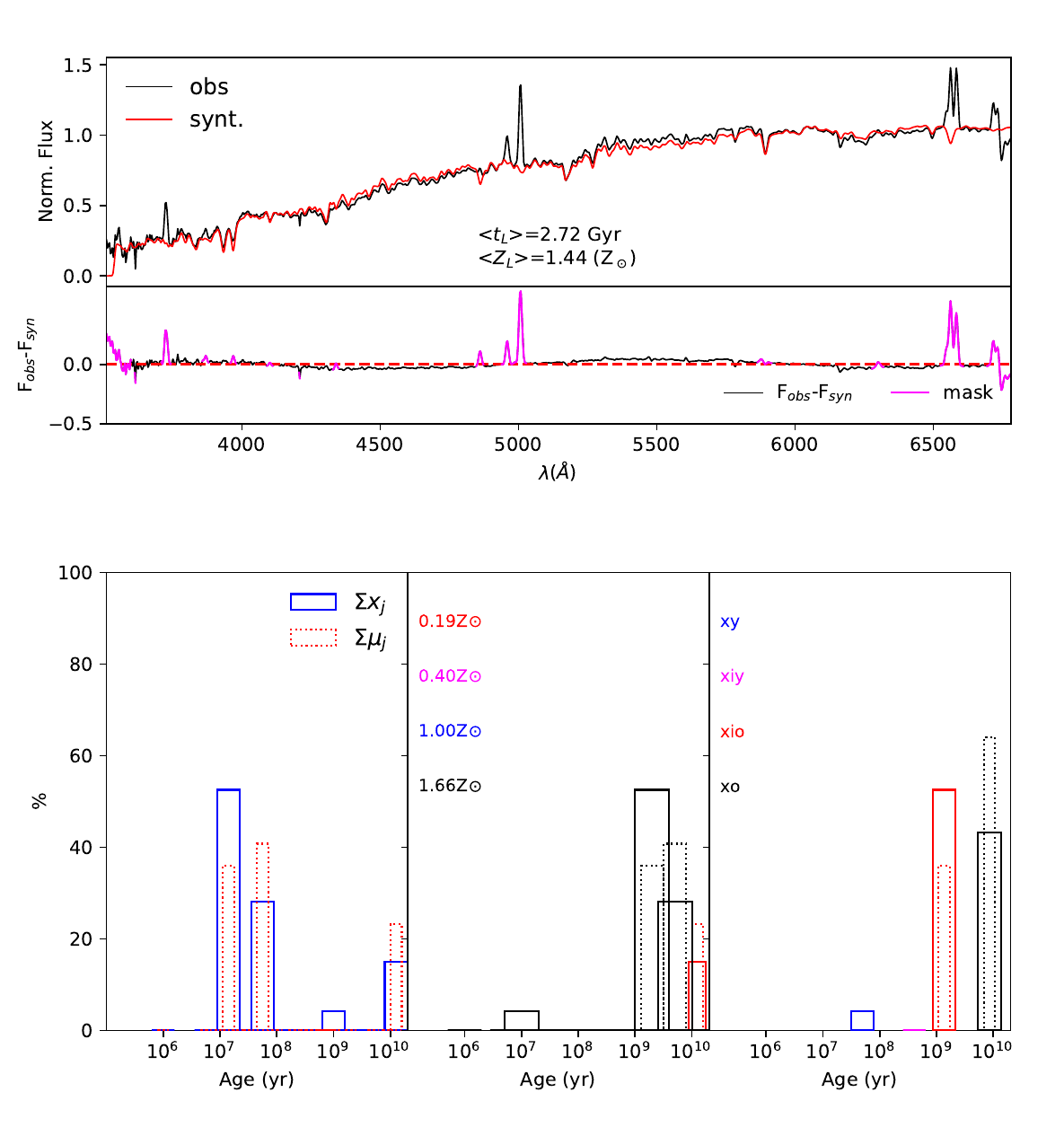}
    \caption{Example of stellar populations synthesis of AM 1933-422 B. Top panel: Observed and stellar populations synthesis fitting (black and red lines, respectively). The inset at the bottom shows the pure emission line spectra (magenta line). Bottom panel: The left histogram shows, in bins of age, the stellar population contribution in light (solid line) and mass (dashed line) fractions. The middle histogram presents, in bins of age, the metallicity of the fitted stellar populations. The right histogram shows the population vectors in light (solid line) and mass (dashed line) fractions.}
    \label{fig:examplesfh}
\end{figure}

The normalization flux at $\lambda_0$ was chosen to be the mean value between 5650\AA\ and 5750\AA. The reddening law used was that of \citet{Cardelli+89}, and the synthesis was performed in the spectral range from $3510$\,{\AA} to $6930$\,{\AA} (rest frame). An example of stellar population fitting is shown in Fig.~\ref{fig:examplesfh} for the galaxy AM 1933-422 B.



\subsection{Emission lines fitting}\label{section:emissionfitting}

\begin{figure}
    \centering
    \includegraphics[width=\columnwidth]{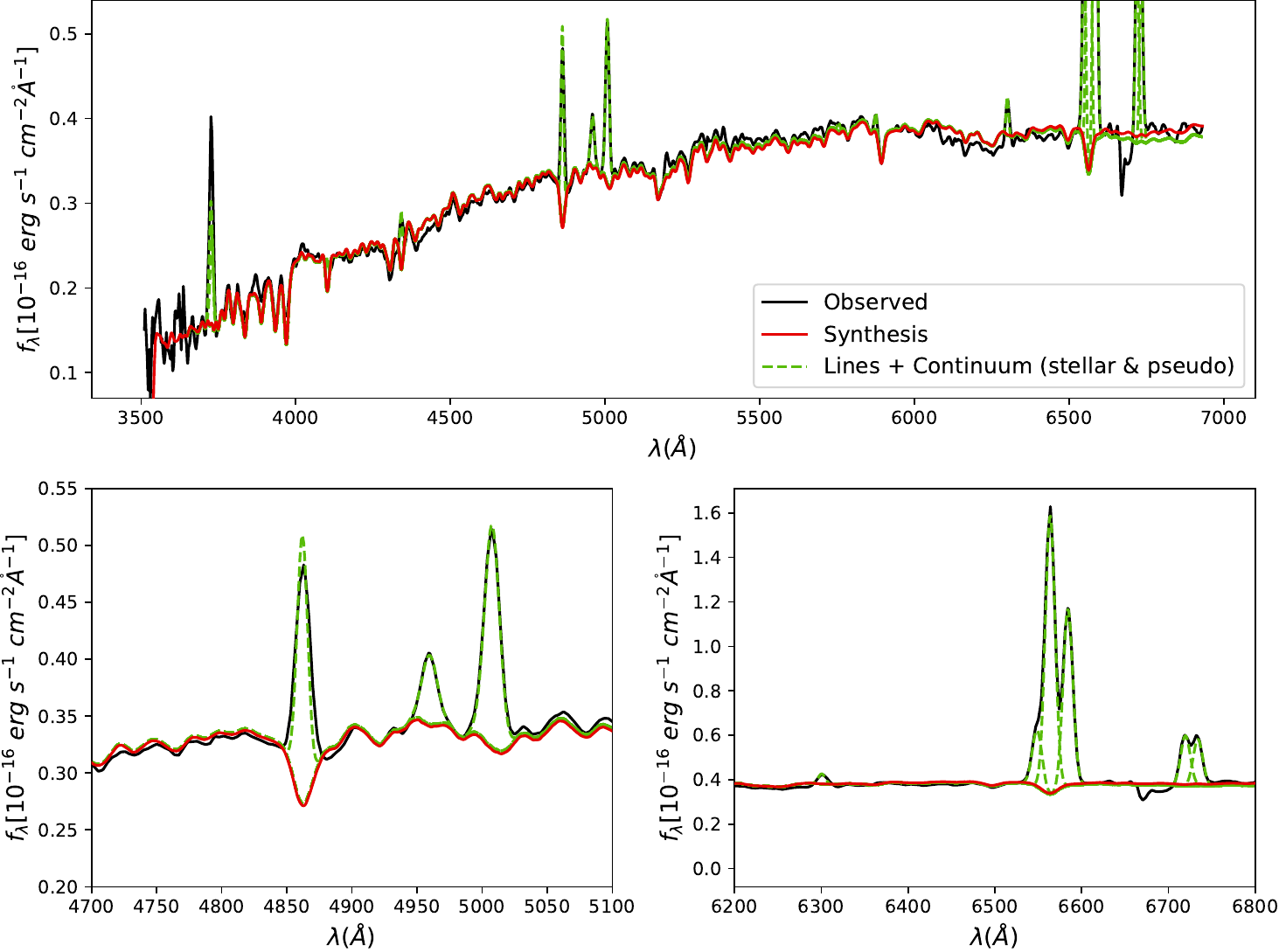}
    \caption{Example of stellar populations synthesis and emission line fitting for the galaxy AM 22217-490 G1. The modelled emission lines can be seen as dashed green lines and the stellar synthesis as a red line. The bottom left and right panels correspond to the spectral regions around H$\beta$ and H$\alpha$, respectively.}
    \label{fig:emissionfit}
\end{figure}

The rest-frame emission-line spectra, free from stellar absorption contribution, have been used to fit the strongest emission lines in the optical region. To this purpose we use the {\sc ifscube}\footnote{https://github.com/danielrd6/ifscube} Python package \citep{Ruschel-Dutra+20,Ruschel-Dutra+21}. The fitted emission lines are [O\,{\sc ii}]$\lambda\lambda$3726,3729, H$\delta$, H$\gamma$, [O\,{\sc iii}]$\lambda$4363, H$\beta$, [O\,{\sc iii}]$\lambda\lambda$4959,5007, He\,{\sc i}$\lambda$5876, [O\,{\sc i}]$\lambda$6300, H$\alpha$, [N\,{\sc ii}]$\lambda\lambda$6548,6583, and [S\,{\sc ii}]$\lambda\lambda$6716,6731. The emission-line profiles are fitted with Gaussians by adopting the following constraints: (i) the width and centroid velocities of emission lines from the same parent ion are constrained to the same value; (ii) the centroid velocity is allowed to vary from $\pm$350 km\,s$^{-1}$ relative to the velocity obtained from the redshift of each galaxy; (iii) if necessary a broad component was added in the case of the HI lines to account for the broad line region emission. In addition, we include a first-order polynomial to reproduce the local continuum. We also applied a cut in signal-to-noise, requiring that the emission line amplitudes must be larger than the noise level computed in the line region after the fit in order to have reliable emission line measurements. The emission line flux errors were estimated using the empirical relation described in \citep{Lenz+92,Wesson+16}. An example of the emission line fitting is shown in Fig.~\ref{fig:emissionfit}.

The extinction correction was made using Balmer decrement, considering case B for recombination, with $N_e$ = 100 $cm^{-3}$ and $T_e$ = 10 000 K, given an intrinsic ratio of H$\alpha$/H$\beta$ = 2.86 \citep{Osterbrock+06} and the extinction law of \cite{Cardelli+89}. In summary, the adopted ionized gas optical extinction was computed as equation 7 of \cite{Riffel+21}. The extinction corrected emission lines fluxes for the INT and CTR galaxies are listed in Tables~\ref{tab:INTforbflux},~\ref{tab:INTHflux},~\ref{tab:CTRforbflux}, and ~\ref{tab:CTRHflux} in Appendix~\ref{AP_A}. 
\section{Stellar populations analysis}\label{section:stellarpop}

\begin{figure*}
    \centering
    \includegraphics[width=2\columnwidth]{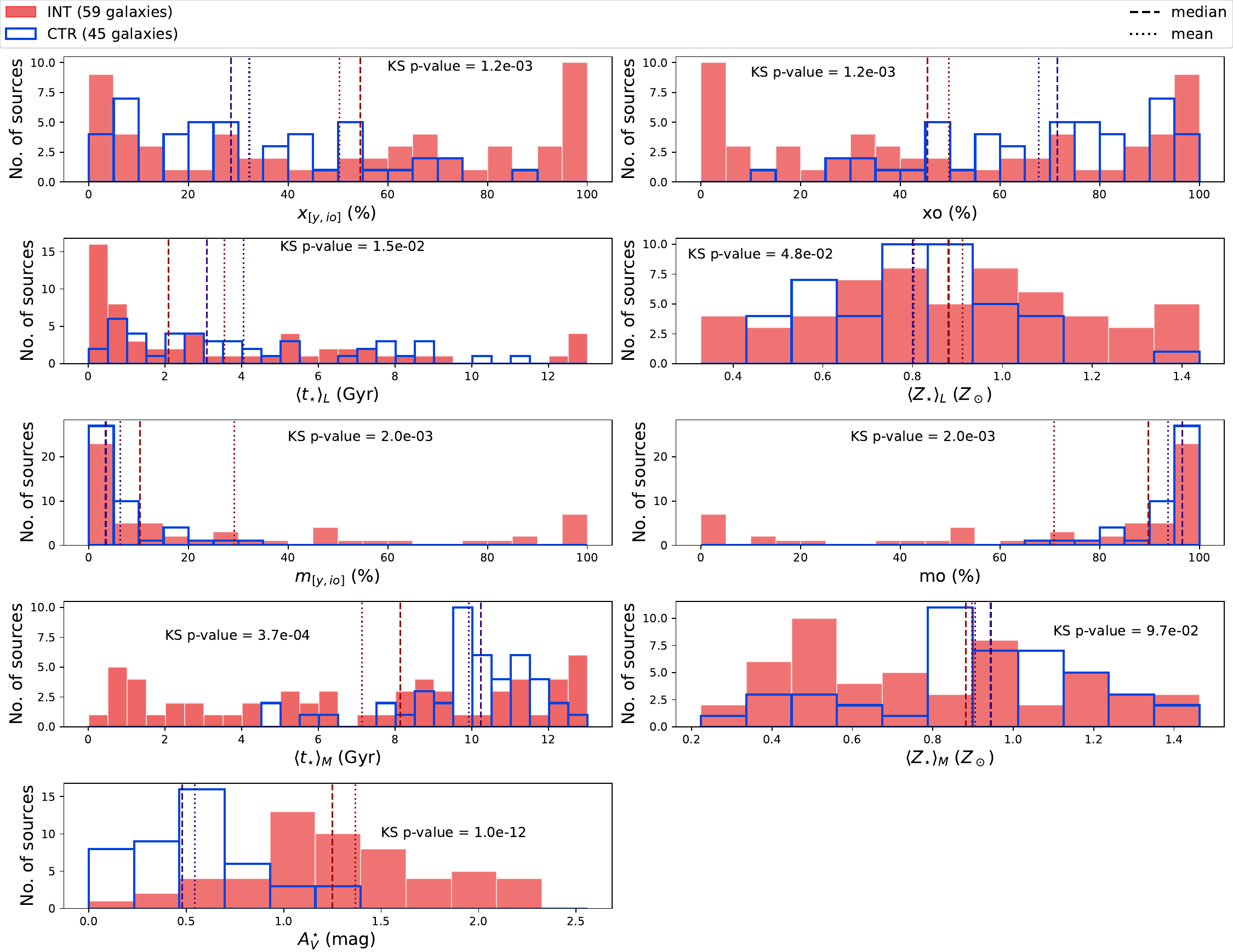}
    \caption{Histograms of the stellar populations synthesis modelling for INT (filled red bars) and CTR (full blue lines) samples. The histograms also show the mean (dotted vertical lines) and median (dashed vertical lines) values for both samples and the KS test p-value. The first two rows of histograms present the light-weighted properties. They present the sum of light fractions younger than 2 Gyr ($x_{[y,io]}$ = xy+xiy+xio), the light fractions in the older stellar populations (xo), the light-weighted mean ages, and mean metallicities. The following two rows present the mass-weighted results of the same properties. The lower right histogram shows the stellar extinction of the samples.} 
    \label{fig:stellarpop}
\end{figure*}

\begin{table*}[ht]
\centering
\small
\caption{Stellar population synthesis main properties derived for the INT and CTR samples. The columns list the mean ($\bar{p}$), standard deviation ($\sigma$), median, and quartile deviation (Q.D) for both samples. The last two columns correspond to the fraction between the median values of the INT and CTR sample and the KS test p-values.}
\label{tab:stellar_populations}
\begin{tabular}{ccccccc}
\hline
\hline
& \multicolumn{2}{c}{INT (59 galaxies)}       & \multicolumn{2}{c}{CTR (45 galaxies)}       & \multicolumn{2}{c}{INT/CTR} \\
Property            & $\bar{p}~\pm~\sigma$ &$median \pm  Q.D$& $\bar{p}~\pm~\sigma$ &$median \pm  Q.D$& Fraction      & KS p-value     \\
\hline
xy (\%) & 20 $\pm$ 24 & 9 $\pm$ 13 & 10 $\pm$ 9 & 8 $\pm$ 5 & 1.11 & 5.4e-02 \\
xiy (\%) & 5 $\pm$ 9 & 0 $\pm$ 4 & 8 $\pm$ 9 & 3 $\pm$ 8 & 0.01 & 4.5e-02 \\
xio (\%) & 25 $\pm$ 25 & 16 $\pm$ 22 & 14 $\pm$ 13 & 10 $\pm$ 10 & 1.63 & 1.0e-01 \\
$x_{[y,io]}$~(\%)\tablefootmark{a} & 50 $\pm$ 36 & 55 $\pm$ 35 & 32 $\pm$ 22 & 28 $\pm$ 18 & 1.91 & 1.2e-03 \\
xo (\%) & 50 $\pm$ 36 & 45 $\pm$ 35 & 68 $\pm$ 22 & 72 $\pm$ 18 & 0.64 & 1.2e-03 \\
$\langle t_{\star} \rangle_{L}$ & 9.1 $\pm$ 0.9 & 9.3 $\pm$ 0.5 & 9.4 $\pm$ 0.4 & 9.5 $\pm$ 0.3 & 0.98 & 1.5e-02 \\
$\langle Z_{\star} \rangle_{L}$ $(Z_\odot)$ & 0.9 $\pm$ 0.3 & 0.88 $\pm$ 0.19 & 0.8 $\pm$ 0.2 & 0.8 $\pm$ 0.1 & 1.1 & 4.8e-02 \\
my (\%) & 1 $\pm$ 4 & 0.2 $\pm$ 0.6 & 0.2 $\pm$ 0.5 & 0.1 $\pm$ 0.1 & 2.12 & 1.0e-02 \\
miy (\%) & 4 $\pm$ 10 & 0 $\pm$ 1 & 1 $\pm$ 2 & 0.5 $\pm$ 0.9 & 0.02 & 7.2e-02 \\
mio (\%) & 24 $\pm$ 30 & 8 $\pm$ 20 & 5 $\pm$ 7 & 2 $\pm$ 3 & 4.65 & 1.3e-03 \\
$m_{[y,io]}$~(\%)\tablefootmark{a} & 29 $\pm$ 35 & 10 $\pm$ 24 & 6 $\pm$ 8 & 3 $\pm$ 3 & 3.02 & 2.0e-03 \\
mo (\%) & 71 $\pm$ 35 & 90 $\pm$ 24 & 94 $\pm$ 8 & 97 $\pm$ 3 & 0.93 & 2.0e-03 \\
$\langle t_{\star} \rangle_{M}$ & 9.7 $\pm$ 0.4 & 9.9 $\pm$ 0.3 & 10.0 $\pm$ 0.1 & 10.01 $\pm$ 0.04 & 0.99 & 3.7e-04 \\
$\langle Z_{\star} \rangle_{M}$ $(Z_\odot)$ & 0.9 $\pm$ 0.4 & 0.88 $\pm$ 0.33 & 0.9 $\pm$ 0.3 & 0.94 $\pm$ 0.2 & 0.93 & 9.7e-02 \\
$A_{V}^{\star}$ (mag) & 1.4 $\pm$ 0.6 & 1.3 $\pm$ 0.4 & 0.5 $\pm$ 0.3 & 0.5 $\pm$ 0.2 & 2.6 & 1.0e-12 \\
\hline
\end{tabular}%
\tablefoot{ 
\tablefoottext{a}{$x(m)_{[y,io]}$ is the sum of x(m)y, x(m)iy, and x(m)io. This is the sum of the contributions of stellar populations younger than 2 Gyr.}
}
\end{table*}

Since small differences in the stellar population ages and metallicities are difficult to characterize, to analyse the stellar synthesis output properly, here we use the following binned population vectors \citep[see][for more details]{CidFernandes+05,Riffel+09}: 

\begin{description} 
     \item[xy (my) - young stellar populations:] Light (mass) binned population vector in the age range  $t \leq$ 100~Myr;
     
     \item[xiy (miy) - intermediate young stellar populations:] Light (mass) binned population vector in the age range 100~Myr $< t \leq$ 700~Myr;
     
     \item[xio (mio) - intermediate old stellar populations:] Light (mass) binned population vector in the age range 700~Myr $< t \leq$ 2.0~Gyr;
     
     \item[xo (mo) - old stellar populations:] Light (mass) binned population vector in the age range 2.0~Gyr $< t \leq$ 15~Gyr.
\end{description}

Additionally, following \citet[][]{CidFernandes+05}, we have also computed the mean logarithm of the ages and the mean metallicities. The mean logarithm of the ages weighted by the stellar light (hereafter light-weighted mean age) and weighted by the stellar mass (hereafter mass-weighted mean age) are defined as follows: 
\begin{equation}\langle t_{\star} \rangle_{L} = \displaystyle \sum^{N_{\star}}_{j=1} x_j {\rm log}t_j, \end{equation} 
\begin{equation}\langle t_{\star} \rangle_{M} = \displaystyle \sum^{N_{\star}}_{j=1} \mu_j {\rm log}t_j. \end{equation}
Futhermore, the light-weighted and mass-weighted mean metallicities are defined as:  
\begin{equation}\langle Z_{\star} \rangle_{L} = \displaystyle \sum^{N_{\star}}_{j=1} x_j Z_j,\end{equation}
\begin{equation}\langle Z_{\star} \rangle_{M} = \displaystyle \sum^{N_{\star}}_{j=1} \mu_j Z_j.\end{equation} 

Note that the definitions above are limited by the ages and metallicities of our stellar population base, as described in Sect.~\ref{section:stellarfitting}.

The results of the stellar populations synthesis modelling are summarized in the histograms shown in Fig.~\ref{fig:stellarpop} and Table~\ref{tab:stellar_populations}. In Table~\ref{tab:stellar_populations}, besides presenting the mean and median properties of the INT and CTR samples, we also show the fraction of the medians of the INT and CTR sample and the p-values of corresponding KS tests. We reject the null hypothesis that the two samples were drawn from the same distribution if the p-value is $<$ 0.01.

The $x_{[y,io]}$\footnote{$x_{[y,io]}$ = $xy+xiy+xio$. The $x_{[y,io]}$ is the sum of light-fraction contributions of stellar populations younger than 2 Gyr.} and xo histograms in Fig.~\ref{fig:stellarpop} show a substantial difference between the samples stellar populations. The INT galaxies present higher mean and median contribution from $x_{[y,io]}$ than CTR galaxies, as can be also seen in Table~\ref{tab:stellar_populations}. This population dominates ($x_{[y,io]}>$ 50\%) in more than half of the INT sample (median INT $x_{[y,io]}$ of 55 $\pm$ 35 \%). The INT median contribution of $x_{[y,io]}$ is nearly twice that of the CTR sample (28 $\pm$ 18 \%). On the other hand, 3/4 of the CTR sample galaxies are dominated by the old stellar populations (median CTR xo of 72 $\pm$ 18 \%). Furthermore, the INT light-weighted mean ages have a median value of 1 Gyr younger than the CTR sample (see Table~\ref{tab:stellar_populations} and Fig.~\ref{fig:stellarpop}). These findings suggest that the INT sample has a larger content of stellar populations younger than 2 Gyr than the CTR sample.

These differences between the properties of the stellar populations of the INT and CTR samples are also seen in the mass-weighted properties. Table~\ref{tab:stellar_populations} shows that the INT median $m_{[y,io]}$\footnote{$m_{[y,io]}$ = $my+miy+mio$. The $m_{[y,io]}$ is the sum of mass-fraction contributions of stellar populations younger than 2 Gyr.} is more than three times the CTR, albeit with a large scatter. This result is mainly driven by the intermediate old population (mio), which exhibits an INT median mass fraction contribution nearly five times higher than in the CTR. Consequently, the INT mass-weighted mean ages show a median value 2 Gyr younger than the CTR sample.

The samples also show differences in the mean stellar metallicities of the stellar populations. As seen in Fig.~\ref{fig:stellarpop} and Table~\ref{tab:stellar_populations}, the light-weighted mean metallicities present similar distributions, with the INT sample presenting a slightly higher median value than the CTR. When considering the mass-weight mean metallicities, the opposite occurs, with the INT median value being smaller than that of the CTR sample. According to the KS test however, we cannot rule out that they come from the same distribution.

 The property that more clearly distinguishes the samples is the stellar extinction ($A_{V}^{\star}$). Fig.~\ref{fig:stellarpop} and Table~\ref{tab:stellar_populations} show that the INT sample distribution is shifted towards high $A_{V}^{\star}$ values compared to the CTR sample. The median INT sample $A_{V}^{\star}$ is 2.6 times higher than in the CTR. Besides the higher mean and median values, the KS test confirms that the distributions are unlikely to come from the same initial one.

As seen in Table~\ref{tab:stellar_populations} and Fig.~\ref{fig:stellarpop}, the INT sample often presents much wider distributions (greater standard and quartile deviations\footnote{The quartile deviation is defined as half of the interquartile range (IQR), used as a measure of the dispersion for median values \citep{Kenney+39}.}) in stellar populations properties compared to the CTR sample, showing that the INT galaxies span a wider range of stellar population properties.

Finally, we did the test of analysing the stellar population properties by removing the 14 INT galaxies without a CTR (see Section~\ref{section:samples} for more details on the CTR sample selection), and found no significant changes in the overall distributions, median, and mean values.

\section{Reddening effects}\label{section:reddening}

\begin{figure*}
    \centering
\includegraphics[width=1.07\columnwidth]{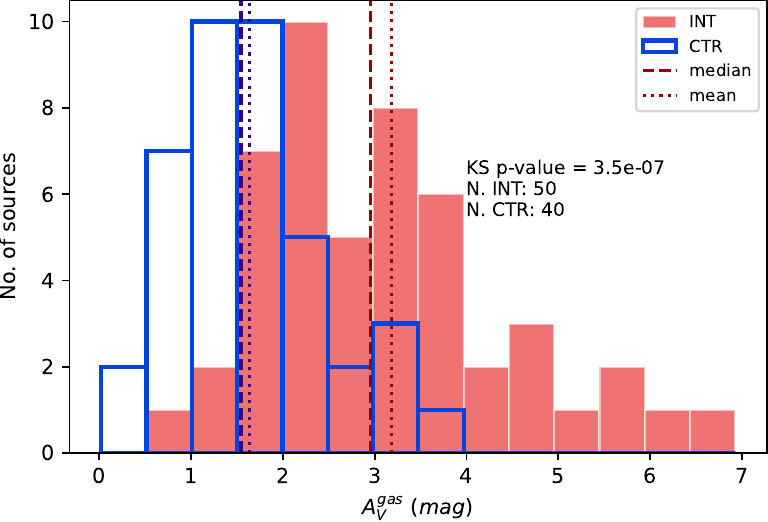}
\includegraphics[width=0.7\columnwidth]{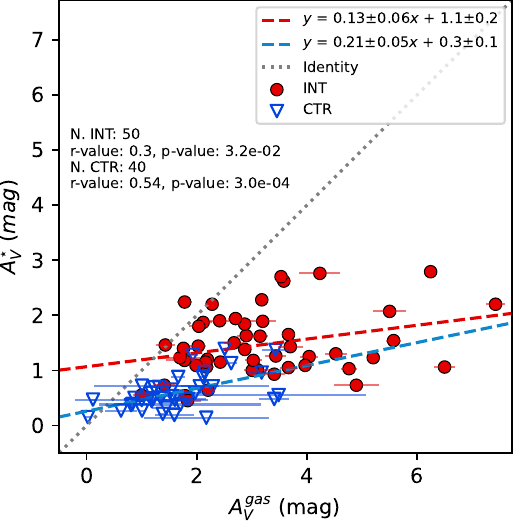}
    \caption{Left-panel: Histogram of the gas extinction obtained via Balmer decrement for the INT (in red) and CTR sample galaxies (in blue). The mean and median values of both samples as in Fig.~\ref{fig:starforming}. Right-panel: Relation between stellar extinction ($A_{V}^{\star}$) and nebular extinction ($A_{V}^{gas}$) for the INT (red circles) and CTR (blue triangles) samples. Dashed lines show the best linear regression fit obtained in both cases. The linear equations, Pearson correlation coefficients (r-value) and p-values of the fits are indicated in the inset. The grey dotted line corresponds to the case when $A_{V}^{gas}$ = $A_{V}^{\star}$.}
    \label{fig:extinction}
\end{figure*}

\begin{table*}[ht]
\centering
\small
\caption{Ionized gas and stellar extinction properties of the INT and CTR samples. The measurements were computed for galaxies where an extinction calculation was possible. The columns include the mean ($\bar{p}$), standard deviation ($\sigma$), median, and quartile deviation (Q.D) for both the INT and CTR samples. The last two columns correspond to the fraction between the median values of the INT and CTR sample and the KS test p-values.}
\label{tab:Gas}
\begin{tabular}{ccccccc}
\hline \hline
                    & \multicolumn{2}{c}{INT (50 galaxies)}       & \multicolumn{2}{c}{CTR (40 galaxies)}       & \multicolumn{2}{c}{INT/CTR} \\
Property            & $\bar{p}~\pm~\sigma$ &$median \pm  Q.D$& $\bar{p}~\pm~\sigma$ &$median \pm  Q.D$& Fraction      & KS p-value     \\
\hline
$A_{Vgas}$ (mag) & 3 $\pm$ 1 & 3.0 $\pm$ 0.8 & 1.6 $\pm$ 0.8 & 1.5 $\pm$ 0.4 & 1.91 & 3.5e-07 \\
$A_{Vgas}$/$A_{V*}$ & 2 $\pm$ 1 & 2.0 $\pm$ 0.9 & 3 $\pm$ 3 & 2.5 $\pm$ 0.7 & 0.8 & 4.4e-02 \\
\hline
\end{tabular}%
\end{table*}

In the left panel of Fig.~\ref{fig:extinction} we present the $A_{V}^{gas}$ distribution, which is the reddening obtained via the H$\alpha$ over H$\beta$ emission line flux ratios. As shown in Table~\ref{tab:Gas}, the nebular extinction of the INT sample has higher mean and median values than the CTR. The median ionized gas extinction of the INT sample (median $\pm$ Q.D. = 3.0 $\pm$ 0.8 mag) is nearly twice the one on the CTR sample (median $\pm$ Q.D. = 1.5 $\pm$ 0.4 mag). The KS test p-value suggests that the samples are unlikely to come from the same original distribution.

On the right panel of Fig.~\ref{fig:extinction}, we show $A_{V}^{gas}$ vs $A_{V}^{\star}$ plot. The INT sample presents a higher dispersion on both extinctions than the CTR. Consequently, the CTR shows a stronger correlation between the two quantities than the INT galaxies. 
Some INT galaxies appear to be offset towards high $A_{V}^{\star}$ and closer to the 1:1 relation. This could suggest that some of these galaxies have a significant fraction of stars in more gas-rich and star-forming regions. This will be further explored in Sec.~\ref{disc:stellargasprop}.     
\begin{figure*}[htbp]
    \centering
\includegraphics[width=12cm]{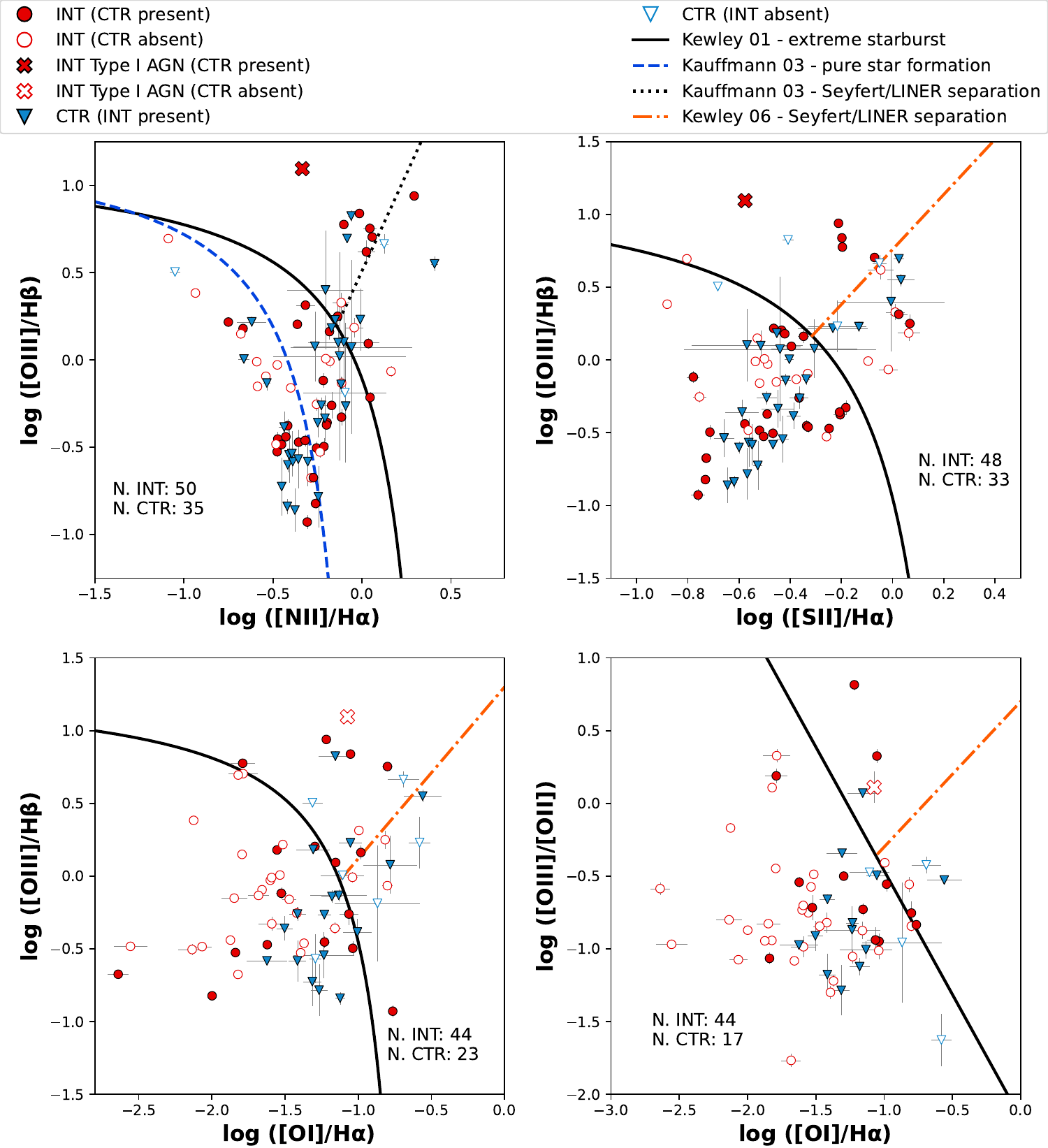}   
    \caption{Emission-line diagnostic diagrams used to compare the ionization mechanisms of the INT galaxies (red points) and the CTR sample (blue triangles). Filled symbols correspond to INT (CTR) galaxies whose corresponding CTR (INT) galaxy is also included on that diagram, and open symbols when it is not. Lines in the diagrams represent the theoretical \citep{Kewley+01} and empirical \citep{Kauffmann+03, Kewley+06} divisions applied in this work to classify the ionization sources. Each diagram shows the number of sources included in each diagram. The red cross marks the position on the diagrams of the only type I AGN found in the INT sample.}
    \label{fig:BPT}
\end{figure*}

\section{Excitation mechanisms}\label{section:excitation}

\begin{figure}
\centering
\includegraphics[width=\columnwidth]{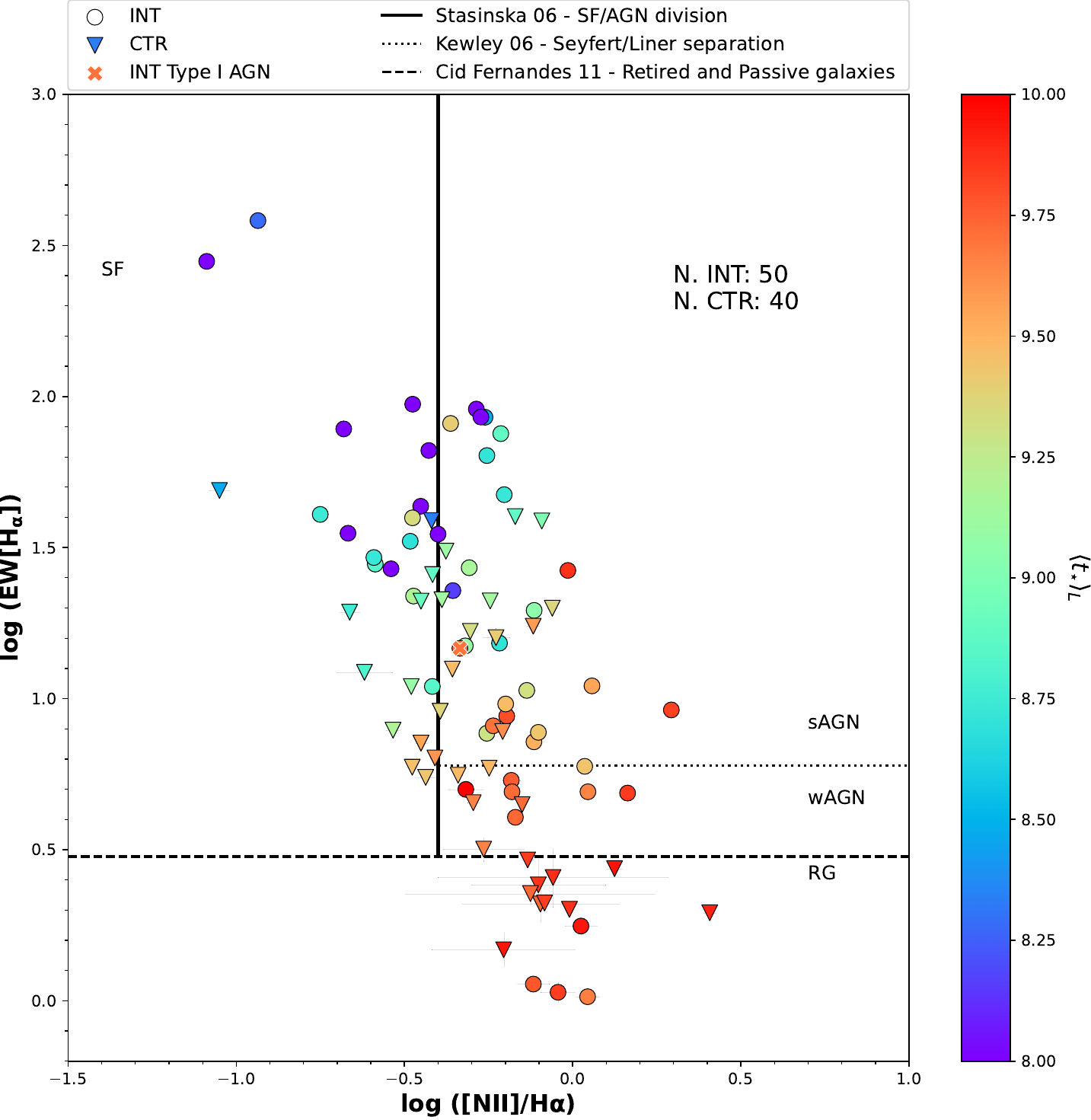}   
\caption{The diagram of [NII]/H$\alpha$ versus the equivalent width of H$\alpha$ (WHAN diagram, \citealt{Cid+10,Cid+11}). INT galaxies are shown as circles, and CTR galaxies are shown as triangles. Symbol colours indicate the light-weighted mean ages of the stellar populations, showing the youngest mean ages in the star-forming galaxies and the oldest in the retired galaxies. Lines represent the divisions applied in this work to obtain the sources of ionization \citep{Stasinska+06,Kewley+06, Cid+11}. The cross marks the position on the diagram of the only type I AGN found in the INT sample.}
\label{fig:WHAN}
\end{figure}

Emission-line diagnostic diagrams are essential tools used to characterize the main ionization mechanisms of the gas. We applied them to compare the sources of ionization that drive the emission on the INT and CTR samples. Figure~\ref{fig:BPT} presents the traditional BPT diagram (hereafter, [NII] BPT), followed by the diagrams involving [SII]$\lambda\lambda$ 6717, 6731$~\mathring{A}$ (hereafter, [SII] BPT), [OI]$\lambda$ 6300$~\mathring{A}$ (hereafter, [OI] BPT) and [OII]$\lambda\lambda$ 3727, 3729$~\mathring{A}$ (hereafter, [OII] BPT) \citep{BPT+81,Veilleux+87}. We also use the WHAN diagram \citep{Cid+10,Cid+11} presented in Fig.~\ref{fig:WHAN}. In all the diagrams, we plot empirical and theoretical curves \citep{Kewley+01,Kauffmann+03,Kewley+06,Stasinska+06,Cid+11} to enable us to classify the main sources of ionization, which is indicated in Tables~\ref{tab:INT_bptclass} and~\ref{tab:CTR_bptclass} in Appendix~\ref{AP_A}.


In Fig.~\ref{fig:NucAct}, we show the histograms with the fractions of activity type classification derived from the diagrams in Figs.~\ref{fig:BPT} and \ref{fig:WHAN}. In all diagrams of Fig.~\ref{fig:BPT}, the fraction of galaxies for which we could not measure one or more emission lines is higher in the CTR sample. On the BPT diagram, 22\% (10/45 sources) of CTR galaxies were not classified, against 15\% (9/59 sources) of the INT sample. A similar fraction was found on the [SII] BPT. In the last two diagrams, the absence of classification is higher. The CTR galaxies present a fraction of 49\% (22/45 objects) and 62\% (28/45 objects) of galaxies unclassified in the [OI] BPT and [OII] BPT, respectively. On the same diagrams, the fraction of INT galaxies that are not classified are 25\% in both (15/59 sources).

The WHAN diagram, which includes a higher number of CTR galaxies than the BPTs since several galaxies with H$\alpha$ equivalent width (EW) < 3 $\mathring{A}$ may lack the other emission lines involved in the latter, suggests an excess of retired galaxies 
in the CTR sample, as seen in Figs.~\ref{fig:WHAN} and \ref{fig:NucAct}. In this diagram, 33\% (15/45 galaxies) of the CTR sources were classified as retired, while in the INT sample the fraction was 22\% (13/59 galaxies).         

The histograms of Fig.~\ref{fig:NucAct} show that more INT galaxies are classified as star-forming than in the CTR. In the BPTs and WHAN diagrams, between 27-59\% (16-35/59 galaxies) of the INT sample are classified as star-forming, whilst in the case of the CTR sample, they represent between 20-55\% (9-25/45 sources). The difference is more evident in the [OI] and [OII] BPT diagrams, where 54\% and 63\% (32 and 37 out of 59 galaxies) of the INT sample, respectively, are classified as star-forming versus 33\% and 31\% (15 and 14 out of 45 galaxies) of them in the CTR sample. Our findings suggest that interactions may be responsible for the enhancement in star formation in the INT sample.

When considering the AGN classification in Fig.~\ref{fig:NucAct}, the Seyfert or strong AGN (sAGN) classification of the INT sources outnumbers the CTR ones in all five diagrams. The diagrams with the highest excess are the [NII] BPT and WHAN, where the INT sample contains 14-39\% (8-23/59 sources) of AGN, respectively, versus 9\%-29\% (4-13/45 galaxies) of the CTR sample. Although the difference in the fraction of AGN is small, it indicates that nuclear activity might be enhanced due to galaxy interactions in the INT sample.

\begin{figure*}
    \centering
\includegraphics[width=1.5\columnwidth]{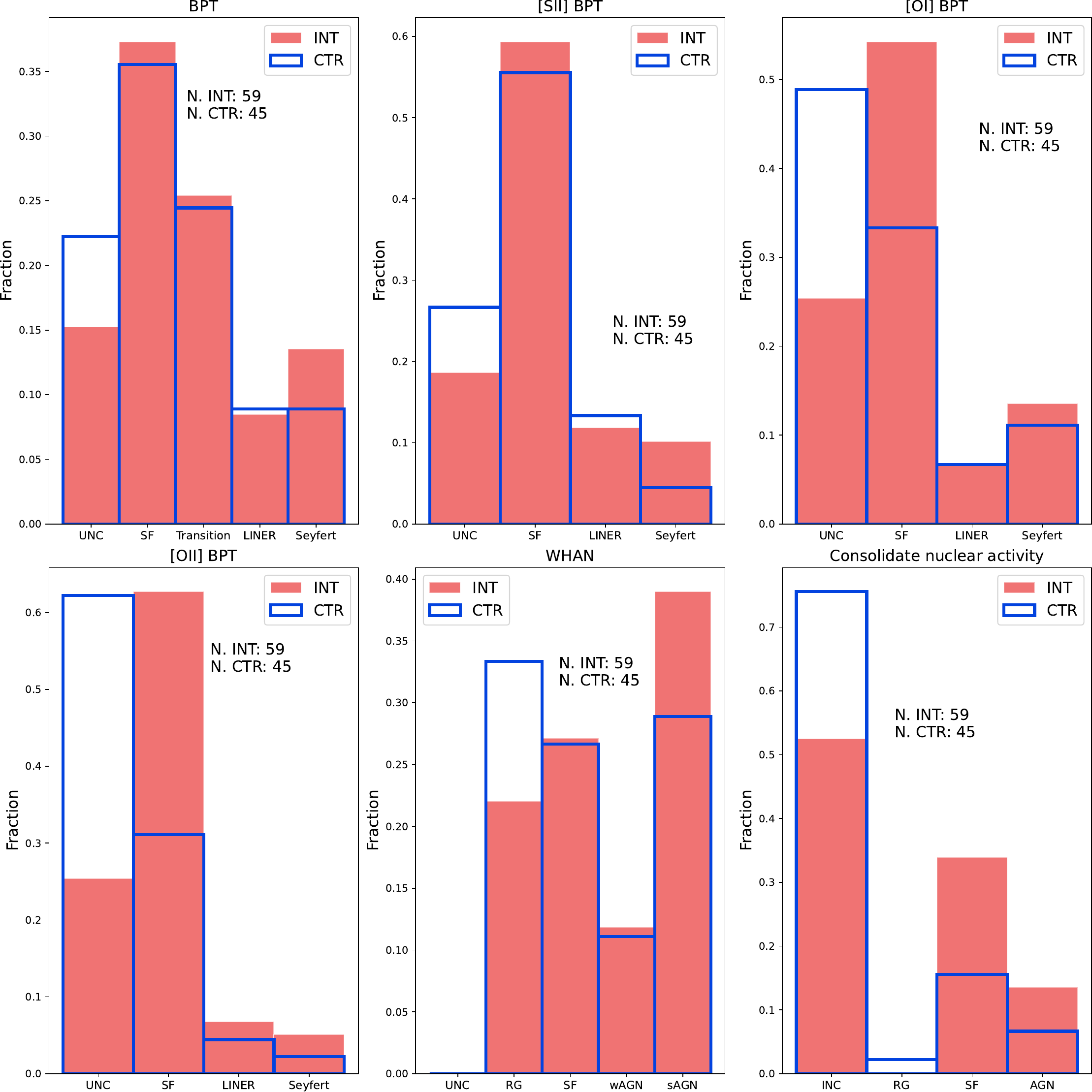}   
    \caption{Histograms of the dominant gas ionization mechanisms found from each diagnostic diagram. The three top and the bottom left histograms show the fractions of galaxies classified from the BPT diagrams as star-forming (SF), AGN (LINER or Seyfert) and composite ([NII] BPT). For the WHAN diagram, the galaxies are classified as retired (RG), SF, strong-AGN (sAGN), and weak-AGN (wAGN). Galaxies that could not be classified in a particular diagram are shown in the histograms as unclassified. The bottom right histogram summarizes the information obtained from all the diagrams, which we divided into four categories: SF, AGN, RG, and inconclusive (INC). Galaxies included in the SF, AGN, and RG groups are those having the same classification on at least four out of the five BPT and WHAN diagrams.}
    \label{fig:NucAct}
\end{figure*}

To find a consensus about the dominant source of ionization of the central region of the INT and CTR galaxies, all five diagrams were combined on the bottom right histogram of Fig.~\ref{fig:NucAct}. This final classification was done by considering only galaxies with consistent classification in at least four out of the five diagrams. Then, these galaxies were classified into one of the following groups:

\begin{enumerate}
    \item Star-forming (SF): Galaxies classified as star-forming.
    \item Retired galaxies (RG): Galaxies classified as retired in the WHAN diagram and as LINER in at least three BPT diagrams.
    \item AGN: Galaxies classified as LINER, Seyfert, wAGN, sAGN or composite.
\end{enumerate}

It is important to note that this is a conservative approach since we require the same classification in four out of five diagrams to classify a galaxy as SF, RG, or AGN. The remaining galaxies are classified as inconclusive (INC) in the bottom right histogram of Fig.~\ref{fig:NucAct}. As can be seen in the first rows of Table~\ref{tab:nuclear_activity}, we find that 47\% (28/59 objects) of the INT sample and 24\% (11/45 galaxies) of the CTR sample could be included in each of the three categories listed above. Only one RG galaxy was consistently found in the CTR sample. 34\% (20/59 sources) of the INT sample and 16\% (7/45 objects) of the CTR sample were consistently classified as SF. These results suggest that, even after using our restrictive criteria, there is an excess of 18\% of SF galaxies in the INT sample relative to the CTR sample. Regarding the AGN classification, we found that they are more common in the INT sample (8/59 sources, 14\%) than in the CTR sample (3/45 sources, 7\%). 


\begin{table}[!ht]
\renewcommand{\tabcolsep}{.50mm}
\tiny
\caption{The classification of the galaxies in the INT and CTR samples using four distinct combinations of diagnostic diagrams. The INT and CTR columns include the percentual and the number of galaxies between parentheses.}
\label{tab:nuclear_activity}
\begin{tabular}{cccc}
\hline \hline
Properties & $\%$ INT (N gal.) & $\%$ CTR (N gal.) & $\%$ INT/$\%$ CTR \\
\hline
\bf 4/5 DGs\tablefootmark{a} &  &  &  \\
INC & 53 (31) & 76 (34) & 0.7 \\
RG & 0 (0) & 2 (1) & 0.0 \\
SF & 34 (20) & 16 (7) & 2.18 \\
AGN & 14 (8) & 7 (3) & 2.03 \\
\bf 3/5 DGs\tablefootmark{b} &  &  &  \\
INC & 22 (13) & 49 (22) & 0.45 \\
RG & 2 (1) & 7 (3) & 0.25 \\
SF & 54 (32) & 38 (17) & 1.44 \\
AGN & 22 (13) & 7 (3) & 3.31 \\
\bf $[NII]+[SII]$+WHAN\tablefootmark{c} &  &  &  \\
INC & 58 (34) & 73 (33) & 0.79 \\
RG & 2 (1) & 4 (2) & 0.38 \\
SF & 25 (15) & 18 (8) & 1.43 \\
AGN & 15 (9) & 4 (2) & 3.43 \\
\bf $[NII]$+WHAN\tablefootmark{d} &  &  &  \\
INC & 31 (18) & 51 (23) & 0.6 \\
RG & 2 (1) & 9 (4) & 0.19 \\
SF & 27 (16) & 20 (9) & 1.36 \\
AGN & 41 (24) & 20 (9) & 2.03 \\
\hline
\end{tabular}
\tablefoot{All the classifications require agreement in:
\tablefoottext{a}{at least four out of the five diagnostic diagrams}
\tablefoottext{b}{at least three out of the five diagnostic diagrams}
\tablefoottext{c}{the [NII], [SII], and WHAN diagrams}
\tablefoottext{d}{the [NII] BPT and WHAN diagrams.}
}
\end{table}

Relaxing our criteria can reduce the number of galaxies without classification and increase the number of galaxies in both the INT and CTR samples. Table~\ref{tab:nuclear_activity} presents the consistent classifications found using three out of the five diagrams (3/5 DGs), [NII] and [SII] BPTs combined to the WHAN diagram ([NII]+[SII]+WHAN), and [NII] BPT with WHAN diagram ([NII]+WHAN). We were able to find a classification for up to 78\% (46/59 galaxies) of the INT sources and 51\% (23/45 galaxies) of the CTR sources. We find that the INT sample has between 1.4 to 2.2 times higher fraction of star-forming galaxies than the CTR sample (0.9 to 1.7 if we exclude the 14 INT galaxies without CTR galaxies). The fraction of AGN in the INT sample is between 2 and 3.4 times greater than the CTR sample (2.3 to 4 if we exclude the 14 INT galaxies without CTR galaxies). Although the fractions change with the criteria and depending on whether we consider INT galaxies with controls or not, we always obtain an excess of AGN galaxies in the INT sample relative to the CTR sample, showing the reliability of this result. 

In order to further test how strong the previous results are, we only considered INT galaxies whose CTR is also classified in the same diagram(s). This has the problem that the number of galaxies is then reduced (between 4 and 19 in the four combinations of diagrams reported in Table \ref{tab:nuclear_activity}). By doing this, we measure a fraction of AGN between 1.6 and 2.3 times greater in the INT sample than the CTR sample, confirming the previous result. However, for the fraction of SF the trend reverts, and we find between 1.2 and 1.4 times higher fraction in the CTR sample than in the INT sample. Therefore, based on the analysis done here, we conclude that the number of AGN in the interacting galaxies is larger than in the non-interacting ones, while for the SF galaxies we cannot either confirm or discard any trend.



\section{Do interactions enhance star-formation?}\label{section:starforming}

\begin{figure*}
 \centering
 \includegraphics[width=\columnwidth]{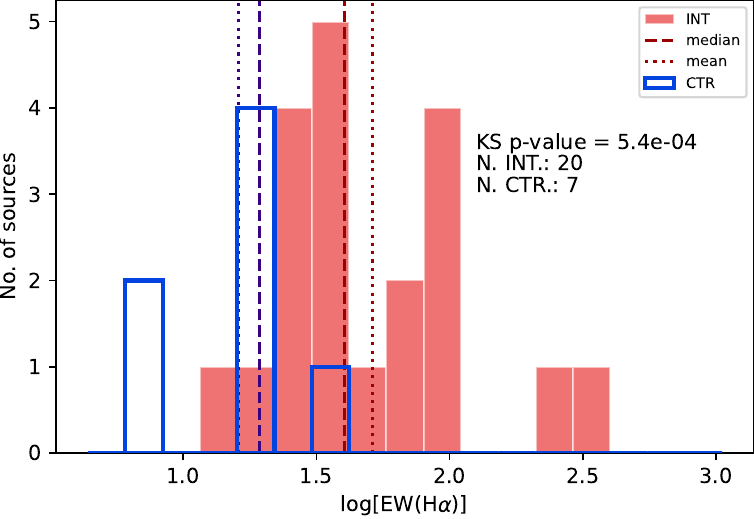} 
\includegraphics[width=\columnwidth]{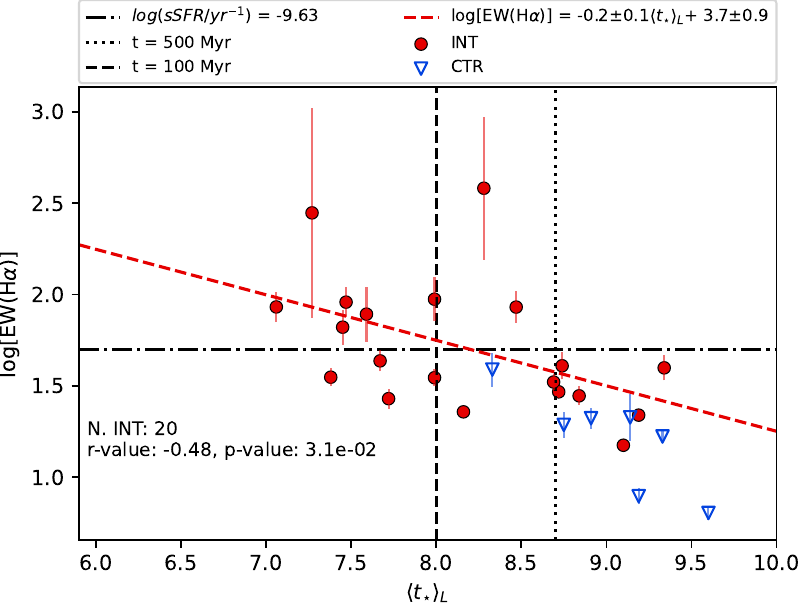}   
    \caption{Left panel: Distribution of EW(H$\alpha$) for the bona-fide SF galaxies. Right panel: The relation between light-weighted mean ages $\langle t_{\ast}\rangle_L$ and H$\alpha $ equivalent width for SF galaxies. SF INT galaxies are shown as red circles, and the SF CTR as blue triangles. The red dashed line corresponds to the linear regression fit obtained for the SF INT galaxies. The linear equation, Pearson correlation coefficient (r-value), and p-value of the fit are provided in the inset. The black dashed and dotted lines indicate the mean stellar ages of 100 Myr and 500 Myr, respectively. The dot-dashed line indicates an EW(H$\alpha$) = 50 $\mathring{A}$ or log(sSFR/yr$^{-1})$ = -9.63.}
    \label{fig:starforming}
\end{figure*}

To evaluate the strength of star formation in both samples, here we consider the INT and CTR sample galaxies classified as SF in the bottom right histogram of Fig.~\ref{fig:NucAct} (34\% of the INT and 16\% of the CTR sample).

\begin{table*}[ht]
\centering
\small
\caption{Ionized gas and stellar population properties of SF galaxies of the INT and CTR samples, as classified in Sec.~\ref{section:excitation}. The columns include the mean ($\bar{p}$), standard deviation ($\sigma$), median, and quartile deviation (Q.D) for both the INT and CTR samples. The last two columns correspond to the fraction between the median values of the INT and CTR sample and the KS test p-values.}
\label{tab:SFgas}
\begin{tabular}{ccccccc}
\hline \hline
& \multicolumn{2}{c}{SF INT (20 galaxies)}       & \multicolumn{2}{c}{SF CTR (7 galaxies)}       & \multicolumn{2}{c}{INT/CTR} \\
Property            & $\bar{p}~\pm~\sigma$ &$median \pm  Q.D$& $\bar{p}~\pm~\sigma$ &$median \pm  Q.D$& Fraction      & KS p-value     \\
\hline
\multicolumn{7}{l}{\bf Ionized gas}\\
EW(H$\alpha$) ($\mathring{A}$) & 77 $\pm$ 90 & 40 $\pm$ 28 & 19 $\pm$ 10 & 19 $\pm$ 4 & 2.08 & 5.4e-04 \\
log(sSFR/yr$^{-1}$)\tablefootmark{a} & -9.6 $\pm$ 0.5 & -9.8 $\pm$ 0.3 & -10.3 $\pm$ 0.3 & -10.2 $\pm$ 0.2 & 0.96 & 5.4e-04 \\
sSFR~(yr$^{-1}$)\tablefootmark{a} & 5e-10 $\pm$ 8e-10 & 2e-10 $\pm$ 2e-10 & 7e-11 $\pm$ 5e-11 & 7e-11 $\pm$ 2e-11 & 2.63 & 5.4e-04 \\
1/sSFR$~$(Gyr)\tablefootmark{a} & 6 $\pm$ 5 & 6 $\pm$ 3 & 26 $\pm$ 20 & 15 $\pm$ 10 & 0.38 & 5.4e-04 \\
\hline
\multicolumn{7}{l}{\bf Stellar population modelling}\\
$\langle t_{\star} \rangle_{L}$\tablefootmark{b} & 8.2 $\pm$ 0.7 & 8.0 $\pm$ 0.6 & 9.0 $\pm$ 0.4 & 9.1 $\pm$ 0.2 & 0.88 & 1.2e-02 \\
SFR$^{20Myr}~$(M$_\odot$/yr)\tablefootmark{b} & 3 $\pm$ 4 & 2 $\pm$ 1 & 0.6 $\pm$ 0.6 & 0.3 $\pm$ 0.4 & 4.85 & 6.3e-02 \\
log(sSFR$^{20Myr}$/yr$^{-1}$)\tablefootmark{b} & -9.3 $\pm$ 0.6 & -9.2 $\pm$ 0.5 & -10.0 $\pm$ 0.4 & -10.0 $\pm$ 0.3 & 0.92 & 1.6e-02 \\
sSFR$^{20Myr}~$(yr$^{-1}$)\tablefootmark{b} & 1e-09 $\pm$ 1e-09 & 6e-10 $\pm$ 8e-10 & 1e-10 $\pm$ 1e-10 & 9e-11 $\pm$ 8e-11 & 6.55 & 1.6e-02 \\
1/sSFR$^{20Myr}~$(Gyr)\tablefootmark{b} & 4 $\pm$ 5 & 2 $\pm$ 2 & 13 $\pm$ 10 & 11 $\pm$ 7 & 0.15 & 1.6e-02 \\
\hline
\end{tabular}%
\tablefoot{ 
\tablefoottext{a}{The specific star formation rate (sSFR) was obtained from the equation $\rm log(sSFR/yr^{-1})=-11.87 + 1.32 \ log(EW(H\alpha)$ from \citet{Belfiore+18}.}\\
\tablefoottext{b}{These properties were obtained from the stellar population synthesis modelling. The star-formation rate over the last 20 Myr (SFR$^{20Myr}$) was calculated following \cite{Riffel+21,Riffel+23}.}
}
\end{table*}

The histogram on the left panel of Fig.~\ref{fig:starforming} shows the distribution of EW(H$\alpha$), of the SF galaxies and Table~\ref{tab:SFgas} summarizes their star formation properties. SF galaxies in the INT sample have significantly higher EW(H$\alpha$) values (median$\pm$ QD. = 40$~\pm~$ 28$~\mathring{A}$) compared to the SF CTR sample (median$\pm$ QD. = 19$~\pm~$4$~\mathring{A}$). This result suggests the presence of younger stellar populations photoionizing the gas and contributing to this enhancement of EW(H$\alpha$) in the INT systems when compared with CTR ones. 

In the right panel of Fig.~\ref{fig:starforming}, we combine the stellar population synthesis modelling results with EW(H$\alpha$). Only SF INT galaxies exhibit light-weighted mean stellar ages younger than 100 Myr and EW(H$\alpha$) $>$ 50$~\mathring{A}$. Furthermore, as shown in Table~\ref{tab:SFgas}, the light-weighted mean stellar ages of SF INT galaxies have a median value of 100 Myr, while the SF CTR galaxies exhibit a median value of 1.2 Gyr. The fit of the SF INT galaxies shown in the right panel of Fig.~\ref{fig:starforming} suggests an anti-correlation (Pearson r = -0.48), which is expected if there is a connection between the enhancement of EW(H$\alpha$) and the age of the stellar populations.

We can use EW(H$\alpha$) to estimate the specific star formation rate (sSFR = $SFR/M_{\star}$). Using equation $\rm log(sSFR/yr^{-1})=-11.87 + 1.32 \ log(EW(H\alpha))$ from \citet{Belfiore+18}, we found, as shown in Table~\ref{tab:SFgas}, that SF INT galaxies have median sSFR 2.6 times higher than SF CTR galaxies. At this rate, SF INT galaxies will need less than half of the time of SF CTR galaxies to double their stellar mass. We also calculated the sSFR (see Table~\ref{tab:SFgas}) from the stellar populations modelling, using the SFR$^{20 Myr}$\footnote{The star-formation rate over the last 20 Myr (SFR$^{20Myr}$) was calculated following \cite{Riffel+21,Riffel+23}.}, and found similar results as with EW(H$\alpha$).

\section{Discussion}
\label{section:disc}
\subsection{The stellar populations and recent star formation of interacting galaxies}\label{disc:stellarpop}

It is well-known that interactions and mergers can induce star formation \citep{Sanders+96,Ellison+13,Knapen+15,Krabbe+17}. We used stellar population synthesis modelling to confirm or discard the presence of recent star formation in our samples. Our findings reveal that more than half of the INT sample is dominated (see Table~\ref{tab:stellar_populations}) by stellar populations younger than 2 Gyr (i.e., higher light-weighted fractions). The median contribution from these young stellar populations in the INT sample is nearly 2 times that of the CTR sample. This difference between the stellar populations of the INT and CTR samples is also detected in the mass-weighted fractions.
 

Regarding the stellar populations light-weighted stellar population fractions, our results agree with the findings of previous studies, mostly done in small samples or individual pairs. For instance, \cite{Krabbe+17} analysed 15 galaxies in 9 close pairs using long-slit spectroscopy. They also used {\sc starlight} but with BC03 models \citep{Bruzual+03}. In agreement with our findings, the stellar population synthesis modelling revealed that 11 out of the 15 galaxies in their sample were dominated in light fractions by young to intermediate stellar populations (< 2 Gyr).

\cite{Bessiere+17} presented optical spectroscopy observations of 21 type-II quasars. Their study investigates the time scales of starburst and AGN triggering associated with merger events. They built stellar population models with {\sc Starburst99} \citep{Leitherer+99} and fitted the spectra using the routine {\sc CONFIT} \citep{Robinson+00}. For most quasars, they were able to fit an 8 Gyr plus a young stellar population (YSP). They found YSP light fractions between 15\%-100\% and maximum YSP ages below 200 Myr. These fractions are much greater than the median values found in our INT sample for our youngest stellar populations. Despite the different models and fitting procedures, most of their quasars are in a later stage in the merging process, and have larger gas masses, both having an impact on star formation.

We also analysed the star formation enhancement using nebular indicators. Since the H$\alpha$ equivalent width is a stellar age indicator in HII regions \citep{Copetti+86,Kennicutt+98, Kennicutt+12,Tacconi+20,Sanchez+20,Sanchez+21} and it is used as a proxy to derive the sSFR, we used it to compare the bona-fide star-forming galaxies in our INT and CTR samples. As shown in Table~\ref{tab:SFgas}, the median EW(H$\alpha$) for SF INT galaxies is 2 times higher than for the CTR sample. Higher equivalent widths in interacting systems were also found in \cite{Donzelli+97}. The authors analysed the spectra of 27 galaxies in pairs and found mean values of the equivalent width of H$\alpha$+[NII] ranging from 37-54 $\mathring{A}$ against a mean of 27 $\mathring{A}$ in visual pairs (i.e., discarded as pairs spectroscopically).

From EW(H$\alpha$) we measure a median sSFR of the SF INT galaxies 2.6 times higher than that of the SF CTR galaxies, which means that at this star formation rate, SF INT galaxies will need less than half of the time of SF CTR galaxies to double their stellar mass. The sSFRs values obtained here are compatible with the ones found in interacting gas-rich spiral galaxies in \cite{Knapen+15}. In that work, the authors analysed 1500 nearby interacting galaxies using estimates of SFR and sSFR from \cite{Querejeta+15} based on far-infrared emission and selected a control sample of isolated galaxies with similar masses and morphologies. They found that the interacting galaxies have SFRs and sSFRs up to twice higher than those measured for the control sample. \cite{Hwang+11} also studied the sSFR in interacting galaxies and found an enhancement in spiral pairs. They found the sSFR to be 1.8 up to 4.0 times higher than in non-interacting galaxies. Their sSFRs are similar to the ones from our work.

\subsection{Stellar and nebular reddening.}\label{disc:stellargasprop}

The analysis of the stellar and ionized gas extinction revealed that the INT sample presents higher median stellar and ionized gas reddening (1.3 $\pm$ 0.4 mag and 3 $\pm$ 1 mag) than the CTR sample (0.5 $\pm$ 0.2 mag and 1.5 $\pm$ 0.4 mag, see Tables~\ref{tab:stellar_populations} and \ref{tab:Gas}). These results are in agreement with the previous ones that found that the line-emitting gas is in dustier regions than the stars \citep{Calzetti+94, Asari+07,Riffel+08,Riffel+09,Riffel+21,Riffel+23,Martins+13,Dametto+14,Krabbe+17}.

\cite{Calzetti+94} interpreted this result as due to the fact that the ionized gas is more often associated with dust than the old stellar populations that largely contribute to the stellar continuum. The extinction ratios (see Table~\ref{tab:Gas}) are consistent with those reported in the literature for star-forming \citep{Calzetti+94, Asari+07, Riffel+08} and AGN \citep{Riffel+21} galaxies, which often report nebular extinctions around twice the stellar. The same was found in interacting pairs \citep{Krabbe+17}. 

The interacting galaxies in our sample exhibit higher median stellar and emission-line gas reddening relative to control galaxies. This result suggests that the interacting galaxies are more frequently dustier. This result agrees with \citet[][see also \citealt{Hwang+11} for more details on the effect of interactions in dust properties.]{Yuan+12} that analyzing GALEX ultraviolet and Spitzer infrared data, found that interacting galaxies, especially spiral pairs, present higher dust attenuation than isolated galaxies. This is also predicted from simulations \citep[e.g.;][]{Yutani+22} where an increase of dust in the central regions of interacting galaxies could, in the late stages of the merging process, lead to dust-obscured galaxies (DOGs).

\begin{figure}
    \centering
    \includegraphics[width=1\columnwidth]{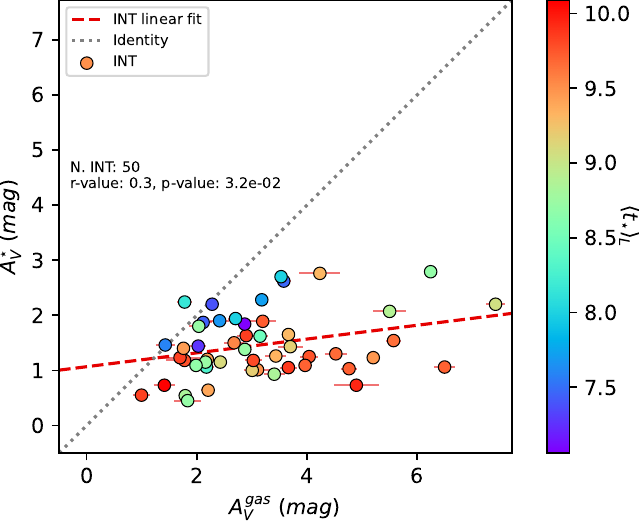}
    \caption{Relation between $A_{V}^{gas}$ versus $A_{V}^{\star}$ for the INT sample. The colors indicate the light-weighted mean ages of the stellar populations. As in Fig.~\ref{fig:extinction}, the plot also includes the linear fit for the INT sample and the 1:1 relation.}
    \label{fig:equalAV}
\end{figure}

Some interacting galaxies in our sample were found closer to the $A_{V}^{gas}$ = $A_{V}^{\star}$ relation shown in Fig.~\ref{fig:extinction} than others. This could indicate that the stellar populations there are associated with dusty star-forming regions. To confirm this hypothesis, in Fig.~\ref{fig:equalAV} we show the relation between the stellar and emission line extinctions in the INT sample, with colours indicating the light-weighted mean ages. The interacting galaxies above the linear fit have median light-weighted mean age of 220 $\pm$ 870 Myr (the median value is 3.1 $\pm$ 2.3 Gyr in the INT galaxies below the linear fit) 
meaning that they have a high amount of very hot massive stars. 
We checked if the distance between the points in Fig.~\ref{fig:equalAV} and the 1:1 relation correlates with their light-weighted mean stellar ages, but we did not find any correlation.


\subsection{Triggering of nuclear activity and star formation in interacting galaxies.}\label{disc:nucact}

In our analysis of the main sources of ionization, five different diagnostic diagrams were applied (see Sec.~\ref{section:excitation}). We combine the results of the five diagnostics and require agreement in their classification on at least four diagrams for the sources to be classified as star-forming (SF), retired galaxies (RG), and AGN. Because of the lack of information in some diagnostics, combined with the inconsistencies between the classifications, it was not possible to classify the dominant ionization mechanism for 34 galaxies (76\%) in the CTR sample and 31 galaxies (53\%) in the INT sample.
Considering the galaxies that we could classify, we measure 34\% of SF, 14\% of AGN, and 0\% RGs in the INT sample and 16\% of SF, 7\% of AGN, and 2\% of RGs in the CTR sample. 

There are similar works in the literature addressing the classification of galaxies in terms of the dominant ionizing source using optical diagnostic diagrams using large survey data, such as the SDSS-III \citep{York+00,Eisenstein+11} and the SDSS-IV integral field spectroscopy (hereafter IFS) survey of Mapping Nearby Galaxies at Apache Point Observatory (MaNGA,~\citealt{Bundy+15}). It is not straightforward to compare with these works since they differ in sample and control sample selection, pair mass ratio and mergers stages, redshifts and analysis methods. Nevertheless, in the next paragraphs, we try to compare our findings with some studies. 

\citet{Sabater+13} studied the dependence of the dominant ionizing source of the gas on the environment and interactions using single fiber observations of a sample of 270.000 galaxies from SDSS-III. The authors used the traditional [NII] BPT diagram to classify their sources and applied a cut on [OIII] luminosity to classify the passive galaxies. They found that the incidence of galaxy interactions decreases from star-forming galaxies to Seyferts to LINERs to passive galaxies.

\citet{Ellison+13} presented a study of 10.800 close pairs and 97 post-mergers using single fiber observations from the SDSS-III. A control sample of isolated galaxies was selected, taking control of the redshift, mass, and density of the environment. As in the previous case, the study used the [NII] BPT diagram, considering as AGN just Seyfert and composite galaxies. They found fractions of star-forming galaxies of 20\% in close pairs (8\% in the controls) and 40\% in post-merger stages (14\% in their controls). They calculated the ratio between the fractions found in the two samples and found that their interacting galaxies present a factor $\sim$~2-3 higher in star-forming galaxies relative to their controls. The authors also found an increasing AGN excess with decreasing pair separation, presenting a fraction of 2.5 times more AGN in close pairs and 3.75 in post-mergers relative to the control galaxies.

\cite{Jin+21} selected a sample of 1156 galaxies in pairs and 2317 isolated galaxies from the SDSS-IV MaNGA survey. They separated the interacting galaxies in different stages of the merging process and used the [NII] and [SII] BPTs. They used the WHAN diagram only for removing the retired galaxies from their samples. Galaxies were classified as AGN if they fell in the Seyfert or LINER regions of either of the BPT diagrams. They reported 5.5\% of AGN, 6.6\% of composite galaxies, 27.8\% of SF, 28.1\% of RGs, and 32\% of lineless galaxies in the sample of pairs. For the controls, they find 5\% of AGN, 10.2\% of composite galaxies, 42.3\% of SF, 26.7\% of RG and 15.8\% of lineless galaxies. The authors argue that they found no significant AGN excess in the pairs compared to the controls. In close pairs in strong interactions, however, they reported a fraction of 19.5\% (AGN + composite) against 15.2\% in the isolated sample.

\cite{Steffen+23} used the last SDSS-IV MaNGA data release and found 391 pairs (213 minor and 178 major mergers). The control sample was built from 7811 MaNGA galaxies without companions with $\Delta$v < 2000 km/s. The dominant ionization source was classified based on the [NII] BPT and on the WHAN diagram to remove the retired galaxies but with a more restricted cutoff in EW(H$\alpha$) of 6 $\mathring{A}$. The galaxy pairs exhibit 13.4\% of AGN (62 composites, 18 LINERs, and 25 Seyferts) versus 11.1\% in their control sample (613 composites, 70 LINERs, 126 Seyferts, and 63 broad line AGN).


In our work, we find that the INT sample has between 1.4 and 2.2 times higher fraction of star-forming galaxies than the controls, depending on the diagnostic diagrams used to classify the galaxies as star-forming (see Section \ref{section:excitation}). These fractions are compatible with those reported by \citet{Ellison+13}. However, this excess of SF galaxies in the INT sample decreases when we exclude the 14 INT galaxies without controls (fractions from 0.9 to 1.7) and it becomes a deficit if we only consider INT (CTR) galaxies with having their twin CTR (INT) galaxy classified in the same diagram (between 1.2 and 1.4 times higher fractions of SF galaxies in the CTR than in the INT sample).

\begin{figure*}
    \centering
    \includegraphics[width=14cm]{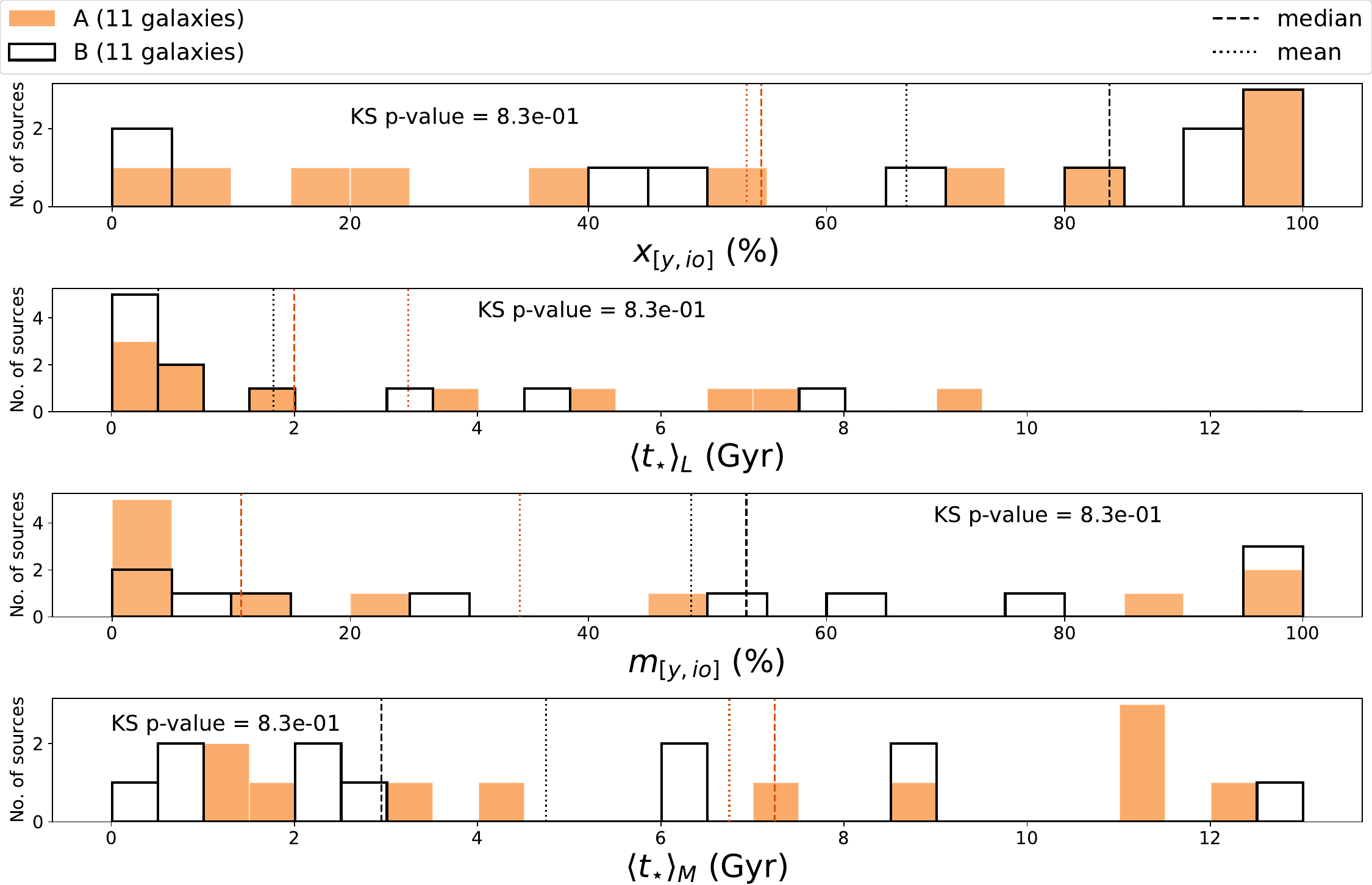}
    \caption{Histograms of the light and mass fractions younger than 2 Gyr ($x_{[y,io]}$ and $m_{[y,io]}$) and light- and mass-weighted mean stellar ages ($\langle t_{\star} \rangle_{L}$ and $\langle t_{\star} \rangle_{M}$) of the A (filled histograms) and B members (empty histograms) of INT pairs with $M[A]>2 M[B]$. The histograms also show the mean (dotted vertical lines), median (dashed vertical lines), and KS p-values.}
    \label{fig:Mover2m}
\end{figure*}

Considering the bona-fide 
AGN that we classified here (including LINER/wAGN, Seyfert/sAGN and Composite galaxies), we found a fraction of 14\% (8/59 objects). 
This is similar to the fractions reported in the works by \cite{Jin+21} (AGN + transition galaxies: 12\%) and \cite{Steffen+23} (13\%).
Using different diagnostic diagrams (BPTs and WHAN) we find between 2 and 3.4 times more AGN in the INT sample than in the CTR sample, which agrees with the factor of 2.5 reported by \cite{Ellison+13} for close pairs. Even when we restrict the sample to include 1) only INT galaxies with CTR and 2) INT (CTR) galaxies having their twin CTR (INT) galaxy classified in the same diagram, we still find a higher fraction of AGN in the INT sample relative to the CTR sample, ranging from 1.6 to 4. 

\subsection{Interacting pairs properties}

Many studies found younger stellar populations in the smaller member of galaxy pairs (e.g., \citealt{Krabbe+08,Krabbe+11,Alonso-Herrero+12}). In order to check if this is also the case for our INT pairs, we calculated their mass ratios using the stellar masses obtained from the stellar population modelling and selected the ones where one of the galaxies is more than twice the mass of the other (M$_*^A$ $>$ 2 M$_*^B$). We found that 11 of the 18 pairs with spectra satisfy this criterion. Figure~\ref{fig:Mover2m} shows the sum of light and mass fractions younger than 2 Gyr ($x_{[y,io]}$ and $m_{[y,io]}$) and the light- and mass-weighted mean stellar ages ($\langle t_{\star} \rangle_{L}$ and $\langle t_{\star} \rangle_{M}$) of the A and B members of the pairs.  
We find that the least massive galaxies in the pairs (B) have a median $x_{[y,io]}$ = 84 $\pm$ 26\% and $\langle t_{\star} \rangle_{L}$ = 8.7 $\pm$ 0.5, whilst the most massive (A) have $x_{[y,io]}$ = 54 $\pm$ 35\% and $\langle t_{\star} \rangle_{L}$ = 9.1 $\pm$ 0.7. In mass fractions, for the B galaxies in the pairs we measure a median $m_{[y,io]}$ = 53 $\pm$ 40\% and $\langle t_{\star} \rangle_{M}$ = 9.5 $\pm$ 0.4, and for the A galaxies a median $m_{[y,io]}$ = 11 $\pm$ 32\% and $\langle t_{\star} \rangle_{M}$ = 9.9 $\pm$ 0.3. Therefore, we find a higher fraction of young stellar populations (ages $\le$ 2 Gyr) in the least massive galaxies of galaxy pairs with M$_*^A$ $>$ 2 M$_*^B$, although we cannot ensure that the distributions are different based on the KS-test p-values that we find (see Fig. \ref{fig:Mover2m}).

Several publications also reported that both star formation and nuclear activity are enhanced with decreasing pair separation (e.g., \citealt{Ellison+13,Satyapal+14,Steffen+23}). Therefore we measured the projected separation between the two members of the 18 pairs using the r' band acquisition images (see Figs. \ref{fig:pairsimages} and \ref{fig:pairsimages2}) and looked for possible trends in light- and mass-weighted stellar ages, SFRs, and/or the different BPT and WHAN diagrams considered in previous sections. Although in the case of our pairs we do not find any trend with projected distance between the pair members, we believe that this is due to the narrow range that they span, between 10 and 90 kpc, with 32 out of the 36 galaxies having projected separations $\le$ 40 kpc (i.e., close pairs; \citealt{Ellison+13}).


\section{Summary and conclusions}
\label{sec:conclusions}

In this work we analysed spectroscopic observations of 95 galaxies from a parent sample of 100 galaxies in 43 interacting system candidates from the \cite{ArpMadore+87} catalogue. The selected galaxies are pairs and small groups. Most of them had no previous spectroscopic data. 

We found that 60 galaxies are in real interacting systems, mainly in galaxy pairs and groups. A control sample of isolated galaxies was built using SDSS spectra of galaxies matched in redshift, absolute J magnitude, and galaxy morphology to the interacting sample. We compared the stellar populations and ionized gas properties of the two samples, INT and CTR, and found the following.

\begin{enumerate}
    \item More than half of the INT sample is dominated in light fractions by stellar populations younger than 2 Gyr. Interacting galaxies present a median fraction of these young stellar populations of 55\%, compared to 28\% found in the CTR sample. This difference is also found in mass fractions: 10\% in the INT sample versus 3\% in the CTR sample.  
    \item We find higher stellar and nebular extinction in the interacting galaxies relative to the control sample. The median $A_{V}^{gas}$ and $A_{V}^{\star}$ of the interacting sample are 2 and 2.6 times the ones found in the control sample. This suggests that our interacting galaxies are dustier.
    \item We find $A_{V}^{gas}$ > $A_{V}^{\star}$ in both the INT and CTR samples, although some INT galaxies 
    have $A_{V}^{gas} \approx A_{V}^{\star}$. These galaxies are the ones having the youngest stellar populations in the INT sample, which could 
 explain their high $A_{V}^{\star}$ values. 
    \item We found an excess of AGN in the INT sample relative to the CTR sample. Depending on the diagnostic diagrams employed and the number of INT and CTR sample galaxies considered, the fraction of AGN in the INT sample relative to the CTR sample ranges between 1.6 and 4. 

    \item Considering the galaxy pairs in the INT sample having M$_*^A > 2\times$M$_*^B$, we find that the least massive galaxies in the pairs present higher fractions of stellar populations younger than 2 Gyr than the most massive galaxies. 
    
\end{enumerate}  

Our study provides further observational evidence that interactions enhance star formation and nuclear activity in galaxies and therefore can have a significant impact on galaxy evolution.


    
\begin{acknowledgements}
     PHC and CRA acknowledge support from project ``Tracking active galactic nuclei feedback from parsec to kiloparsec scales'', with reference PID2022-141105NB-I00. PHC thanks to Conselho Nacional de Desenvolvimento Científico e Tecnológico (CNPq). RR acknowledges support from the Fundaci\'on Jes\'us Serra and the Instituto de Astrof{\'{i}}sica de Canarias under the Visiting Researcher Programme 2023-2025 agreed between both institutions. RR, also acknowledges support from the ACIISI, Consejer{\'{i}}a de Econom{\'{i}}a, Conocimiento y Empleo del Gobierno de Canarias and the European Regional Development Fund (ERDF) under grant with reference ProID2021010079, and the support through the RAVET project by the grant PID2019-107427GB-C32 from the Spanish Ministry of Science, Innovation and Universities MCIU. This work has also been supported through the IAC project TRACES, which is partially supported through the state budget and the regional budget of the Consejer{\'{i}}a de Econom{\'{i}}a, Industria, Comercio y Conocimiento of the Canary Islands Autonomous Community. RR also thanks to Conselho Nacional de Desenvolvimento Cient\'{i}fico e Tecnol\'ogico  ( CNPq, Proj. 311223/2020-6,  304927/2017-1, 400352/2016-8, and  404238/2021-1), Funda\c{c}\~ao de amparo \`{a} pesquisa do Rio Grande do Sul (FAPERGS, Proj. 16/2551-0000251-7 and 19/1750-2), Coordena\c{c}\~ao de Aperfei\c{c}oamento de Pessoal de N\'{i}vel Superior (CAPES, Proj. 0001). ACK thanks Fundação de Amparo à Pesquisa do Estado de São Paulo (FAPESP) for the support grant 2020/16416-5 and the Conselho Nacional de Desenvolvimento Científico e Tecnológico (CNPq). The authors thank Jose Andres Hernandez Jimenez for helpful discussions. The authors thank the anonymous referee for useful and constructive suggestions.
\end{acknowledgements}

%
%
\bibliographystyle{aa}
\bibliography{References}
\begin{appendix} 
\section{INT and CTR samples additional material\label{AP_A}}
The additional material in this appendix comprises the complementary images and spectra of the INT sample presented in Figs.~\ref{fig:pairsimages2} and ~\ref{fig:pairsspectra2}. Furthermore, the following pages present tables of the INT and CTR samples. Table~\ref{tab:Sample} shows the information of the initial interacting candidates selected from the \cite{ArpMadore+87} catalogue. Table~\ref{tab:radial_vel} presents the confirmed INT galaxies radial velocities and their corresponding CTR galaxies selected from the SDSS. Tables~\ref{tab:INTforbflux} and ~\ref{tab:INTHflux} show the fluxes of the INT sample forbidden and permitted emission lines, respectively. Tables~\ref{tab:CTRforbflux} and ~\ref{tab:CTRHflux} present the same for the CTR sample. Finally, Tables~\ref{tab:INT_bptclass} and ~\ref{tab:CTR_bptclass} present the classifications with different diagnostic diagrams for the INT and CTR samples, respectively. 

    \begin{figure*}[ht]
    \centering
    \includegraphics[width=0.64\columnwidth]{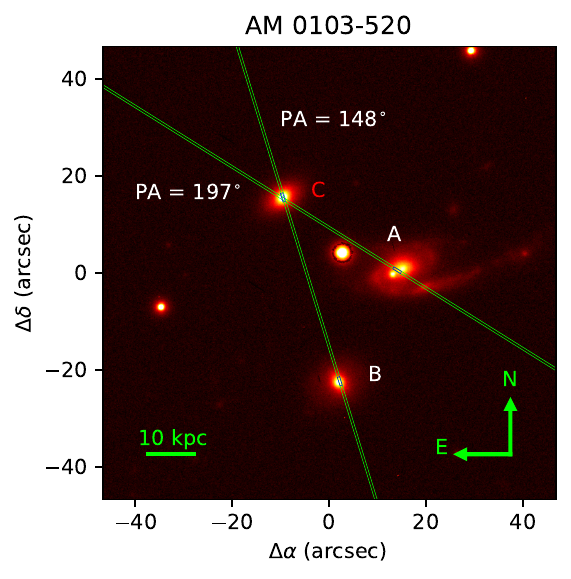}
    \includegraphics[width=0.64\columnwidth]{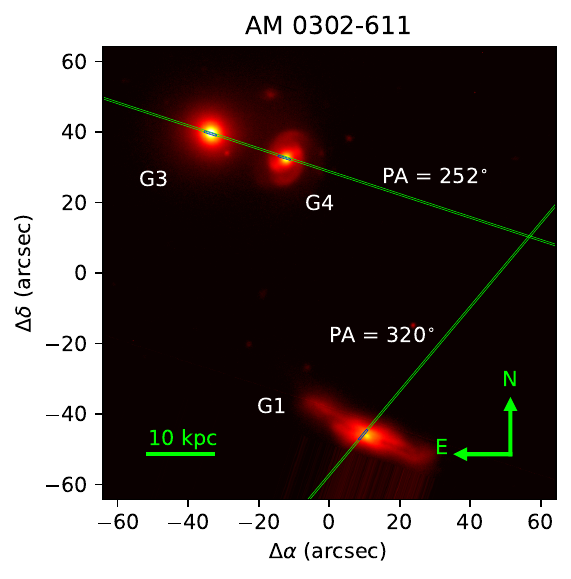}
    \includegraphics[width=0.64\columnwidth]{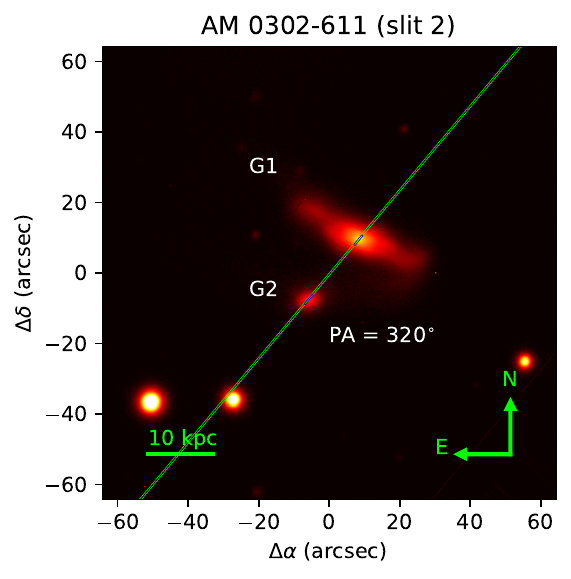}
    \includegraphics[width=0.64\columnwidth]{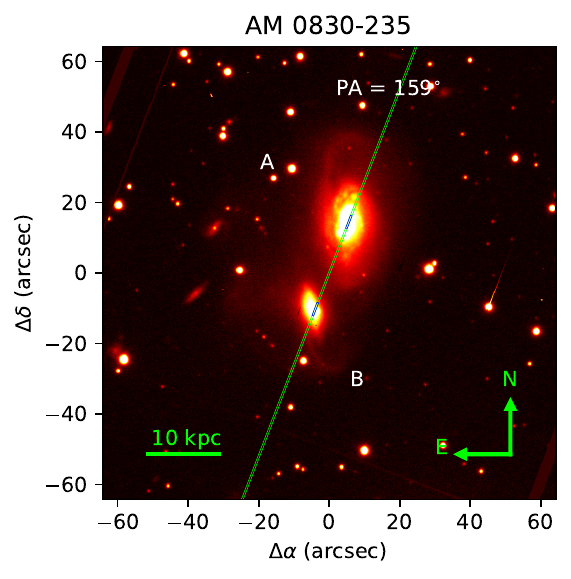}
    \includegraphics[width=0.64\columnwidth]{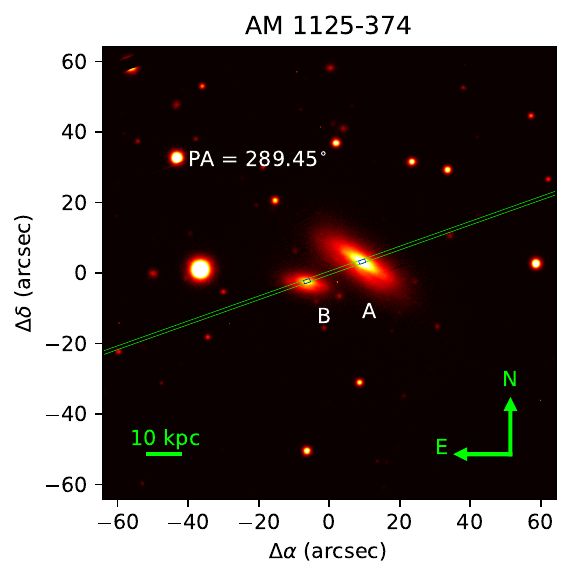}
    \includegraphics[width=0.64\columnwidth]{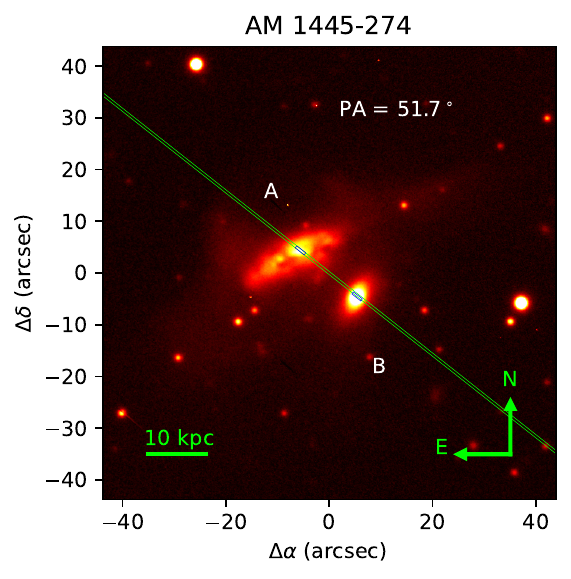}
    \includegraphics[width=0.64\columnwidth]{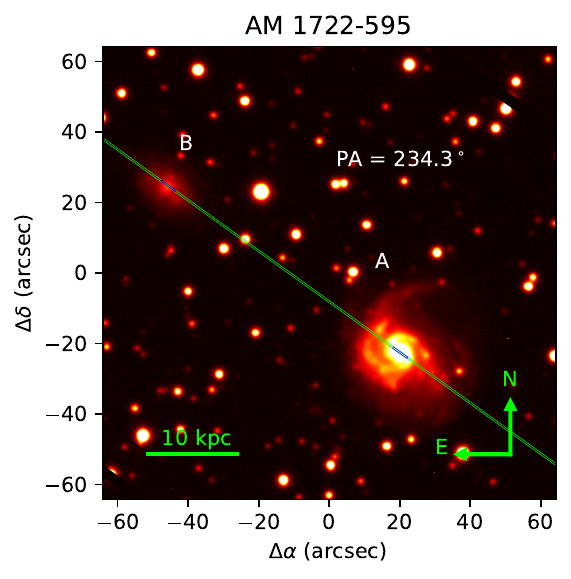}
    \includegraphics[width=0.64\columnwidth]{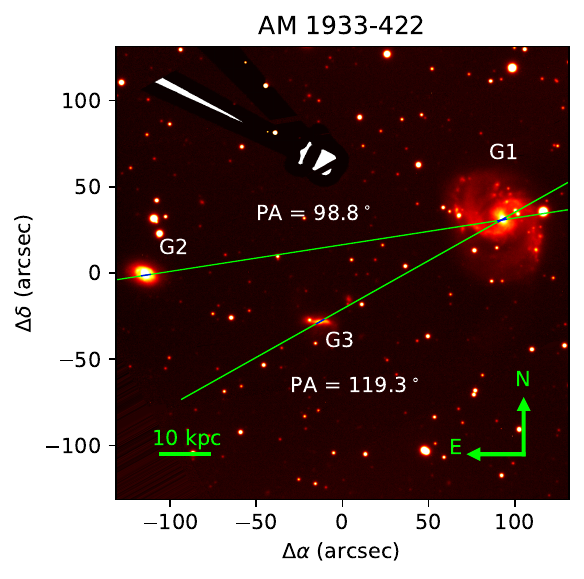}
    \includegraphics[width=0.64\columnwidth]{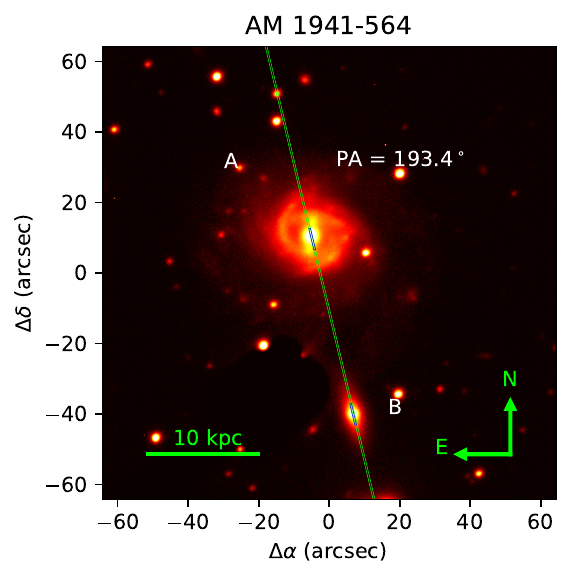}
    \caption{Same as in Fig.~\ref{fig:pairsimages}, but for the INT galaxies not included there.}
    \label{fig:pairsimages2}
\end{figure*}

\setcounter{figure}{0}
    \begin{figure*}[ht]
    \centering
    \includegraphics[width=0.64\columnwidth]{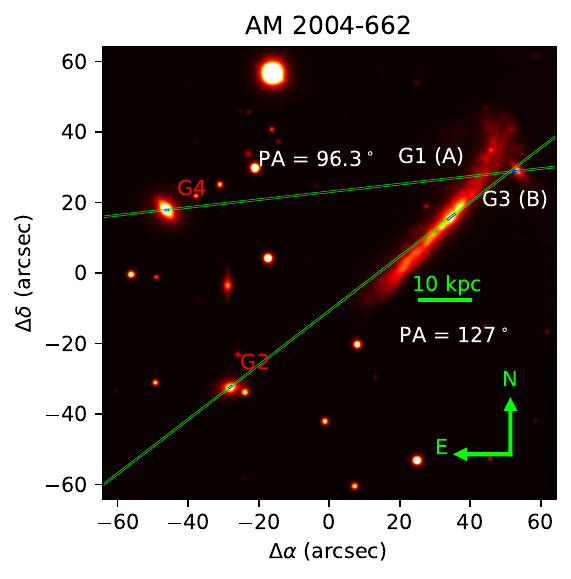}
    \includegraphics[width=0.64\columnwidth]{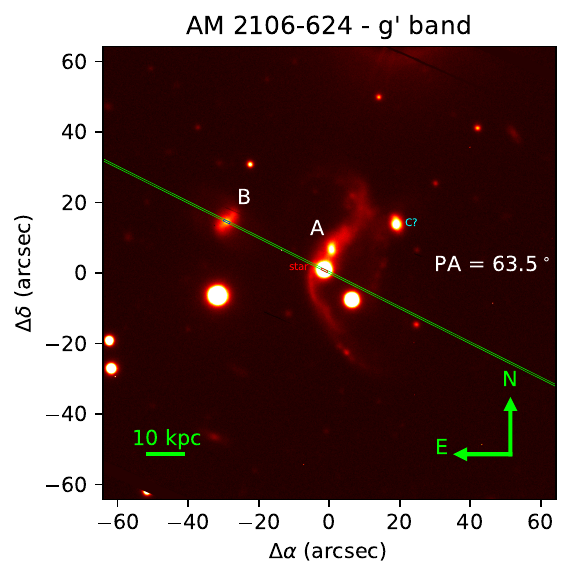}
    \includegraphics[width=0.64\columnwidth]{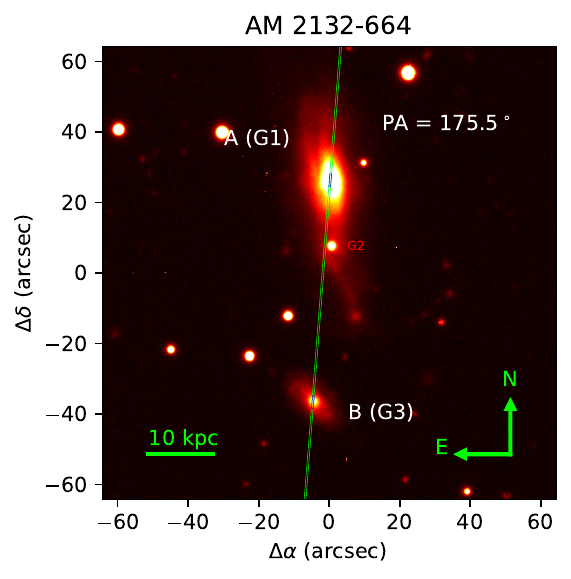}
    \includegraphics[width=0.64\columnwidth]{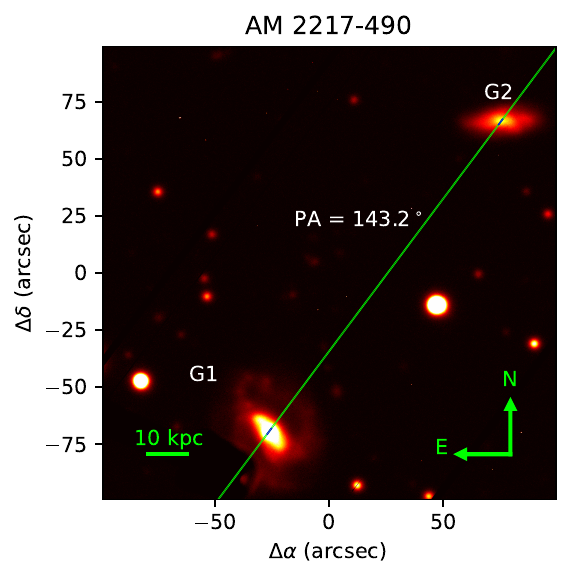}
    \includegraphics[width=0.64\columnwidth]{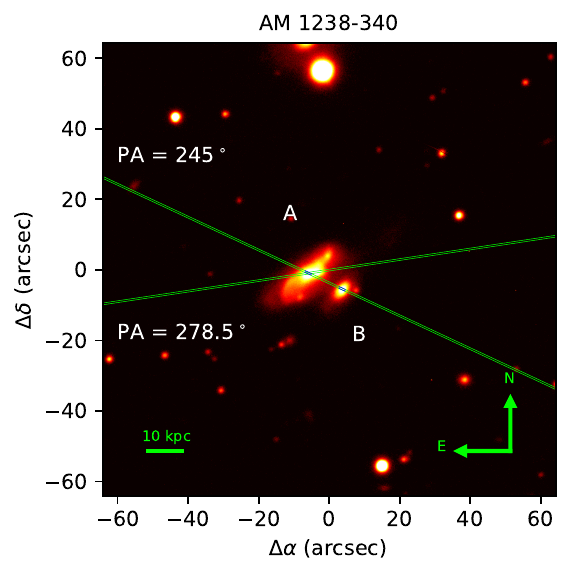}
    \caption{continued.}
\end{figure*}

\begin{figure*}
    \centering
    \includegraphics[width=0.97\columnwidth]{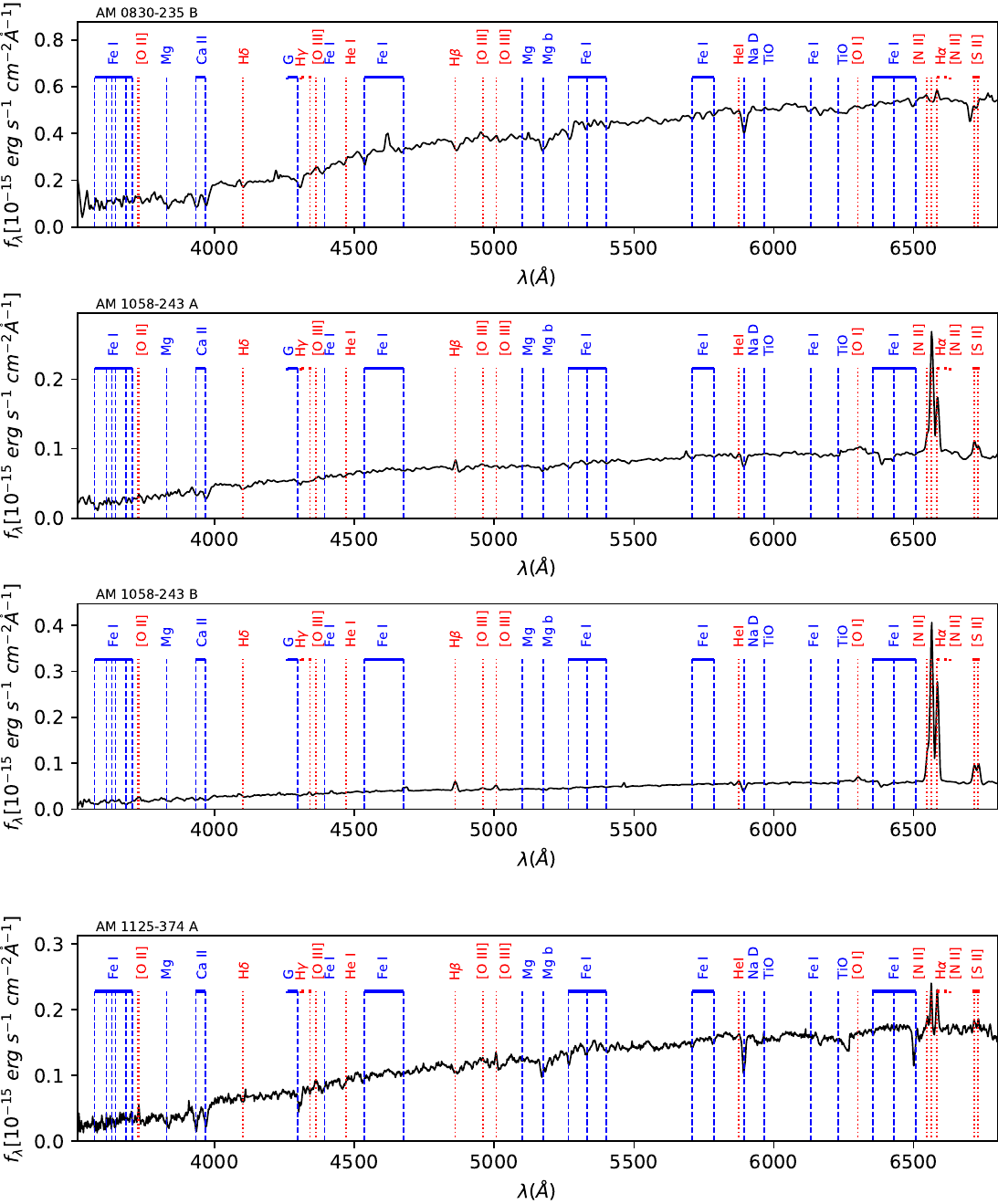}
    \includegraphics[width=0.97\columnwidth]{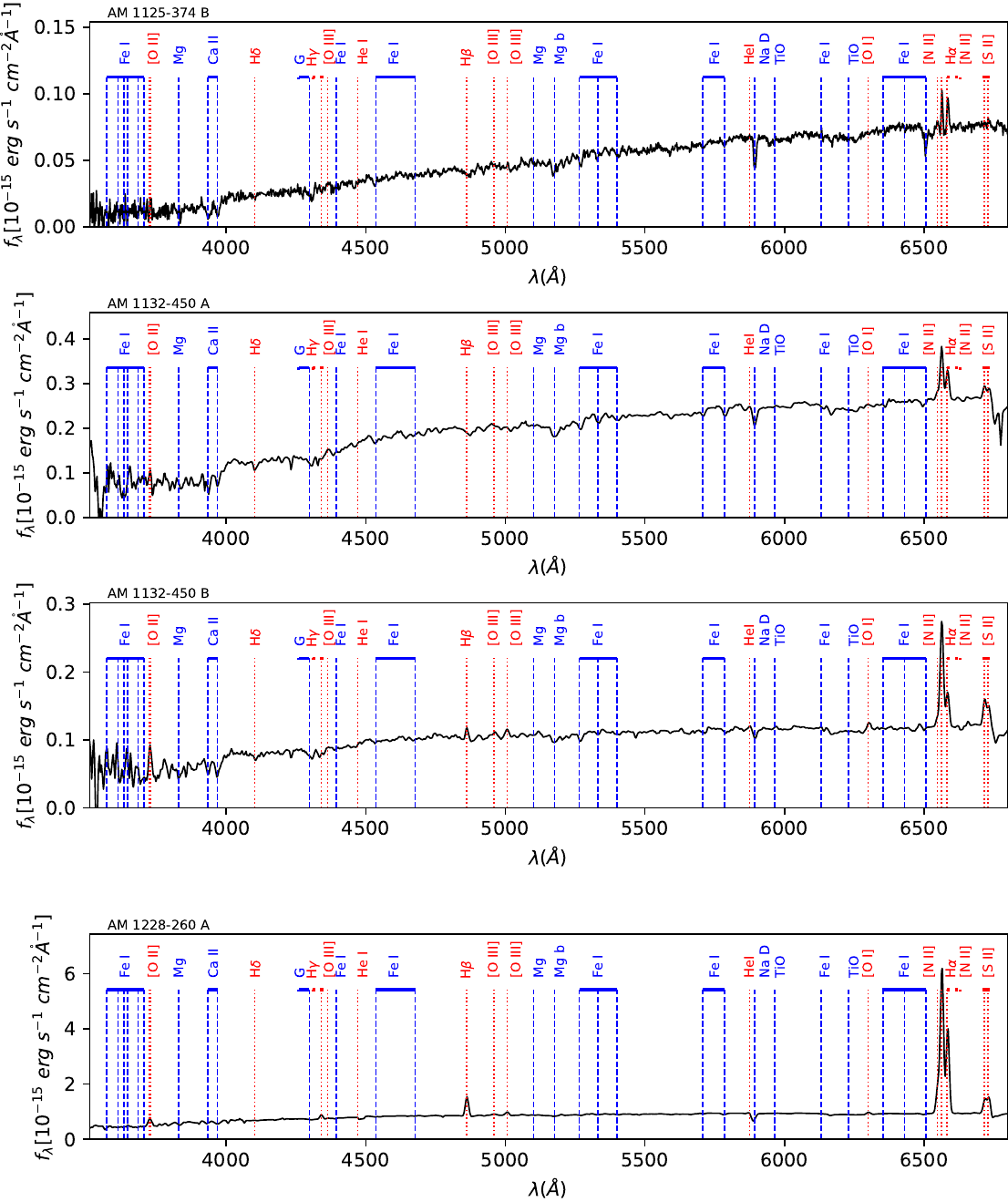}   
    \caption{Same as in Fig.~\ref{fig:pairsspectra}, but for the  INT galaxies not included there. }
    \label{fig:pairsspectra2}
\end{figure*}

\setcounter{figure}{1}
\begin{figure*}
    \centering
    \includegraphics[width=0.97\columnwidth]{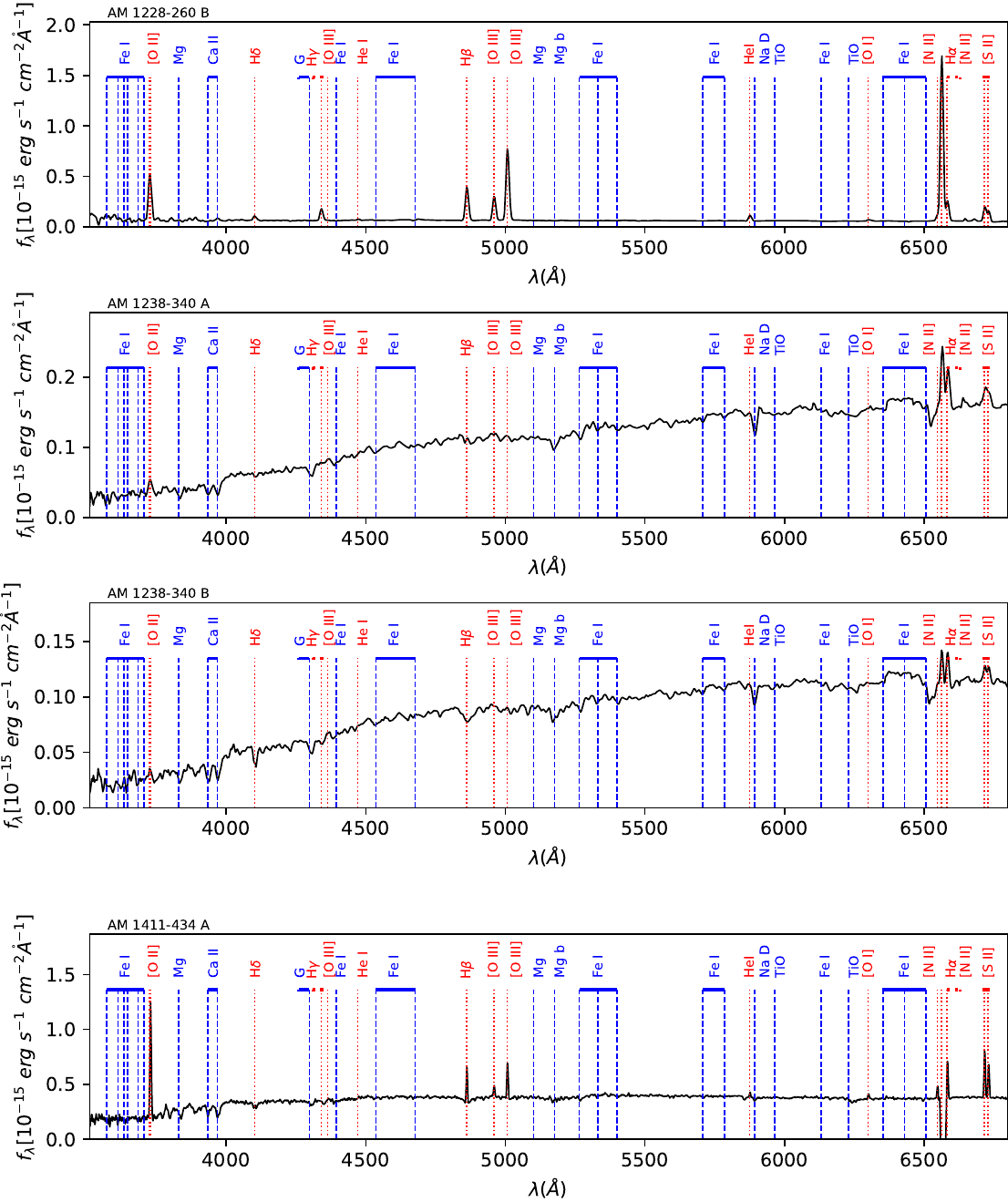}
    \includegraphics[width=0.97\columnwidth]{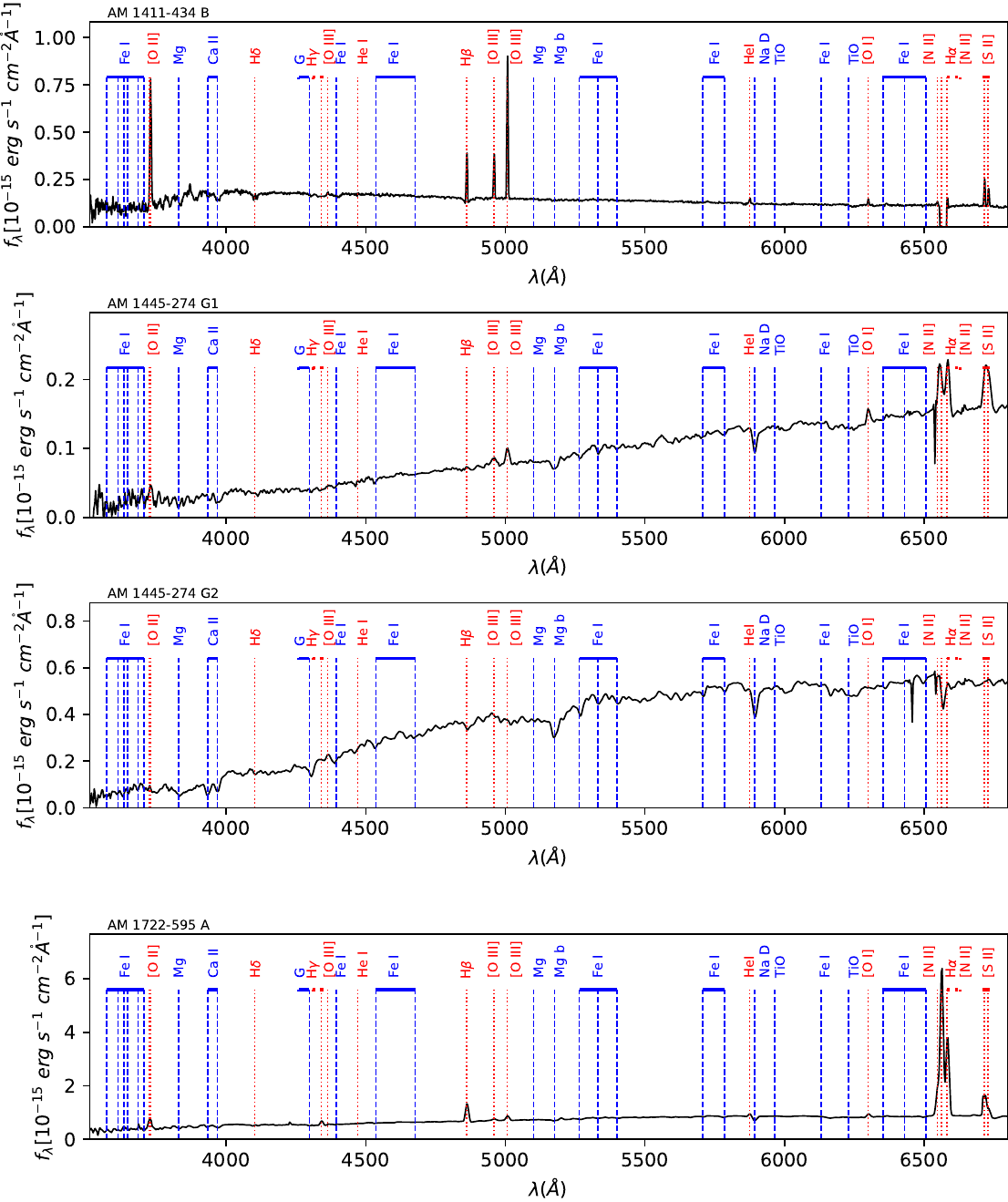}
    \includegraphics[width=0.97\columnwidth]{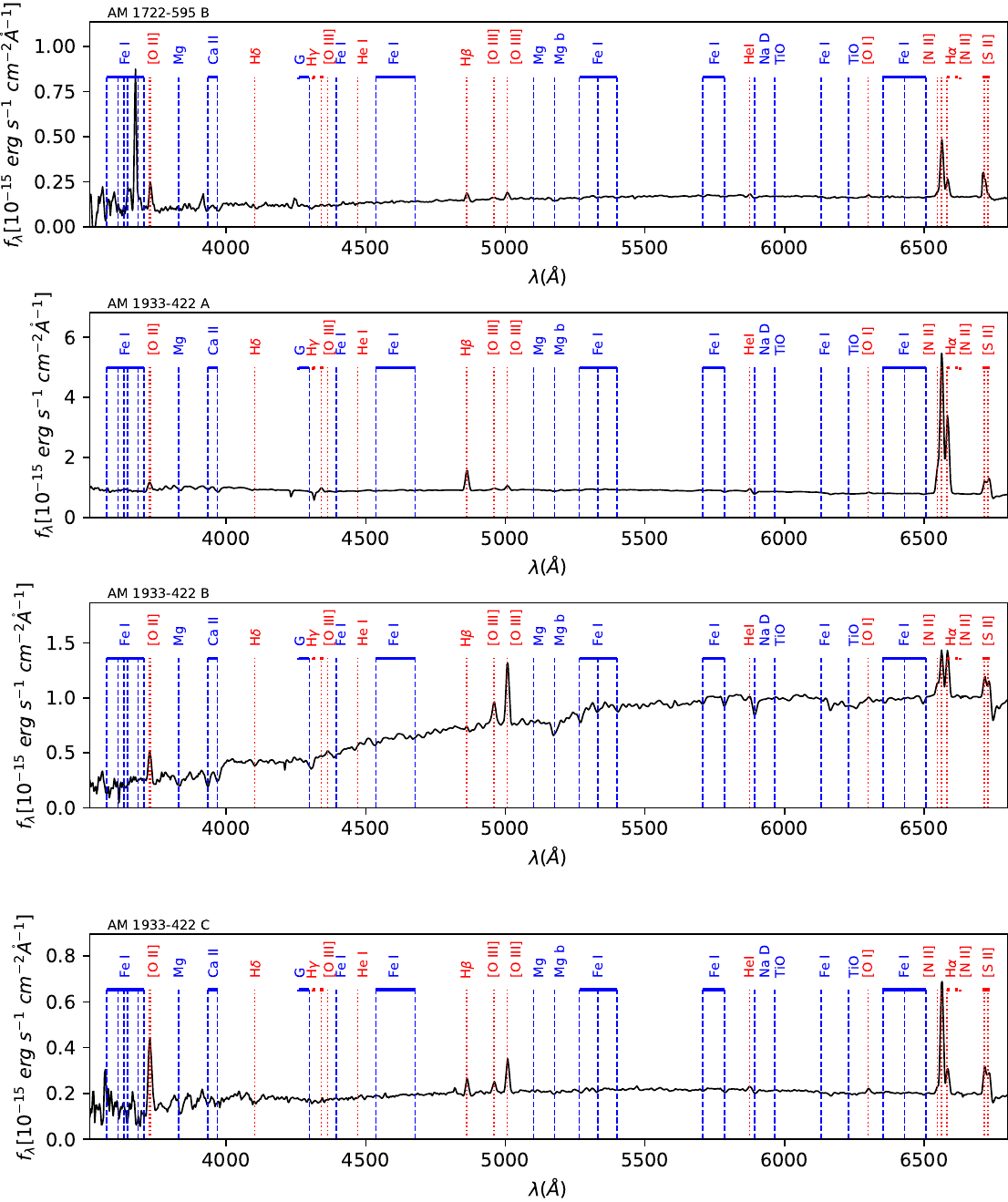}
    \includegraphics[width=0.97\columnwidth]{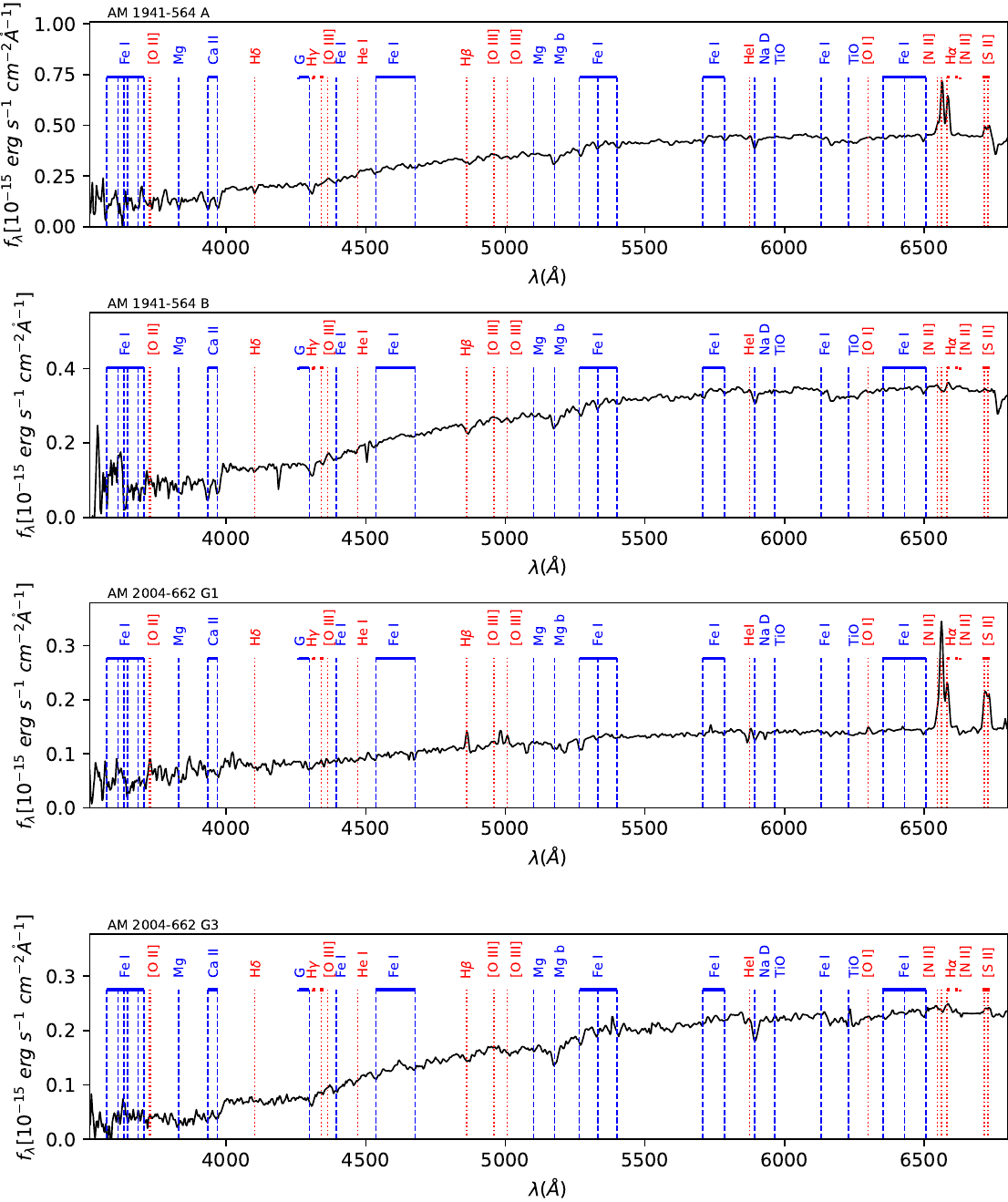}  
    \caption{continued.}
\end{figure*}

\setcounter{figure}{1}
\begin{figure*}
    \centering
    \includegraphics[width=0.97\columnwidth]{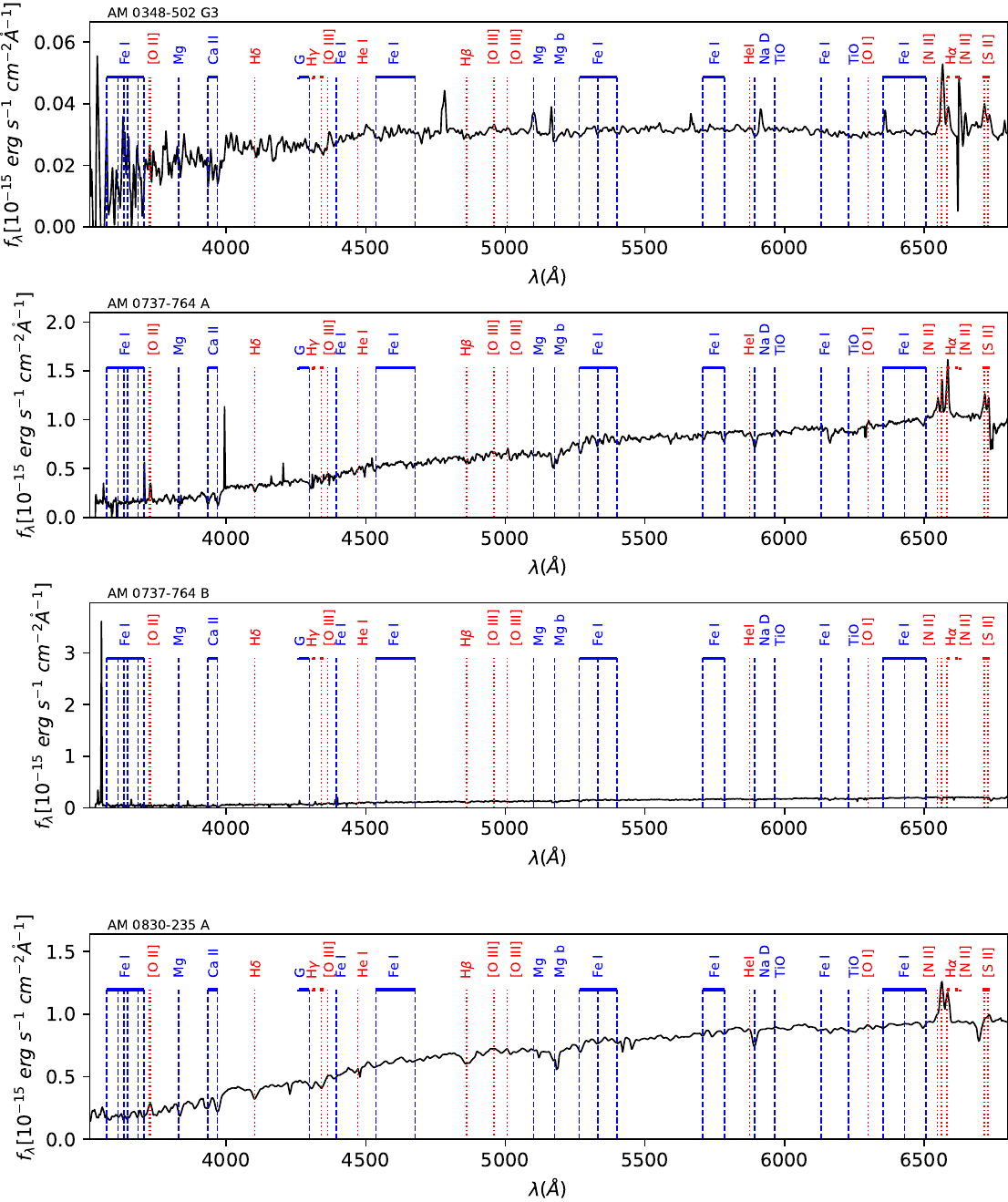}
    \includegraphics[width=0.97\columnwidth]{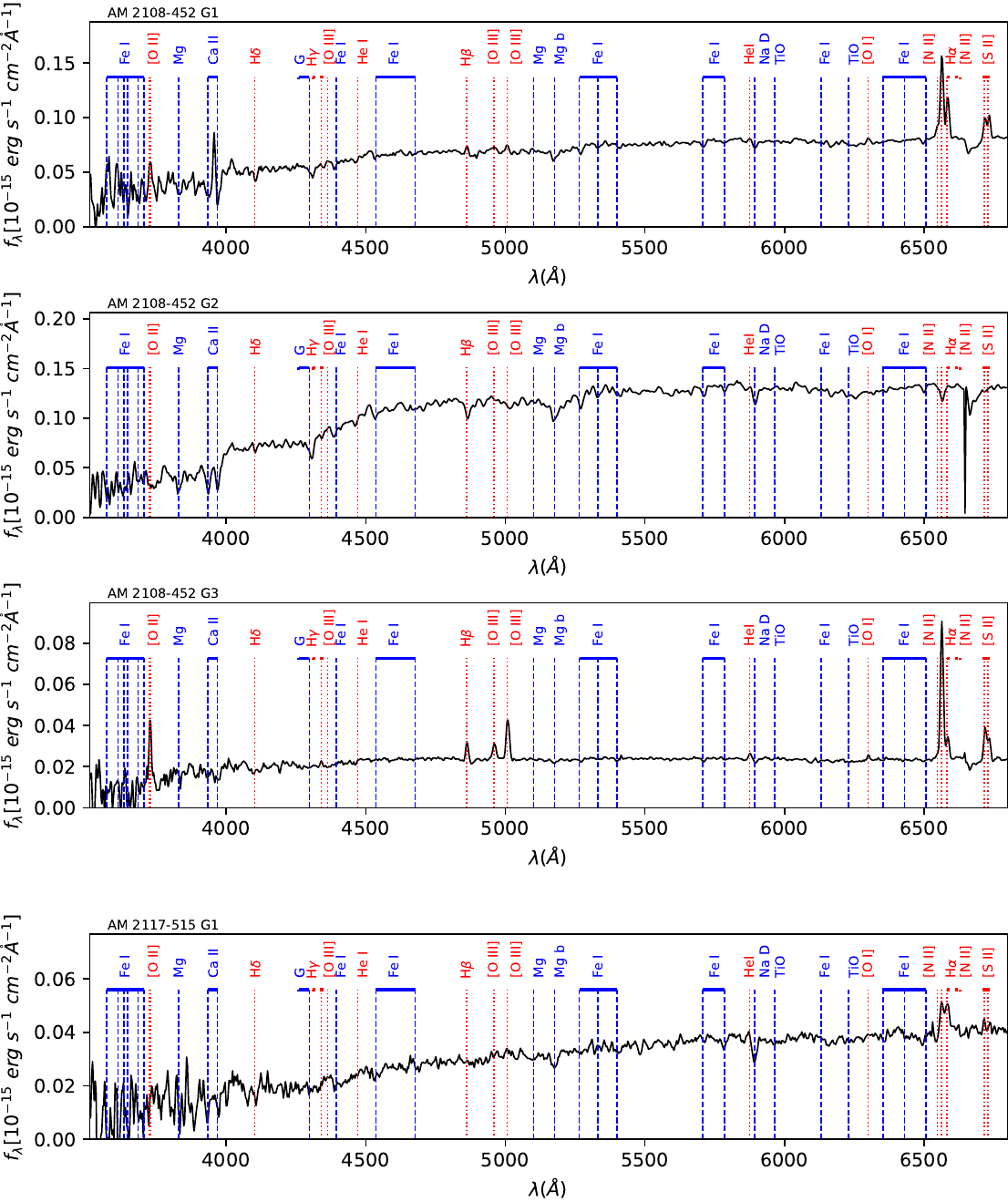}
    \includegraphics[width=0.97\columnwidth]{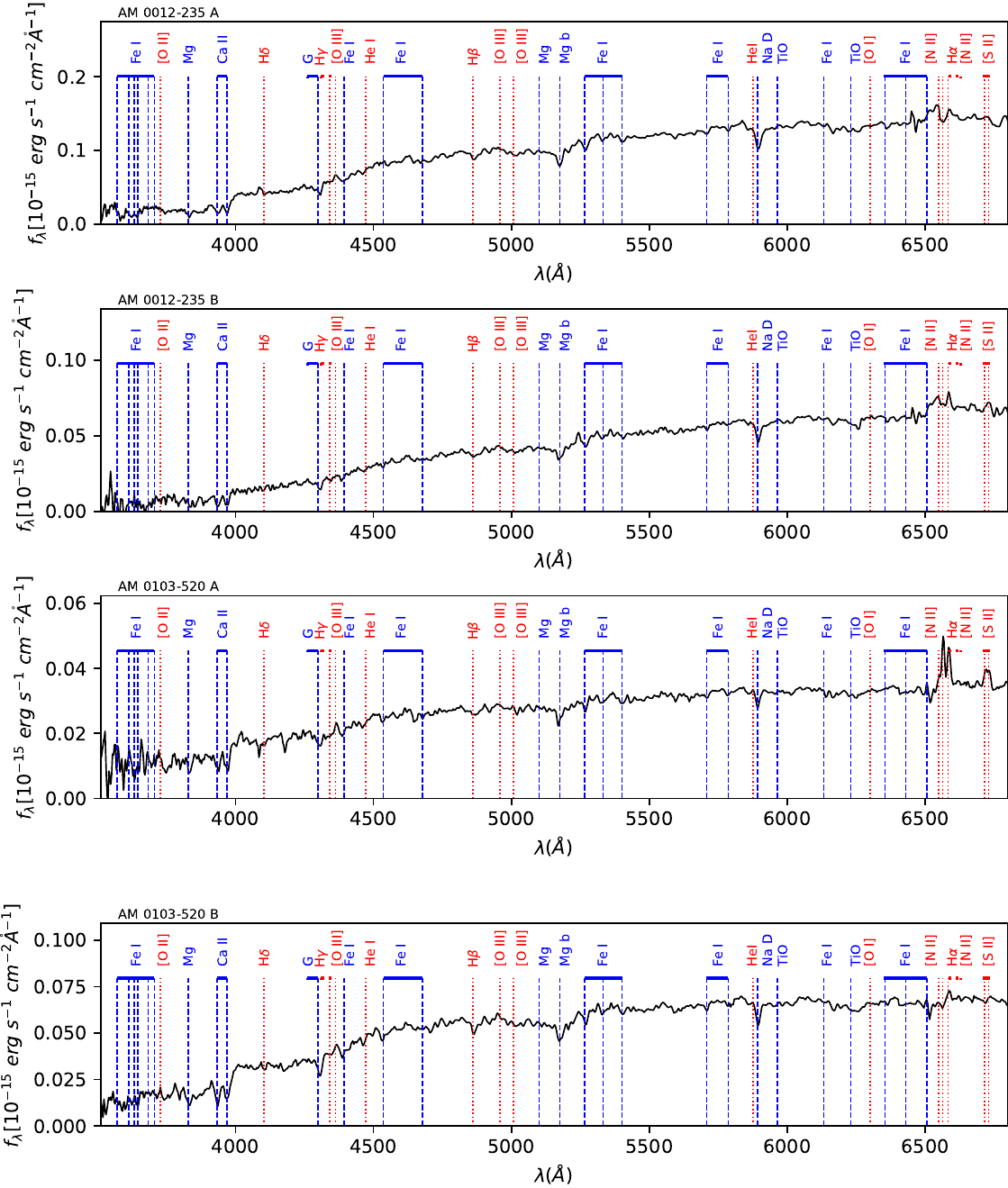}
    \includegraphics[width=0.97\columnwidth]{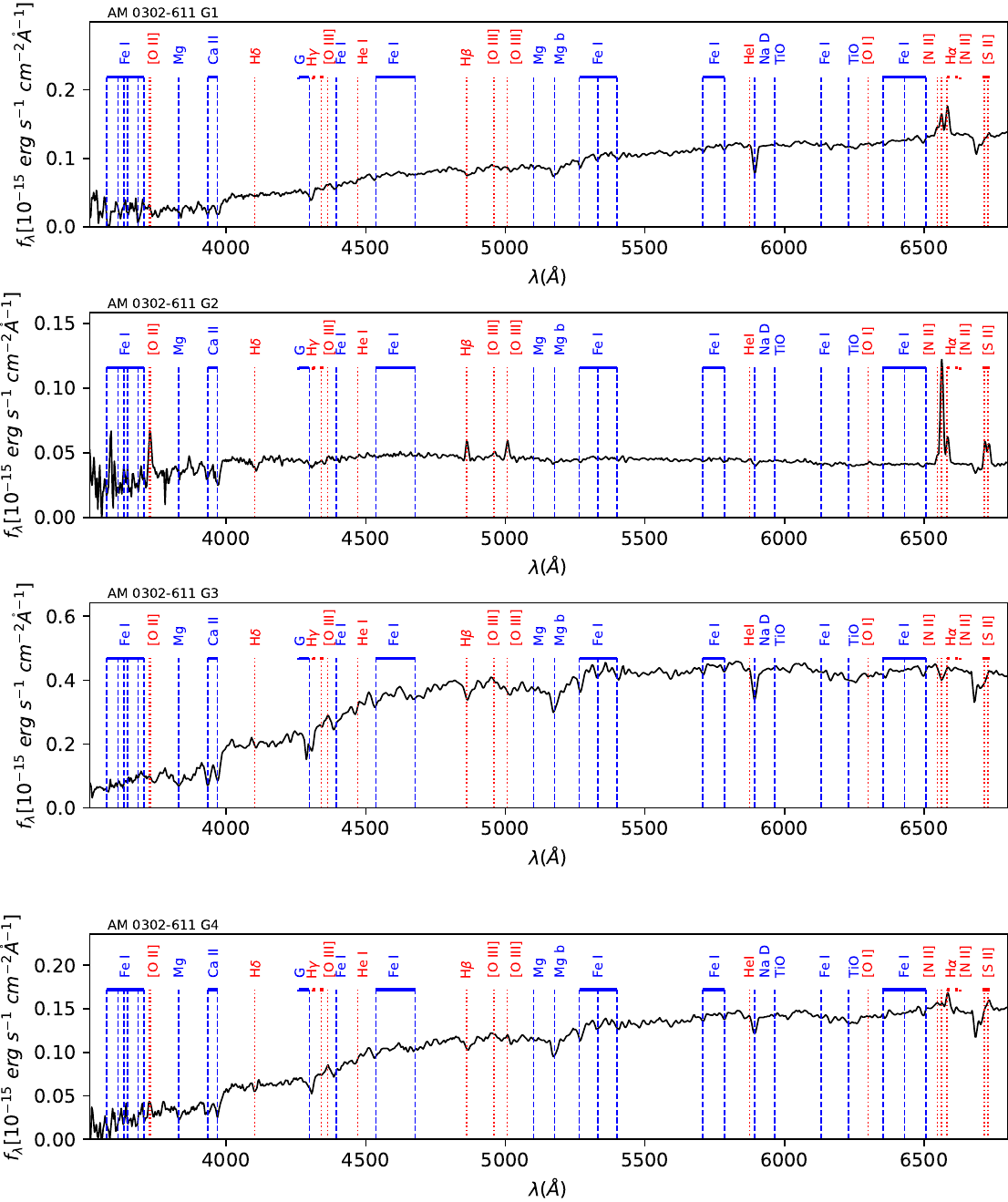}ç
    \caption{continued.}
\end{figure*}

\setcounter{figure}{1}
\begin{figure*}
    \centering
    \includegraphics[width=0.97\columnwidth]{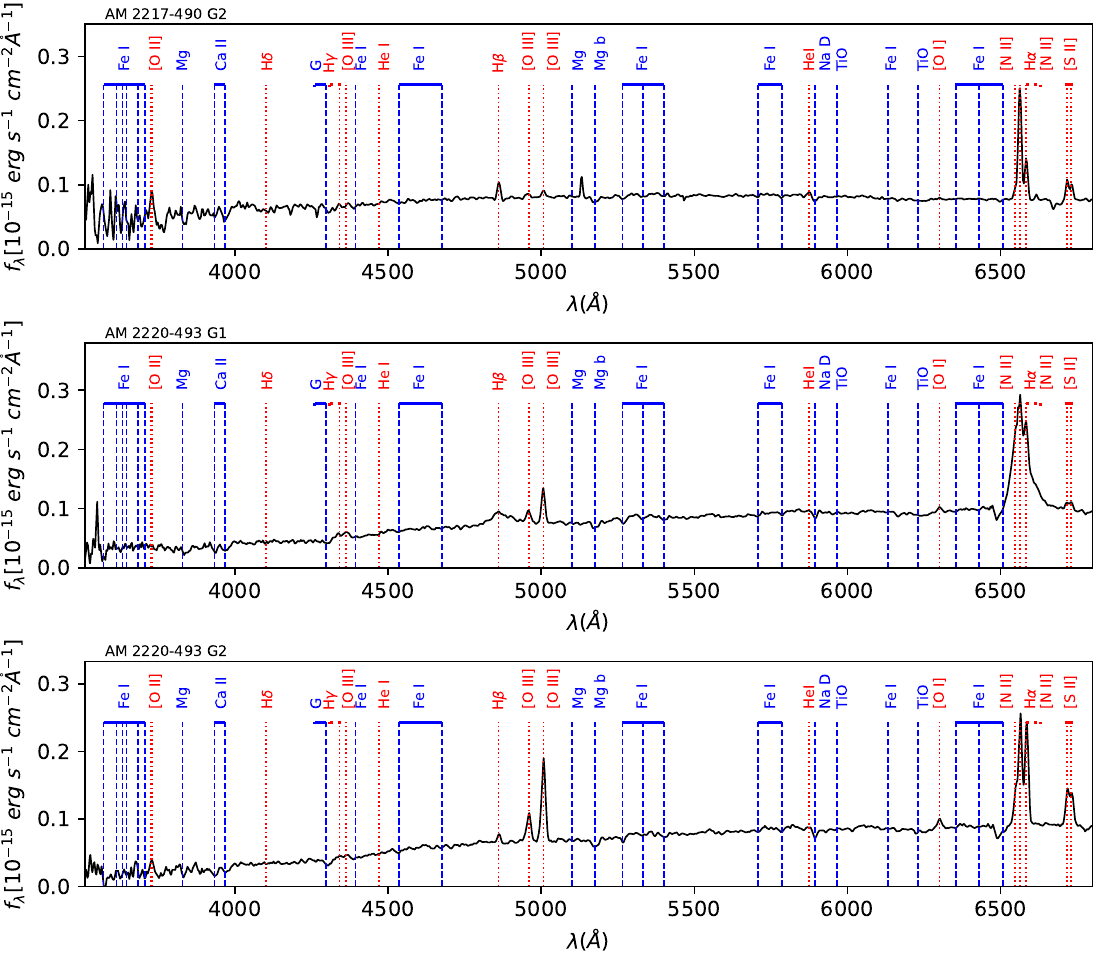}
    \caption{continued.}
\end{figure*}


\onecolumn
\begin{longtable}{ccccccc}
\caption{Interacting galaxies sample selected from the Arp \& Madore Catalogue.}
\label{tab:Sample}
\\
\hline\hline
System Name & Component & $\alpha$(2000) & $\delta$(2000) & PA & Systemic velocity\\ (1) & (2) & (hh:mm:ss) & (dd:mm:ss) & [$^{\circ}$] (3) & [km/s] (4)\\
\hline
\endfirsthead
\caption{continued.}\\
\hline\hline
System Name & Component & $\alpha$(2000) & $\delta$(2000) & PA & Systemic velocity\\ (1) & (2) & (hh:mm:ss) & (dd:mm:ss) & [$^{\circ}$] (3) & [km/s] (4)\\
\hline
\endhead 
\hline 
\multicolumn{6}{c}{\tablefoot{\footnotesize (1) System name according to \cite{ArpMadore+87} catalogue. (2) Galaxy identifier from Simbad (and NASA/IPAC Extragalactic Database - NED in case of divergency) when available. (3) Position angle of the observed slits. (4) NED systemic velocities for the galaxies with previous information.}} \\ \hline
\endfoot
\hline 
\multicolumn{6}{c}{\tablefoot{\footnotesize (1) System name according to \cite{ArpMadore+87} catalogue. (2) Galaxy identifier from Simbad (and NASA/IPAC Extragalactic Database - NED in case of divergency) when available. (3) Position angle of the observed slits. (4) NED systemic velocities for the galaxies with previous information.}} \\ \hline
\endlastfoot
\object{AM 0012-235}  & A = \object{LEDA 73318}     & 00:15:10.230             & -23:42:58.50             & 195.5           & 19076                         \\
& B = \object{2MASX J00150913-2343541}      &                          &                          &                 & $>$30000                      \\
\object{AM 0103-520}  & A = \object{ESO 195-30}                & 01:05:51.320             & -51:45:04.50             & 58/197          & -                             \\

            & B = \object{LEDA  452921}                        &                          &                          &                 & -              \\
            & C = \object{Gaia DR3 4927464765788769536}/                          &                          &                          &                 & -                             \\
&\object{ESO 195-IG 030 NED02} &&&&\\
\object{AM 0237-525}  & A = \object{LEDA 128584}                  & 02:38:44.420             & -52:46:39.10             & 220             & -                             \\
& B = \object{[MDS99] F154-017} &&&& - \\
&\object{LSBG F154-017} &&&&\\
\object{AM 0302-611}  & G1  = \object{ESO 116-6}      & 03:03:30.430             & -61:08:19.50             & 252/320         &            -               \\
& G2  = \object{ESO 116-IG 006 NED02} &                          &                          &                 & -                             \\
& G3  = \object{2MASX J03033046-6108201}                &                          &                          &                 &
     -                        \\
 & G4  = \object{2MASX J03032756-6108281}                &                          &                          &                 &
 8553                             \\
\object{AM 0305-824}  & A = \object{LEDA  222825}                  & 03:01:44.9               & -82:35:31                & 328             & -                          \\
\object{AM 0338-320}  & A = \object{6dFGS gJ034029.8-315142}      & 03:40:29.660             & -31:51:41.30             & 237             & -                             \\
& B = \object{2MASX J03402854-3151520}     &                          &                          &                 & -                             \\ 
\object{AM 0348-502}  & G1  = \object{ESO 201-4}                   & 03:50:22.910             & -50:18:09.00             & 343/305.5       & 10766                         \\
& G2  = \object{2MASX J03502377-5018354}     &                          &                          &                 & 10940                         \\
& G3  = \object{2MASX J03501198-5017165}                         &                          &                          &                 & -                             \\
\object{AM 0737-764}  & A = \object{ESO 35-11}                 & 07:35:18.700             & -76:54:52.09             & 200.5           & 5900                          \\
& B = \object{WISEA J073520.20-765437.6} 	                           &                          &                          &                 & -                             \\
\object{AM 0830-235}  & A = \object{ESO 495-16}                 & 08:32:41.1 & -24:02:32    & 159             & 7924                          \\
& B = \object{LEDA 86223}              &                          &                          &                 & -                             \\
\object{AM 0908-674}  & A = \object{NGC  2788}                     & 09:09:03.394             & -67:55:59.61             & 294             & 1641                          \\
\object{AM 0952-273}  & A = \object{LEDA 749047}                  & 09:54:56.750             & -27:50:50.90             & 330             & -                             \\
& B = \object{LEDA 181883}                           &                          &                          &                 & -                             \\
\object{AM 1026-442}  & G1 - A = \object{NGC 3261}                     & 10:29:01.263             & -44:39:27.03             & 3               & 2563                          \\
\object{AM 1058-243}  & A = \object{LEDA 784699}              & 11:00:50.060             & -24:55:07.89             & 276             & 22917                         \\
& B = \object{WISEA J110049.40-245506.4}
              &                          &                          &                 & -                             \\
\object{AM 1125-374}  & A = \object{LEDA 615154} & 11:27:36.207             & -37:57:18.13             & 289.45          & 17392                         \\
& B = \object{2MASX J11273693-3757208}              &                          &                          &                 & -                             \\
\object{AM 1132-450}  & A = \object{ESO 266-5}           & 11:35:09.287             & -45:23:15.97             & 165.7           & 5266                          \\
& B = \object{LEDA 35805}              &                          & -                        &                 & -                         \\
\object{AM 1228-260}  & A = \object{ESO 506-15}                 & 12:31:16.842             & -26:17:29.93             & 2.9             & 5912                          \\
& B = \object{WISE J123116.76-261741.1}/             &                          &                          &                 & -                             \\
& \object{AM 1228-260 NED01} &&&&\\
\object{AM 1238-340}  & A = \object{LEDA 88654}       & 12:40:41.124             & -34:21:47.75             & 245             & 16247                         \\
& B = \object{2MASS J12404057-3421489}                         &                          &                          &                 & -                             \\
\object{AM 1411-434}  & A = \object{IC 4386}                      & 14:15:02.475             & -43:57:39.94             & 184             & 1882                          \\
& B = \object{IC 4387}                      &                          &                          &                 & -                             \\
\object{AM 1438-400}  & G1  = \object{ESO 327-8}                 & 14:41:20.089             & -40:19:50.69             & 180             & 7825                          \\
& G2 = \object{WISEA J144118.20-401956.4}    &                          & -                        & 202             & -                             \\
\object{AM 1445-274}  & G1  = \object{ESO 448-2}                 & 14:48:52.770             & -27:52:38.17             & 51.7            & 14292                         \\
& G2  = \object{WISEA J144852.47-275242.2}                          &                          &                          &                 & -                             \\
\object{AM 1722-595}  & A = \object{ESO 138-27}                 & 17:26:43.352 & -59:55: 55.17              & 234.3           & 6230                          \\
& B = \object{LEDA 367658}           &                          &                          &                 & -                             \\
\object{AM 1933-422}  & G1  = \object{NGC 6806}                    & 19:37:05.0 & -42:17:47            & 119.3           & 5733                          \\
& G2  = \object{2MASX J19372362-4218194}                 &                          &                          & 98.8            & -                             \\
& G3  = \object{LEDA 3098348}            &                          &                          &                 & 5996                          \\
\object{AM 1941-564}  & A = \object{IC 4890}                      & 19:45:34.844             & -56:33:04.92             & 193.4           & 5239                          \\
& B = \object{WISEA J194534.04-563334.2}      &                          &                          &                 & -                             \\
\object{AM 2004-662}  & G1 = \object{NAME McL A}/    & 20:09:27.332             & -66:12:55.63             & 127.6           & 11299                         \\
& \object{ESO 105-G026} &&&&\\
& G2 = \object{WISEA J200937.54-661343.8}     &                          &                          &                 & -                             \\
& G3 = \object{NAME McL B}/                  &                          &                          & 96.3            & -                             \\
&\object{WISEA J200924.36-661241.8} &&&&\\
& G4 = \object{LEDA 306574}     &                          &                          &                 & -                             \\
\object{AM 2024-525}  & G1 = \object{ESO 186-51}          & 20:28:16.548             & -52:40:32.00             & 215.8           & 4912                          \\
& G3 = \object{LEDA 64756}          &                          &                          &                 & -                             \\
& G2 = \object{2MASX J20281941-5240587}          &                          &                          & 283.5           & 16084                         \\
& G4 = \object{LEDA 440336}             &                          &                          &                 & -                             \\
&\object{WISEA J202809.53-524037.6} &&&&\\
\object{AM 2031-500}  & A = \object{ESO 234-49}                & 20:35:20.907             & -49:51:15.29             & 41              & 2579                          \\
& B = \object{LEDA 65018}              &                          &                          &                 & 4561                          \\
\object{AM 2036-510}  & G1 = \object{FRL 908}     & 20:40:03.800             & -50:55:51.19             & 145.7           & -                             \\
& G2 = \object{ESO 234-62}               &                          &                          &                 & 4800                          \\
& G3 = \object{2MASS J20401226-5057017}     &                          &                          &                 & -                             \\
\object{AM 2037-550}  & A = \object{ESO 187-1} & 20:41:27.90              & -54:58:47.3              & 70              & 11446                         \\
\object{AM 2038-323}  & G1 = \object{NGC 6947}                    & 20:41:12.795             & -32:30:13.22             & 24.8            & 5592                          \\
& G2 =  \object{LEDA 690531}/             &                          &                          &                 & 37171                             \\
&\object{WISEA J204110.83-323106.8} &&&&\\
\object{AM 2057-650}  & A = \object{ESO 107-2}                 & 21:01:17.64 & -64:57:53.0             & 108.7           & 13720                         \\
& B = \object{LEDA 317523}               &                          &                          &                 & -                             \\
\object{AM 2100-503}  & G1 = \object{ESO 235-45}               & 21:03:48.159             & -50:21:52.19             & 289.4           & 4934                          \\
& G2 = \object{6dFGS gJ210341.7-502203}              &                          &                          & 69              & 15233                         \\
& G3 = \object{WISEA J210346.19-502146.2}              &                          &                          &                 &                               \\
\object{AM 2106-624}  & G1 = \object{ESO 107-9}                  & 21:10:06.84 & -62:35:52.5            & 63.5            & 15761\\
& G2 = \object{2MASX J21101120-6235442}                  &              &              &            & -\\
\object{AM 2108-452}  & G1 = \object{2dFGRS TGS876Z088}               & 21:11:34.35                           & -45:15:08.0                          & 137,113.1,112.5 & -                             \\
& G2 =  \object{ESO 286-73}       &                          &                          &                 & 9569                          \\
& G3 = \object{LEDA 528634}       &                          &                          &                 & -                             \\
&\object{LCRS B210819.1-452815} &&&&\\
\object{AM 2117-371}  & G1 = \object{6dFGS gJ212026.9-370005}            & 21:20:27.514             & -37:00:04.88             & 274             &                        \\
&\object{APMUKS(BJ) B211720.70-371250.1} &&&&\\
& G2 = \object{6dFGS gJ212028.1-370005}               &                          &                          &                 & 31109                         \\
\object{AM 2117-515}  & G1 - A = \object{ESO 236-IG 005 NED01} & 21:21:14.09         & -51:43:57.7                          & 83              & 15471                         \\
& G2 - B = \object{ESO 236-IG 005 NED02}         &  &       & 5               & -                             \\
& G3 - E = \object{ESO 236-IG 005 NED03}                            &                          &                          & 285.5           & -                             \\
& G4 - C = \object{2MASS J21211536-5143581}         &                          &                          & 83                & -                             \\
& G5 = \object{2MASS J21211361-5143535}/&                          &                          & 285.5                & -                             \\
&(tadpole at NW of G3)&&&&\\
& G6 - D = \object{LEDA 453396}         &                          &                          &  5               & -                             \\
& knots (See Fig.~\ref{fig:pairsimages})                         &                          &                          &   285.5              & -                             \\
\object{AM 2117-682}  & G1 = \object{LEDA 127236}                  & 21:21:38.623             & -68:10:05.06             & 273.8           & 25987                         \\
& G2 = \object{2MASS J21215262-6810066}                  &                          &                          &                 & -                             \\
& G3 = \object{LEDA 287882}                  &                          &                          &                 & -                             \\
\object{AM 2132-664}  & G1 - A = \object{ESO 107-38}                & 21:36:24.618             & -66:28:50.10             & 175.5           & 8490                          \\
& G2  = \object{2MASS J21362448-6628496}               &                          &                          &                 & -                             \\
& (Our images show that it is a star).&&&&\\
& G3 - B = \object{LEDA 304337}               &                          &                          &                 & -                             \\
\object{AM 2137-515} & A = \object{ESO 236-31}               & 21:41:17.04 & -51:45:48.2             & 209           & 16765                          \\
& B = \object{LEDA 453034}             &                          &                          &                 & -                             \\
&\object{WISEA J214120.56-514450.9} &&&&\\
\object{AM 2147-544}  & G1 = \object{ESO 189-5}                & 21:50:58.202             & -54:33:08.56             & 229             & 8623                          \\
& G2 = \object{LEDA 190556} & 21:51:00.125             & -54:33:37.35             &              & -\\
\object{AM 2217-490}  & G1 = \object{ESO 238-2}                & 22:20:46.258             & -48:45:17.43             & 143.2           & 9212                          \\
& G2 = \object{2MASX J22204151-4844140}             &                          &                          &                 & 9060                          \\
& G3 = \object{LEDA 484571}/ &&& & 9081\\
& \object{GALEXASC J222040.91-484242.0} &&&&\\
\object{AM 2220-493}  & G1 = \object{ESO 238-6}             & 22:23:13.043             & -49:17:25.78             & 168             & -\\
& G2  = \object{LEDA 95417}            &                          &                          &                 & 17946                         \\
\object{AM 2220-661}  & A = \object{LEDA 127657}                & 22:24:14.320             & -66:02:15.70             & 75 & 10556\\
\object{AM 2246-490}  & G1 = \object{ESO 239-2}                 & 22:49:40.219             & -48:51:01.92             & 322             & 12901\\
& G2 = \object{LEDA  482697}/                  &                          &                          &                 & -\\
& \object{PRC D-41} &&&&\\
\end{longtable}
\twocolumn

\begin{table*}[ht]
\caption{Confirmed interacting pairs and groups. The second column presents the INT galaxies measured radial velocities. The third and fourth columns present the SDSS and astrophysical catalogues information of their control galaxies.}
\label{tab:radial_vel}
\centering
\tiny
\begin{tabular}{cccc} 
System name & Radial velocity & SDSS DR9 Plate-MJd-Fiberid & CTR galaxy name \\ (1) & [km/s] (2) & (3) & (4)\\
\hline\hline



AM 0012-235~A  & 18961 $\pm$   92 & 2124-53770-0272& \object{2MASX J14030448+2446252}\\
~B                  & 18830 $\pm$   95  & 2613-54481-0304& \object{2MASX J12340843+1929256}\\ \hline
AM 0103-520~\tablefootmark{[I]}~A  & 16345 $\pm$   83 & 1693-53446-0629 & \object{2MASX J17053367+2506193}\\
~B                  & 16441 $\pm$   65  & 1601-53115-0280 & \object{LEDA 1389965}\\ \hline
AM 0302-611~G1~(A)  & 8321 $\pm$   97 & 2746-54232-0165 & \object{UGC 9185}\\
~G2~(D)                  & 8509 $\pm$  129  & -&-\\
~G3~(C)                  & 8543 $\pm$   54  & 1237663234989424850~\tablefootmark{[*]} & \object{UGC 29}\\
&&(SDSS ObjId, no spectrum)&\\
~G4~(B)                  & 8419 $\pm$   83  & 2776-54554-0098 & \object{2MASX J14451344+1745338}\\ \hline
AM 0338-320~A  & 29034 $\pm$  102     & 0420-51871-0517& \object{2MASX J00550203+1516012}\\
~B                  & 29116 $\pm$  179      & 0421-51821-0003& \object{2MASX J01035856+1346065}\\ \hline
AM 0348-502~G1~(A)  & 10707 $\pm$  54 & 2774-54534-0448& \object{IC 1021}\\
~G2~(B)                  & 10759 $\pm$   63 & 1607-53083-0452 & \object{LEDA 3090857}\\
~G3~(C)                  & 10581 $\pm$ 116  & 1652-53555-0578 & \object{KISS F1530-273}/\object{WISEA J153423.63+295110.4}\\ \hline
AM 0737-764~A  & 5899 $\pm$   35      & 2352-53770-0006 & \object{NGC 3327} \\
~B                  & 5875 $\pm$   33       & 2515-54180-0046 & \object{LEDA 1557424}\\ \hline
AM 0830-235~A  & 7838 $\pm$   92       & 2234-53823-0120 & \object{NGC  4514} \\
~B                  & 7628 $\pm$   95       & 2130-53881-0553 & \object{2MASX J14230749+2835418}\\ \hline
AM 1058-243~A  & 22782 $\pm$  100      & 1937-53388-0453 & \object{LEDA 1853840}\\
~B                  & 22881 $\pm$   42      & 0731-52460-0345 & \object{LEDA 1396403}\\ \hline
AM 1125-374~A  & 17075 $\pm$   30      & 1771-53498-0483 & \object{2MASX J12593825+1510458}\\
~B                  & 16851 $\pm$   26      & 1588-52965-0534 & \object{LEDA 1847759}\\ \hline
AM 1132-450~A  & 5226 $\pm$   85       & 2132-53493-0224 & \object{IC 1006}\\
~B                  & 5078 $\pm$   63       & 2277-53705-0378 & \object{UGC  4504}\\ \hline
AM 1228-260~A  & 5737 $\pm$   83       & 1852-53534-0298 & \object{NVSS J160707+220337}\\
~B                  & 5835 $\pm$   76       & -                   &-\\ \hline
AM 1238-340~A       & 15961 $\pm$   94      & 1693-53446-0629& \object{2MASX J17053367+2506193}\\
~B                  & 16255 $\pm$   82      & -                   &-\\ \hline
AM 1411-434~A       & 1877 $\pm$   45       & 1237680245198487603~\tablefootmark{[*]} & \object{NGC 7497}\\
&&(SDSS ObjId, no SDSS spectrum)&\\
~B                  & 1788 $\pm$   38       & 2273-53709-27 & \object{UGC 4444}\\
AM 1445-274~G1~(A)  & 14008 $\pm$   83      & 2170-53875-0347 & \object{Z 107-41}\\
~G2~(B)             & 13847 $\pm$   91 & - & -\\ \hline
AM 1722-595~A       & 6119 $\pm$   60       & 1929-53349-0372 & \object{IC 509}\\
~B & 6162 $\pm$   69 & -&-\\ \hline
AM 1933-422~G1~(A)  & 5562 $\pm$  128       & 0275-51910-0320 & \object{NGC 3339}\\
~G2~(B)    		   & 5582 $\pm$  139       & 1776-53858-0098& \object{LEDA 1429669}\\
~G3~(C)             & 5609 $\pm$   84       & 2580-54092-0232& \object{MCG+02-25-009} \\\hline
AM 1941-564~A & 5130 $\pm$ 86 & 2350-53765-0539 & \object{NGC 3204}\\
~B                  & 4787 $\pm$   91       & - &-\\ \hline
AM 2004-662~\tablefootmark{[II]}~A~(G1) & 11100 $\pm$  220   & 2426-53795-0583 & \object{LEDA 1427577}\\
~B~(G3) & 10876 $\pm$  101 & - &-\\ \hline
AM 2036-510~G2~(A)  & 11792 $\pm$   48 & 2523-54572-0428 & \object{LEDA 1390992}\\
~G1~(B)                  & 12224 $\pm$   92 & 1266-52709-0142 & \object{SDSS J081524.39$+$254729.7}\\
~G3~(C)                  & 12018 $\pm$   90 &-&-\\\hline
AM 2106-624~\tablefootmark{[III]}~G1~(A)   & 15761$\pm$ 44 & 2759-54534-0160 & \object{IC 4410}\\
~G2~(B)                  & 15382 $\pm$  71 & 2227-53820-0018 & \object{LEDA 1785836}\\ \hline
AM 2108-452~G2~(A)  & 9400 $\pm$   77                & 2782-54592-0240 & \object{LEDA 1468533}\\
~G1~(B)                  & 9633 $\pm$   28                & 2126-53794-0075 & \object{2MASX J14164251+2854406}\\
~G3~(C)& 9436 $\pm$   50 & 0420-51871-0285& \object{SDSS J004923.00+135600.9}\\ \hline
AM 2117-515~G1~(A)  & 15053 $\pm$  103               & 0860-52319-0120 & \object{LEDA 1866838}\\
~G2~(B)                  & 15443 $\pm$  178               & 2268-53682-0424 & \object{LEDA 1448216}\\
~G4~(C)                  & 15293 $\pm$  104               & -                   &-\\
~G5~(D)                  & 15535 $\pm$   56               & -                  & -\\
~G6~(E)                  & 15396 $\pm$   18               & 2765-54535-0129 & \object{2MASX J15065663+1426145}\\
~Central Knots      & 15305 $\pm$  123               & -                   &-\\ \hline
AM 2132-664~\tablefootmark{[IV]}~G1~(A)& 8469 $\pm$  130 & 2094-53851-0261 & \object{MCG+05-32-055}\\
~G3~(B)                  & 8222 $\pm$   43                & - &-\\ \hline
AM 2217-490~\tablefootmark{[V]}~G1~(A)       & 9092 $\pm$   57                & 2496-54178-0330 & \object{Z 126-20}\\
~G2~(B)                  & 9003 $\pm$  102                & 2599-54234-0571 &\object{IC 3615}\\ 
\hline
AM 2220-493~G1~(A)& 17832 $\pm$  101               & 2764-54535-0355 & \object{MCG+03-38-035}\\
~G2~(B)                  & 17746 $\pm$   43               & 2518-54243-0066 & \object{2MASX J15503870+1530524}\\ \hline
\hline
\end{tabular}
\tablefoot{\footnotesize (1) System name and component according to \cite{ArpMadore+87} catalogue and Table~\ref{tab:Sample}. (2) Systemic velocity obtained as described in Sec.~\ref{section:samples}. (3) Corresponding SDSS DR9 control partner. (4) Control galaxy identifier from Simbad (and NASA/IPAC Extragalactic Database - NED in case of divergency) when available. 
\tablefoottext{[I]}{AM 0103-520: C is not a member of the system.}
\tablefoottext{[II]}{AM 2004-662: G2 and G4 are foreground galaxies.}
\tablefoottext{[III]}{AM 2106-624: A = ESO107-009 is a member with radial velocity 15761 $\pm$ 44 km/s \citep{Strauss+92}, but we do not have the source spectra, because the slit was centred in a star. The A component CTR galaxy was not included in the next comparisons since we do not observe component G1. C (indicated in Fig.~\ref{fig:pairsimages2}) could be a member; however, it was not observed.}
\tablefoottext{[IV]}{AM 2132-664: Component G2 it is a star.}
\tablefoottext{[V]}{AM 2217-490: G3 (6dFJ2220410) is a member but was not observed. There is also a possible member towards the east of G1 (not in the image of Fig.~ \ref{fig:pairsimages2}).}
\tablefoottext{[*]}{The control galaxy is not included in the INT versus CTR sample comparison as the spectrum of the control is unavailable in the SDSS DR9.}
}
\end{table*}

\onecolumn
\begin{landscape}
\centering
\renewcommand{\tabcolsep}{.5mm}
\tiny
\begin{longtable}{ccccccccccccc}
\caption{Extinction corrected emission lines measurements for forbidden and helium transitions in the interacting sample.}\label{tab:INTforbflux}\\
\hline\hline
Object & [O\,{\sc ii}]$\lambda$3726 & [O\,{\sc ii}]$\lambda$3729 & [O\,{\sc iii}]$\lambda$4363 & [O\,{\sc iii}]$\lambda$4959 & [O\,{\sc iii}]$\lambda$5007 & [N\,{\sc ii}]$\lambda$5755 & He\,{\sc i}$\lambda$5876 & [O\,{\sc i}]$\lambda$6300 & [N\,{\sc ii}]$\lambda$6548 & [N\,{\sc ii}]$\lambda$6583 & [S\,{\sc ii}]$\lambda$6716 & [S\,{\sc ii}]$\lambda$6731 \\
& flux x $\rm 10^{-15}$ & flux x $\rm 10^{-15}$ & flux x $\rm 10^{-15}$ & flux x $\rm 10^{-15}$ & flux x $\rm 10^{-15}$ & flux x $\rm 10^{-15}$ & flux x $\rm 10^{-15}$ & flux x $\rm 10^{-15}$ & flux x $\rm 10^{-15}$ & flux x $\rm 10^{-15}$ & flux x $\rm 10^{-15}$ & flux x $\rm 10^{-15}$
\\
& $[\ erg\ s^{-1}\ cm^{-2}]$ & $[\ erg\ s^{-1}\ cm^{-2}]$ & $[\ erg\ s^{-1}\ cm^{-2}]$ & $[\ erg\ s^{-1}\ cm^{-2}]$ & $[\ erg\ s^{-1}\ cm^{-2}]$ & $[\ erg\ s^{-1}\ cm^{-2}]$ & $[\ erg\ s^{-1}\ cm^{-2}]$ & $[\ erg\ s^{-1}\ cm^{-2}]$ & $[\ erg\ s^{-1}\ cm^{-2}]$ & $[\ erg\ s^{-1}\ cm^{-2}]$ & $[\ erg\ s^{-1}\ cm^{-2}]$ & $[\ erg\ s^{-1}\ cm^{-2}]$
\\
\hline
\endfirsthead

\caption{continued.}\\
\hline\hline
Object & [O\,{\sc ii}]$\lambda$3726 & [O\,{\sc ii}]$\lambda$3729 & [O\,{\sc iii}]$\lambda$4363 & [O\,{\sc iii}]$\lambda$4959 & [O\,{\sc iii}]$\lambda$5007 & [N\,{\sc ii}]$\lambda$5755 & He\,{\sc i}$\lambda$5876 & [O\,{\sc i}]$\lambda$6300 & [N\,{\sc ii}]$\lambda$6548 & [N\,{\sc ii}]$\lambda$6583 & [S\,{\sc ii}]$\lambda$6716 & [S\,{\sc ii}]$\lambda$6731 \\
& flux x $\rm 10^{-15}$ & flux x $\rm 10^{-15}$ & flux x $\rm 10^{-15}$ & flux x $\rm 10^{-15}$ & flux x $\rm 10^{-15}$ & flux x $\rm 10^{-15}$ & flux x $\rm 10^{-15}$ & flux x $\rm 10^{-15}$ & flux x $\rm 10^{-15}$ & flux x $\rm 10^{-15}$ & flux x $\rm 10^{-15}$ & flux x $\rm 10^{-15}$
\\
& $[\ erg\ s^{-1}\ cm^{-2}]$ & $[\ erg\ s^{-1}\ cm^{-2}]$ & $[\ erg\ s^{-1}\ cm^{-2}]$ & $[\ erg\ s^{-1}\ cm^{-2}]$ & $[\ erg\ s^{-1}\ cm^{-2}]$ & $[\ erg\ s^{-1}\ cm^{-2}]$ & $[\ erg\ s^{-1}\ cm^{-2}]$ & $[\ erg\ s^{-1}\ cm^{-2}]$ & $[\ erg\ s^{-1}\ cm^{-2}]$ & $[\ erg\ s^{-1}\ cm^{-2}]$ & $[\ erg\ s^{-1}\ cm^{-2}]$ & $[\ erg\ s^{-1}\ cm^{-2}]$
\\
\hline
\endhead 

\hline \multicolumn{13}{l}{\tablefoot{\footnotesize (1) System name and component according to \cite{ArpMadore+87} catalogue and Table~\ref{tab:Sample}.}} \\ \hline
\endfoot

\hline \multicolumn{13}{l}{\tablefoot{\footnotesize (1) System name and component according to \cite{ArpMadore+87} catalogue and Table~\ref{tab:Sample}.}} \\ \hline
\endlastfoot
\hline
AM0012-235 A & - & - & - & - & - & - & - & - & - & - & - & - \\
AM0012-235 B & - & - & - & - & - & - & - & - & - & - & - & - \\
AM0103-520 A & 3.94$\pm$0.57 & - & 0.46$\pm$0.03 & 0.11$\pm$0.01 & 0.41$\pm$0.03 & 0.14$\pm$0.02 & 0.14$\pm$0.02 & 0.06$\pm$0.01 & 0.6$\pm$0.04 & 1.91$\pm$0.04 & 0.84$\pm$0.02 & 0.79$\pm$0.02 \\
AM0103-520 B & - & - & - & - & - & - & - & - & - & - & - & - \\
AM0302-611 G1 & 236.8$\pm$32.61 & - & - & - & 8.91$\pm$0.51 & 6.13$\pm$0.58 & 4.54$\pm$0.31 & - & 14.92$\pm$0.26 & 46.5$\pm$0.43 & - & 3.11$\pm$0.29 \\
AM0302-611 G2 & 2.35$\pm$0.17 & 4.78$\pm$0.17 & 0.14$\pm$0.02 & 0.23$\pm$0.02 & 1.06$\pm$0.03 & 0.05$\pm$0.01 & 0.11$\pm$0.01 & 0.06$\pm$0.01 & 0.35$\pm$0.03 & 1.12$\pm$0.01 & 0.8$\pm$0.01 & 0.72$\pm$0.02 \\
AM0302-611 G3 & - & - & - & - & - & - & - & - & - & - & - & - \\
AM0302-611 G4 & 107.98$\pm$21.65 & 79.52$\pm$21.35 & - & 3.38$\pm$0.28 & 14.76$\pm$1.2 & 1.32$\pm$0.19 & 1.13$\pm$0.09 & - & 5.96$\pm$0.52 & 10.76$\pm$0.82 & 2.09$\pm$0.54 & 7.04$\pm$0.54 \\
AM0338-320 A & 4.58$\pm$0.14 & 4.41$\pm$0.13 & 0.09$\pm$0.02 & 0.45$\pm$0.02 & 1.42$\pm$0.06 & 0.03$\pm$0.0 & 0.29$\pm$0.02 & 0.1$\pm$0.02 & 1.69$\pm$0.03 & 7.26$\pm$0.13 & 2.9$\pm$0.12 & 1.55$\pm$0.15 \\
AM0338-320 B & 0.52$\pm$0.15 & 1.37$\pm$0.13 & 0.34$\pm$0.04 & 1.19$\pm$0.04 & 4.04$\pm$0.06 & - & 0.07$\pm$0.01 & 0.04$\pm$0.01 & 0.44$\pm$0.02 & 2.61$\pm$0.06 & 1.02$\pm$0.06 & 0.92$\pm$0.03 \\
AM0348-502 G1 & 0.89$\pm$0.12 & 1.08$\pm$0.11 & 0.51$\pm$0.02 & 4.43$\pm$0.04 & 12.87$\pm$0.11 & 0.12$\pm$0.01 & 0.16$\pm$0.02 & 0.26$\pm$0.01 & 4.63$\pm$0.06 & 8.33$\pm$0.2 & 1.16$\pm$0.02 & 1.44$\pm$0.03 \\
AM0348-502 G2 & 38.54$\pm$0.91 & 38.63$\pm$0.98 & - & 7.69$\pm$0.05 & 24.4$\pm$0.14 & 0.15$\pm$0.01 & 1.17$\pm$0.05 & 2.21$\pm$0.07 & 6.81$\pm$0.2 & 18.96$\pm$0.18 & 9.03$\pm$0.11 & 7.1$\pm$0.08 \\
AM0348-502 G3 & 7.58$\pm$0.44 & - & - & 0.39$\pm$0.01 & 0.63$\pm$0.04 & - & 0.19$\pm$0.02 & - & 0.53$\pm$0.03 & 1.64$\pm$0.04 & 1.72$\pm$0.08 & 0.96$\pm$0.13 \\
AM0737-764 A & 32.13$\pm$4.07 & 55.27$\pm$4.02 & - & 4.27$\pm$0.31 & 12.53$\pm$0.72 & 1.18$\pm$0.36 & - & 6.61$\pm$0.48 & 27.41$\pm$1.01 & 60.81$\pm$2.64 & 24.75$\pm$2.58 & 15.44$\pm$1.98 \\
AM0737-764 B & - & - & - & - & - & - & - & - & - & - & - & - \\
AM0830-235 A & 65.87$\pm$12.79 & 75.95$\pm$13.65 & - & 6.96$\pm$0.45 & 21.49$\pm$1.81 & - & 4.37$\pm$0.3 & 4.21$\pm$0.3 & 18.79$\pm$1.11 & 61.44$\pm$1.42 & 6.68$\pm$0.82 & 12.77$\pm$0.81 \\
AM0830-235 B & 83.5$\pm$21.79 & 100.49$\pm$22.83 & - & 15.18$\pm$2.19 & 32.44$\pm$2.15 & - & 1.89$\pm$0.15 & 2.6$\pm$0.14 & 10.86$\pm$0.72 & 18.18$\pm$1.02 & - & - \\
AM1058-243 A & - & 11.1$\pm$0.54 & - & 0.37$\pm$0.05 & 1.63$\pm$0.09 & 0.26$\pm$0.06 & 0.89$\pm$0.07 & 6.81$\pm$0.37 & 7.63$\pm$0.26 & 19.48$\pm$0.33 & 5.1$\pm$0.32 & 1.78$\pm$0.27 \\
AM1058-243 B & 74.67$\pm$23.0 & 196.87$\pm$18.17 & 11.17$\pm$1.77 & 5.45$\pm$0.19 & 30.59$\pm$1.59 & 1.85$\pm$0.28 & 6.97$\pm$0.26 & 25.25$\pm$0.62 & 56.06$\pm$1.24 & 168.08$\pm$3.7 & 26.07$\pm$0.73 & 27.09$\pm$0.67 \\
AM1125-374 A & - & 33.54$\pm$2.18 & 5.2$\pm$1.25 & 2.9$\pm$0.34 & 9.34$\pm$0.94 & - & 2.14$\pm$0.12 & 1.92$\pm$0.13 & 5.0$\pm$0.47 & 12.11$\pm$0.6 & 4.38$\pm$0.22 & 3.9$\pm$0.18 \\
AM1125-374 B & - & 5.4$\pm$0.63 & 0.99$\pm$0.11 & 0.24$\pm$0.11 & 0.62$\pm$0.1 & - & 0.45$\pm$0.06 & 0.28$\pm$0.04 & 0.74$\pm$0.08 & 2.2$\pm$0.07 & 0.68$\pm$0.09 & 0.72$\pm$0.08 \\
AM1132-450 A & 48.64$\pm$4.59 & 34.87$\pm$5.33 & - & 0.43$\pm$0.05 & 4.19$\pm$0.27 & 1.03$\pm$0.34 & 1.16$\pm$0.05 & 1.64$\pm$0.09 & 7.96$\pm$0.15 & 23.51$\pm$0.29 & 12.15$\pm$0.39 & 10.2$\pm$0.68 \\
AM1132-450 B & 16.53$\pm$1.78 & 16.83$\pm$1.45 & - & 0.82$\pm$0.1 & 2.95$\pm$0.26 & 0.1$\pm$0.03 & 0.83$\pm$0.06 & 1.42$\pm$0.11 & 2.48$\pm$0.11 & 8.09$\pm$0.13 & 6.38$\pm$0.18 & 4.72$\pm$0.24 \\
AM1228-260 A & 195.69$\pm$9.32 & 129.97$\pm$12.64 & 0.51$\pm$0.25 & 11.34$\pm$0.52 & 43.8$\pm$2.4 & 4.97$\pm$0.27 & - & 8.38$\pm$0.29 & 156.57$\pm$1.84 & 458.28$\pm$3.18 & 77.83$\pm$1.45 & 77.23$\pm$1.31 \\
AM1228-260 B & 90.71$\pm$1.29 & 47.26$\pm$0.9 & - & 31.77$\pm$0.28 & 93.53$\pm$0.86 & 0.02$\pm$0.01 & 4.18$\pm$0.12 & 0.83$\pm$0.05 & 2.02$\pm$0.7 & 12.87$\pm$0.17 & 8.74$\pm$0.08 & 5.87$\pm$0.05 \\
AM1238-340 A & 7.06$\pm$1.25 & 7.87$\pm$1.13 & 0.35$\pm$0.07 & 0.72$\pm$0.09 & 1.99$\pm$0.15 & 0.07$\pm$0.03 & 0.47$\pm$0.06 & 0.91$\pm$0.09 & - & 8.32$\pm$0.15 & 4.12$\pm$0.14 & 3.98$\pm$0.16 \\
AM1238-340 B & 34.85$\pm$5.83 & 33.63$\pm$5.82 & 3.18$\pm$0.96 & 0.79$\pm$0.14 & 6.67$\pm$0.43 & - & 1.7$\pm$0.18 & 1.78$\pm$0.14 & - & 12.86$\pm$0.36 & 7.95$\pm$0.28 & 7.68$\pm$0.24 \\
AM1411-434 A & - & - & - & - & - & - & - & - & - & - & - & - \\
AM1411-434 B & - & - & - & - & - & - & - & - & - & - & - & - \\
AM1445-274  G1 & 20.37$\pm$6.11 & 65.3$\pm$6.29 & - & 7.2$\pm$0.54 & 23.77$\pm$1.65 & 1.43$\pm$0.18 & 2.24$\pm$0.19 & 5.86$\pm$0.31 & 3.38$\pm$0.97 & 27.99$\pm$1.21 & 22.68$\pm$0.89 & 22.09$\pm$0.76 \\
AM1445-274  G2 & - & - & - & - & - & - & - & - & - & - & - & - \\
AM1722-595 A & 458.22$\pm$12.15 & 282.51$\pm$12.29 & - & 18.97$\pm$0.5 & 84.73$\pm$2.77 & 11.0$\pm$0.9 & 25.71$\pm$0.84 & 17.36$\pm$0.95 & 201.97$\pm$2.99 & 595.29$\pm$4.25 & 169.69$\pm$8.58 & 45.43$\pm$1.88 \\
AM1722-595 B & 33.16$\pm$4.21 & 131.92$\pm$4.12 & - & 3.64$\pm$0.24 & 13.67$\pm$0.62 & 0.77$\pm$0.07 & 1.69$\pm$0.14 & 1.07$\pm$0.07 & 3.12$\pm$0.37 & 14.03$\pm$0.1 & 19.88$\pm$0.94 & 2.79$\pm$0.21 \\
AM1933-422 A & 120.46$\pm$9.0 & 43.49$\pm$10.92 & - & 14.77$\pm$0.7 & 42.56$\pm$1.92 & 3.51$\pm$0.22 & 11.45$\pm$0.46 & 1.32$\pm$0.22 & 97.48$\pm$2.22 & 307.78$\pm$2.1 & 51.03$\pm$1.93 & 57.03$\pm$0.72 \\
AM1933-422 B & 33.7$\pm$4.26 & 19.63$\pm$4.44 & - & 26.16$\pm$0.15 & 82.48$\pm$0.51 & 2.36$\pm$0.16 & 1.64$\pm$0.19 & 0.64$\pm$0.12 & 9.23$\pm$0.33 & 31.34$\pm$0.45 & 14.08$\pm$0.49 & 11.15$\pm$0.6 \\
AM1933-422 C & 59.01$\pm$2.1 & 56.77$\pm$1.87 & - & 5.83$\pm$0.06 & 20.51$\pm$0.15 & 0.54$\pm$0.05 & 1.61$\pm$0.08 & 1.08$\pm$0.05 & 2.15$\pm$0.21 & 8.36$\pm$0.07 & 8.1$\pm$0.05 & 6.64$\pm$0.22 \\
AM1941-564 A & 163.86$\pm$11.98 & 79.16$\pm$13.85 & - & 11.82$\pm$1.06 & 31.57$\pm$1.51 & 10.23$\pm$0.53 & 8.93$\pm$1.32 & - & 46.73$\pm$1.1 & 134.84$\pm$0.79 & 31.31$\pm$1.66 & 37.74$\pm$1.55 \\
AM1941-564 B & 1.04$\pm$0.26 & 2.33$\pm$0.35 & - & 0.41$\pm$0.04 & 1.08$\pm$0.09 & 0.11$\pm$0.01 & 0.15$\pm$0.02 & - & 0.78$\pm$0.06 & 1.11$\pm$0.09 & 0.85$\pm$0.06 & 0.64$\pm$0.05 \\
AM2004-662  G1 & 2.9$\pm$0.27 & 2.2$\pm$0.22 & - & 0.5$\pm$0.03 & 1.47$\pm$0.06 & 0.27$\pm$0.02 & - & 0.3$\pm$0.02 & 2.05$\pm$0.04 & 5.48$\pm$0.05 & 3.84$\pm$0.07 & 3.19$\pm$0.03 \\
AM2004-662  G3 & 0.43$\pm$0.11 & 0.81$\pm$0.1 & - & 0.34$\pm$0.03 & 0.47$\pm$0.04 & 0.43$\pm$0.03 & 0.09$\pm$0.02 & - & 0.54$\pm$0.05 & 0.8$\pm$0.07 & 0.46$\pm$0.05 & 0.56$\pm$0.04 \\
AM2036-510  G1 & 4.62$\pm$0.24 & 3.22$\pm$0.23 & - & 1.02$\pm$0.04 & 3.07$\pm$0.08 & 0.04$\pm$0.01 & - & 0.43$\pm$0.01 & 0.52$\pm$0.09 & 2.05$\pm$0.22 & 2.53$\pm$0.14 & 1.98$\pm$0.18 \\
AM2036-510  G2 & 22.66$\pm$0.77 & 23.76$\pm$0.79 & - & 1.74$\pm$0.09 & 5.27$\pm$0.17 & 0.34$\pm$0.03 & 1.19$\pm$0.06 & 0.56$\pm$0.03 & 4.18$\pm$0.28 & 15.55$\pm$0.23 & 6.26$\pm$0.08 & 4.77$\pm$0.07 \\
AM2036-510  G3 & 63.07$\pm$1.49 & 39.92$\pm$1.46 & - & 6.18$\pm$0.06 & 19.1$\pm$0.14 & 0.22$\pm$0.04 & 1.9$\pm$0.1 & 1.47$\pm$0.07 & 4.68$\pm$0.41 & 19.56$\pm$0.28 & 11.07$\pm$0.12 & 7.94$\pm$0.1 \\
AM2106-624  G2 & 3048.66$\pm$803.87 & 6996.87$\pm$565.07 & - & 81.7$\pm$3.55 & 171.94$\pm$7.52 & 27.74$\pm$1.46 & 11.35$\pm$3.89 & 13.95$\pm$2.28 & 106.19$\pm$7.63 & 514.08$\pm$6.79 & 171.94$\pm$8.82 & 109.6$\pm$8.67 \\
AM2108-452 G1 & 2.34$\pm$0.54 & 12.21$\pm$0.55 & 0.36$\pm$0.09 & 0.17$\pm$0.02 & 0.88$\pm$0.06 & 0.08$\pm$0.01 & 0.27$\pm$0.02 & 0.31$\pm$0.01 & 1.17$\pm$0.04 & 3.48$\pm$0.05 & 1.7$\pm$0.05 & 1.7$\pm$0.08 \\
AM2108-452 G2 & - & - & - & - & - & - & - & - & - & - & - & - \\
AM2108-452 G3 & 4.12$\pm$0.2 & 4.25$\pm$0.24 & - & 0.97$\pm$0.01 & 2.72$\pm$0.05 & - & 0.21$\pm$0.01 & 0.14$\pm$0.0 & 0.2$\pm$0.04 & 0.84$\pm$0.01 & 0.97$\pm$0.01 & 0.65$\pm$0.01 \\
AM2117-515  G1 & - & 2.03$\pm$0.13 & 0.14$\pm$0.01 & 0.07$\pm$0.02 & 0.38$\pm$0.02 & - & 0.09$\pm$0.02 & 0.06$\pm$0.0 & - & 0.95$\pm$0.03 & 0.22$\pm$0.02 & 0.13$\pm$0.02 \\
AM2117-515  G2 & 2.22$\pm$0.35 & 16.71$\pm$0.48 & - & 0.52$\pm$0.05 & 1.59$\pm$0.07 & - & 0.37$\pm$0.02 & 0.12$\pm$0.02 & 0.87$\pm$0.05 & 4.93$\pm$0.09 & 2.51$\pm$0.03 & 1.7$\pm$0.02 \\
AM2117-515  G4 & - & 29.35$\pm$1.84 & - & 2.0$\pm$0.23 & 4.2$\pm$0.13 & - & - & 0.59$\pm$0.06 & 1.4$\pm$0.07 & 6.93$\pm$0.08 & 3.23$\pm$0.08 & 2.03$\pm$0.07 \\
AM2117-515  G5 & 22.57$\pm$0.32 & 19.95$\pm$0.28 & 0.13$\pm$0.04 & 18.42$\pm$0.17 & 54.62$\pm$0.44 & - & 1.22$\pm$0.05 & 0.48$\pm$0.02 & 0.99$\pm$0.03 & 2.58$\pm$0.1 & 3.09$\pm$0.04 & 1.88$\pm$0.05 \\
AM2117-515  G6 & 7.93$\pm$1.48 & 15.53$\pm$1.36 & - & - & 2.02$\pm$0.08 & 0.17$\pm$0.02 & 0.89$\pm$0.02 & 0.28$\pm$0.02 & 1.06$\pm$0.07 & 6.49$\pm$0.13 & 3.64$\pm$0.12 & 2.45$\pm$0.06 \\
AM2117-515  KNOTS & 5.57$\pm$0.09 & 7.44$\pm$0.11 & - & 1.57$\pm$0.02 & 4.65$\pm$0.05 & 0.01$\pm$0.01 & 0.31$\pm$0.02 & 0.15$\pm$0.01 & 0.54$\pm$0.01 & 1.97$\pm$0.04 & 1.64$\pm$0.02 & 1.16$\pm$0.02 \\
AM2132-664  G1 & - & 1118.5$\pm$195.36 & - & 45.26$\pm$4.25 & 214.86$\pm$19.94 & 23.48$\pm$2.19 & 18.27$\pm$0.69 & 24.11$\pm$2.1 & 160.98$\pm$3.84 & 489.88$\pm$3.33 & 57.75$\pm$6.4 & 76.94$\pm$4.1 \\
AM2132-664  G3 & 10.85$\pm$0.34 & 8.17$\pm$0.37 & - & 1.08$\pm$0.03 & 3.78$\pm$0.04 & 0.31$\pm$0.01 & 0.34$\pm$0.06 & 0.28$\pm$0.02 & 1.17$\pm$0.11 & 2.85$\pm$0.06 & 1.83$\pm$0.03 & 1.4$\pm$0.06 \\
AM2217-490  G1 & 119.29$\pm$1.38 & 86.34$\pm$2.0 & - & 17.55$\pm$0.19 & 54.88$\pm$0.45 & 1.37$\pm$0.07 & 2.57$\pm$0.12 & 4.52$\pm$0.21 & 31.39$\pm$0.51 & 96.85$\pm$0.5 & 25.56$\pm$0.55 & 23.31$\pm$0.32 \\
AM2217-490  G2 & 4.31$\pm$0.08 & 7.69$\pm$0.1 & - & 0.32$\pm$0.03 & 1.29$\pm$0.06 & 0.12$\pm$0.02 & 0.4$\pm$0.02 & 0.03$\pm$0.01 & 0.99$\pm$0.07 & 3.7$\pm$0.05 & 1.75$\pm$0.03 & 1.33$\pm$0.03 \\
AM2220-493  G1 & 323.44$\pm$111.18 & 307.94$\pm$107.54 & 199.97$\pm$8.83 & 313.74$\pm$4.51 & 815.01$\pm$18.33 & 1.36$\pm$0.18 & - & 15.85$\pm$0.57 & 93.82$\pm$1.66 & 86.71$\pm$1.61 & 25.3$\pm$0.59 & 24.47$\pm$0.38 \\
AM2220-493  G2 & 15.04$\pm$3.34 & 26.59$\pm$3.07 & - & 30.26$\pm$0.65 & 88.12$\pm$1.12 & 0.49$\pm$0.1 & 0.91$\pm$0.13 & 3.24$\pm$0.11 & 13.06$\pm$0.13 & 35.43$\pm$0.25 & 12.84$\pm$0.31 & 10.37$\pm$0.38 \\
\end{longtable}
\end{landscape}
\twocolumn

\begin{table*}[ht]
\caption{Extinction corrected permitted emission lines measurements in the INT sample.}
\label{tab:INTHflux}
\centering
\small
\begin{tabular}{cccccccc} 
\hline\hline
Object & Av & H$\delta$ & H$\gamma$ & H$\beta$ & H$\alpha$ & EW H$\alpha$\\
(1) & [mag] & flux x $\rm 10^{-15}$ & flux x $\rm 10^{-15}$ & flux x $\rm 10^{-15}$ & flux x $\rm 10^{-15}$& [$\mathring{A}$]
\\
& & $[\ erg\ s^{-1}\ cm^{-2}]$ & $[\ erg\ s^{-1}\ cm^{-2}]$ & $[\ erg\ s^{-1}\ cm^{-2}]$ & $[\ erg\ s^{-1}\ cm^{-2}]$ &
\\
\hline \hline
AM0012-235 A & - & - & - & - & - & 0.7$\pm$0.1 \\
AM0012-235 B & - & - & - & - & - & 1.0$\pm$0.1 \\
AM0103-520 A & 3.1$\pm$0.2 & 0.72$\pm$0.11 & 0.57$\pm$0.03 & 0.87$\pm$0.07 & 2.49$\pm$0.04 & 7.2$\pm$0.1 \\
AM0103-520 B & - & - & - & - & - & 0.3$\pm$0.1 \\
AM0302-611 G1 & 5.6$\pm$0.1 & 121.45$\pm$5.04 & 59.88$\pm$3.04 & 14.6$\pm$0.61 & 41.9$\pm$0.44 & 4.9$\pm$0.0 \\
AM0302-611 G2 & 1.8$\pm$0.1 & 0.23$\pm$0.05 & 0.69$\pm$0.04 & 1.51$\pm$0.06 & 4.32$\pm$0.03 & 27.9$\pm$0.2 \\
AM0302-611 G3 & - & - & - & - & - & 0.0$\pm$0.0 \\
AM0302-611 G4 & 4.9$\pm$0.4 & - & - & 3.54$\pm$0.44 & 10.17$\pm$0.79 & 1.8$\pm$0.1 \\
AM0338-320 A & 1.8$\pm$0.2 & 1.17$\pm$0.13 & 2.33$\pm$0.2 & 4.55$\pm$0.36 & 13.06$\pm$0.2 & 63.8$\pm$0.4 \\
AM0338-320 B & 2.7$\pm$0.2 & 0.5$\pm$0.06 & 0.65$\pm$0.05 & 0.8$\pm$0.04 & 2.29$\pm$0.05 & 11.0$\pm$0.2 \\
AM0348-502 G1 & 1.0$\pm$0.2 & 0.19$\pm$0.03 & 0.77$\pm$0.04 & 1.48$\pm$0.07 & 4.24$\pm$0.03 & 9.2$\pm$0.0 \\
AM0348-502 G2 & 2.2$\pm$0.1 & 3.68$\pm$0.12 & 6.62$\pm$0.25 & 15.24$\pm$0.59 & 43.75$\pm$0.25 & 81.4$\pm$0.4 \\
AM0348-502 G3 & 3.4$\pm$0.2 & 2.32$\pm$0.59 & 0.33$\pm$0.1 & 1.49$\pm$0.09 & 4.29$\pm$0.04 & 11.0$\pm$0.1 \\
AM0737-764 A & 2.9$\pm$0.2 & 1.33$\pm$0.62 & 6.27$\pm$0.45 & 14.56$\pm$0.64 & 41.78$\pm$1.21 & 4.9$\pm$0.1 \\
AM0737-764 B & - & - & - & - & - & 0.0$\pm$0.0 \\
AM0830-235 A & 3.7$\pm$0.1 & - & 7.15$\pm$0.97 & 38.54$\pm$1.34 & 110.6$\pm$0.88 & 7.7$\pm$0.1 \\
AM0830-235 B & 4.5$\pm$0.2 & - & - & 5.72$\pm$0.16 & 16.42$\pm$1.02 & 1.0$\pm$0.1 \\
AM1058-243 A & 3.7$\pm$0.2 & 3.11$\pm$0.39 & 7.59$\pm$0.44 & 13.79$\pm$1.01 & 39.57$\pm$0.54 & 27.1$\pm$0.2 \\
AM1058-243 B & 5.5$\pm$0.3 & 52.9$\pm$8.28 & 83.73$\pm$9.39 & 95.81$\pm$9.51 & 274.98$\pm$6.05 & 75.4$\pm$0.4 \\
AM1125-374 A & 4.0$\pm$0.2 & 14.44$\pm$1.95 & 10.66$\pm$1.32 & 6.42$\pm$0.25 & 18.44$\pm$0.68 & 5.4$\pm$0.2 \\
AM1125-374 B & 3.2$\pm$0.2 & 1.75$\pm$0.29 & 1.65$\pm$0.26 & 1.13$\pm$0.09 & 3.25$\pm$0.06 & 4.0$\pm$0.1 \\
AM1132-450 A & 4.0$\pm$0.1 & - & - & 14.12$\pm$0.5 & 40.52$\pm$0.28 & 8.1$\pm$0.1 \\
AM1132-450 B & 3.0$\pm$0.1 & 3.1$\pm$0.45 & 1.77$\pm$0.54 & 8.38$\pm$0.19 & 24.05$\pm$0.09 & 21.9$\pm$0.1 \\
AM1228-260 A & 3.1$\pm$0.1 & 76.71$\pm$6.83 & 146.0$\pm$11.56 & 291.02$\pm$12.9 & 835.22$\pm$5.35 & 85.4$\pm$0.4 \\
AM1228-260 B & 2.2$\pm$0.1 & 7.58$\pm$0.6 & 17.55$\pm$0.9 & 38.67$\pm$1.74 & 110.98$\pm$0.86 & 382.3$\pm$2.3 \\
AM1238-340 A & 3.0$\pm$0.2 & 0.86$\pm$0.13 & 3.59$\pm$0.39 & 4.55$\pm$0.22 & 13.07$\pm$0.21 & 8.7$\pm$0.1 \\
AM1238-340 B & 4.8$\pm$0.1 & - & 11.5$\pm$0.54 & 6.79$\pm$0.32 & 19.49$\pm$0.22 & 4.9$\pm$0.1 \\
AM1411-434 A & - & - & - & - & - & 0.0$\pm$0.0 \\
AM1411-434 B & - & - & - & - & - & 0.0$\pm$0.0 \\
AM1445-274  G1 & 4.2$\pm$0.4 & 16.45$\pm$1.83 & - & 13.36$\pm$1.67 & 38.33$\pm$1.62 & 10.6$\pm$0.3 \\
AM1445-274  G2 & - & - & - & - & - & 0.0$\pm$0.0 \\
AM1722-595 A & 3.6$\pm$0.1 & 123.72$\pm$6.24 & 193.49$\pm$7.79 & 401.2$\pm$17.49 & 1151.44$\pm$6.92 & 90.9$\pm$0.4 \\
AM1722-595 B & 3.2$\pm$0.1 & - & 2.06$\pm$0.15 & 16.96$\pm$0.57 & 48.66$\pm$0.35 & 26.9$\pm$0.2 \\
AM1933-422 A & 2.9$\pm$0.1 & 57.96$\pm$3.41 & 74.36$\pm$3.03 & 201.02$\pm$7.77 & 576.94$\pm$3.2 & 85.5$\pm$0.4 \\
AM1933-422 B & 2.2$\pm$0.1 & 3.42$\pm$0.13 & 1.47$\pm$0.54 & 13.81$\pm$0.34 & 39.64$\pm$0.43 & 7.7$\pm$0.1 \\
AM1933-422 C & 2.3$\pm$0.1 & 6.13$\pm$0.49 & 6.18$\pm$0.35 & 13.55$\pm$0.52 & 38.9$\pm$0.24 & 35.3$\pm$0.2 \\
AM1941-564 A & 5.2$\pm$0.1 & - & - & 74.35$\pm$2.53 & 213.38$\pm$1.57 & 9.6$\pm$0.1 \\
AM1941-564 B & 1.8$\pm$0.3 & - & - & 0.51$\pm$0.05 & 1.46$\pm$0.1 & 1.1$\pm$0.1 \\
AM2004-662  G1 & 1.8$\pm$0.1 & 1.76$\pm$0.15 & 1.92$\pm$0.06 & 4.34$\pm$0.2 & 12.45$\pm$0.07 & 22.8$\pm$0.1 \\
AM2004-662  G3 & 1.7$\pm$0.4 & 0.11$\pm$0.03 & - & 0.31$\pm$0.03 & 0.88$\pm$0.07 & 1.1$\pm$0.1 \\
AM2036-510  G1 & 1.4$\pm$0.2 & - & - & 1.49$\pm$0.02 & 4.27$\pm$0.27 & 5.0$\pm$0.3 \\
AM2036-510  G2 & 2.1$\pm$0.2 & 3.66$\pm$0.16 & 5.83$\pm$0.28 & 14.51$\pm$0.72 & 41.64$\pm$0.35 & 66.3$\pm$0.4 \\
AM2036-510  G3 & 2.7$\pm$0.1 & 4.87$\pm$0.45 & 8.19$\pm$0.42 & 20.38$\pm$0.85 & 58.48$\pm$0.42 & 94.3$\pm$0.6 \\
AM2106-624  G2 & 7.4$\pm$0.2 & - & - & 233.27$\pm$13.23 & 669.48$\pm$6.17 & 19.6$\pm$0.1 \\
AM2108-452 G1 & 2.4$\pm$0.1 & 0.83$\pm$0.2 & 2.22$\pm$0.13 & 2.53$\pm$0.09 & 7.27$\pm$0.04 & 14.9$\pm$0.1 \\
AM2108-452 G2 & - & - & - & - & - & 0.0$\pm$0.0 \\
AM2108-452 G3 & 2.2$\pm$0.1 & 0.67$\pm$0.03 & 1.16$\pm$0.08 & 1.65$\pm$0.07 & 4.74$\pm$0.04 & 40.7$\pm$0.3 \\
AM2117-515  G1 & 1.8$\pm$0.1 & - & 0.22$\pm$0.04 & 0.31$\pm$0.01 & 0.88$\pm$0.01 & 6.0$\pm$0.1 \\
AM2117-515  G2 & 2.4$\pm$0.2 & - & 1.53$\pm$0.19 & 4.86$\pm$0.28 & 13.96$\pm$0.13 & 43.3$\pm$0.2 \\
AM2117-515  G4 & 3.5$\pm$0.1 & 6.24$\pm$0.67 & 4.64$\pm$0.37 & 6.07$\pm$0.23 & 17.41$\pm$0.1 & 35.0$\pm$0.2 \\
AM2117-515  G5 & 2.0$\pm$0.1 & 1.92$\pm$0.13 & 4.67$\pm$0.27 & 11.02$\pm$0.49 & 31.62$\pm$0.4 & 280.2$\pm$3.2 \\
AM2117-515  G6 & 3.4$\pm$0.2 & 1.72$\pm$0.08 & 1.2$\pm$0.04 & 6.77$\pm$0.42 & 19.43$\pm$0.21 & 39.7$\pm$0.3 \\
AM2117-515  KNOTS & 1.4$\pm$0.2 & 0.25$\pm$0.02 & 1.2$\pm$0.07 & 3.29$\pm$0.19 & 9.45$\pm$0.11 & 78.1$\pm$0.7 \\
AM2132-664  G1 & 6.2$\pm$0.0 & - & 38.83$\pm$13.17 & 281.77$\pm$3.01 & 808.68$\pm$3.78 & 15.3$\pm$0.1 \\
AM2132-664  G3 & 2.0$\pm$0.2 & 0.74$\pm$0.08 & 2.12$\pm$0.17 & 3.87$\pm$0.21 & 11.11$\pm$0.09 & 29.3$\pm$0.1 \\
AM2217-490  G1 & 2.9$\pm$0.1 & 16.33$\pm$0.73 & 23.13$\pm$0.77 & 53.93$\pm$1.95 & 154.79$\pm$0.78 & 47.3$\pm$0.2 \\
AM2217-490  G2 & 2.0$\pm$0.1 & 1.17$\pm$0.06 & 1.79$\pm$0.06 & 3.92$\pm$0.16 & 11.25$\pm$0.07 & 33.2$\pm$0.2 \\
AM2220-493  G1 & 6.5$\pm$0.2 & - & 157.94$\pm$20.21 & 65.4$\pm$3.77 & 187.69$\pm$5.18 & 14.7$\pm$0.4 \\
AM2220-493  G2 & 3.7$\pm$0.2 & 0.83$\pm$0.06 & 4.55$\pm$0.1 & 12.76$\pm$0.62 & 36.62$\pm$0.64 & 26.6$\pm$0.4 \\  \hline
\end{tabular}
\tablefoot{\footnotesize (1) System name and component according to \cite{ArpMadore+87} catalogue and Table~\ref{tab:Sample}.}
\end{table*}

\begin{sidewaystable*}
\caption{Extinction corrected emission lines measurements for forbidden and helium transitions in the CTR sample.}
\label{tab:CTRforbflux}
\centering
\renewcommand{\tabcolsep}{.5mm}
\tiny
\begin{tabular}{ccccccccccccc}
\hline\hline
Object & [O\,{\sc ii}]$\lambda$3726 & [O\,{\sc ii}]$\lambda$3729 & [O\,{\sc iii}]$\lambda$4363 & [O\,{\sc iii}]$\lambda$4959 & [O\,{\sc iii}]$\lambda$5007 & [N\,{\sc ii}]$\lambda$5755 & He\,{\sc i}$\lambda$5876 & [O\,{\sc i}]$\lambda$6300 & [N\,{\sc ii}]$\lambda$6548 & [N\,{\sc ii}]$\lambda$6583 & [S\,{\sc ii}]$\lambda$6716 & [S\,{\sc ii}]$\lambda$6731 
\\
& flux x $\rm 10^{-15}$ & flux x $\rm 10^{-15}$ & flux x $\rm 10^{-15}$ & flux x $\rm 10^{-15}$ & flux x $\rm 10^{-15}$ & flux x $\rm 10^{-15}$ & flux x $\rm 10^{-15}$ & flux x $\rm 10^{-15}$ & flux x $\rm 10^{-15}$ & flux x $\rm 10^{-15}$ & flux x $\rm 10^{-15}$ & flux x $\rm 10^{-15}$
\\
& $[\ erg\ s^{-1}\ cm^{-2}]$ & $[\ erg\ s^{-1}\ cm^{-2}]$ & $[\ erg\ s^{-1}\ cm^{-2}]$ & $[\ erg\ s^{-1}\ cm^{-2}]$ & $[\ erg\ s^{-1}\ cm^{-2}]$ & $[\ erg\ s^{-1}\ cm^{-2}]$ & $[\ erg\ s^{-1}\ cm^{-2}]$ & $[\ erg\ s^{-1}\ cm^{-2}]$ & $[\ erg\ s^{-1}\ cm^{-2}]$ & $[\ erg\ s^{-1}\ cm^{-2}]$ & $[\ erg\ s^{-1}\ cm^{-2}]$ & $[\ erg\ s^{-1}\ cm^{-2}]$
\\
\hline
%
%
0275-51910-0320 & - & - & - & - & 0.9$\pm$0.09 & - & 1.38$\pm$0.19 & 1.34$\pm$0.1 & 2.25$\pm$0.05 & 6.77$\pm$0.1 & 2.59$\pm$0.06 & 1.69$\pm$0.04 \\
0420-51871-0285 & 1.39$\pm$0.63 & 3.86$\pm$0.86 & - & 0.09$\pm$0.05 & 0.7$\pm$0.05 & - & - & - & - & 0.29$\pm$0.06 & 0.39$\pm$0.14 & 0.32$\pm$0.08 \\
0420-51871-0517 & 0.5$\pm$0.26 & 5.87$\pm$0.27 & 0.41$\pm$0.13 & 0.37$\pm$0.07 & 1.02$\pm$0.1 & 0.33$\pm$0.05 & 0.43$\pm$0.07 & - & 1.45$\pm$0.03 & 4.5$\pm$0.02 & 1.68$\pm$0.03 & 1.27$\pm$0.04 \\
0421-51821-0003 & - & 7.34$\pm$3.57 & - & 1.63$\pm$0.73 & 1.71$\pm$0.64 & 0.63$\pm$0.22 & - & - & 1.55$\pm$0.5 & 3.07$\pm$1.02 & 0.51$\pm$0.21 & 0.54$\pm$0.2 \\
0731-52460-0345 & 2.21$\pm$0.5 & 8.69$\pm$0.62 & - & 1.16$\pm$0.24 & 3.5$\pm$0.14 & - & 0.71$\pm$0.18 & 0.52$\pm$0.09 & 1.57$\pm$0.11 & 4.19$\pm$0.06 & 2.33$\pm$0.11 & 2.04$\pm$0.08 \\
0860-52319-0120 & 0.71$\pm$0.16 & 6.35$\pm$0.19 & - & 0.2$\pm$0.06 & 0.75$\pm$0.04 & - & 0.38$\pm$0.11 & 0.2$\pm$0.06 & 1.06$\pm$0.02 & 3.37$\pm$0.02 & 1.6$\pm$0.01 & 1.2$\pm$0.01 \\
1266-52709-0142 & 3.88$\pm$2.78 & 22.52$\pm$13.23 & - & 1.25$\pm$0.79 & 2.79$\pm$1.55 & - & - & - & 1.0$\pm$0.35 & 1.98$\pm$0.71 & 1.64$\pm$0.59 & 1.5$\pm$0.48 \\
1588-52965-0534 & 0.46$\pm$0.13 & 0.71$\pm$0.06 & - & - & 0.16$\pm$0.06 & 0.19$\pm$0.02 & 0.06$\pm$0.02 & 0.09$\pm$0.04 & 0.18$\pm$0.03 & 0.62$\pm$0.02 & 0.34$\pm$0.02 & 0.25$\pm$0.01 \\
1601-53115-0280 & - & 2.67$\pm$1.78 & 0.36$\pm$0.23 & 0.34$\pm$0.18 & 0.29$\pm$0.2 & - & - & 0.18$\pm$0.1 & 0.22$\pm$0.1 & 1.04$\pm$0.4 & 0.38$\pm$0.21 & - \\
1607-53083-0452 & 18.72$\pm$2.32 & 96.25$\pm$2.68 & 1.22$\pm$0.27 & 5.2$\pm$0.27 & 17.33$\pm$0.56 & 1.33$\pm$0.31 & 2.16$\pm$0.65 & 5.37$\pm$0.57 & 29.88$\pm$0.64 & 74.02$\pm$1.29 & 21.86$\pm$0.34 & 17.81$\pm$0.15 \\
1652-53555-0578 & 0.38$\pm$0.06 & 0.72$\pm$0.0 & - & 0.08$\pm$0.01 & 0.37$\pm$0.01 & - & 0.07$\pm$0.03 & 0.08$\pm$0.03 & 0.09$\pm$0.01 & 0.23$\pm$0.02 & 0.27$\pm$0.01 & 0.15$\pm$0.01 \\
1693-53446-0629 & - & 1.54$\pm$0.19 & - & 0.1$\pm$0.03 & 0.37$\pm$0.11 & - & 0.39$\pm$0.14 & - & 0.5$\pm$0.05 & 1.49$\pm$0.03 & 0.49$\pm$0.03 & 0.32$\pm$0.05 \\
1771-53498-0483 & - & 38.19$\pm$2.28 & 1.43$\pm$0.63 & 1.94$\pm$0.58 & 8.36$\pm$0.45 & - & - & 1.68$\pm$0.18 & 6.44$\pm$0.65 & 25.85$\pm$1.4 & 7.42$\pm$0.46 & 6.65$\pm$0.28 \\
1776-53858-0098 & 2.45$\pm$0.85 & 5.12$\pm$0.21 & 0.06$\pm$0.05 & 0.08$\pm$0.03 & 0.75$\pm$0.06 & - & 0.5$\pm$0.11 & 0.21$\pm$0.1 & 0.36$\pm$0.01 & 0.85$\pm$0.04 & 0.83$\pm$0.01 & 0.51$\pm$0.02 \\
1852-53534-0298 & - & - & 4.74$\pm$1.12 & - & 6.12$\pm$2.5 & 1.25$\pm$0.38 & 7.08$\pm$1.15 & 5.77$\pm$0.75 & 20.42$\pm$0.33 & 60.68$\pm$0.66 & 16.38$\pm$0.14 & 12.5$\pm$0.13 \\
1929-53349-0372 & - & - & 0.41$\pm$0.22 & - & - & 0.42$\pm$0.13 & 0.47$\pm$0.14 & 0.18$\pm$0.06 & 0.58$\pm$0.07 & 1.75$\pm$0.06 & 0.95$\pm$0.17 & 0.45$\pm$0.15 \\
1937-53388-0453 & - & 44.72$\pm$7.5 & 6.46$\pm$1.17 & 2.65$\pm$0.93 & 5.49$\pm$0.95 & 2.36$\pm$0.33 & - & 1.13$\pm$0.14 & 6.67$\pm$0.46 & 20.34$\pm$0.41 & 4.57$\pm$0.67 & 4.74$\pm$0.71 \\
2094-53851-0261 & - & 297.27$\pm$35.83 & - & - & 22.46$\pm$1.59 & 1.2$\pm$0.55 & 5.47$\pm$1.0 & 5.89$\pm$0.97 & 22.63$\pm$0.89 & 67.91$\pm$1.58 & 18.05$\pm$0.5 & 15.93$\pm$0.5 \\
2124-53770-0272 & - & - & - & - & - & - & - & - & - & - & - & - \\
2126-53794-0075 & 3.43$\pm$0.86 & 14.02$\pm$0.6 & 0.44$\pm$0.15 & 2.37$\pm$0.08 & 5.99$\pm$0.2 & 0.45$\pm$0.13 & - & - & 1.17$\pm$0.16 & 2.86$\pm$0.03 & 1.82$\pm$0.12 & 1.85$\pm$0.04 \\
2130-53881-0553 & 23.74$\pm$5.01 & 185.38$\pm$5.91 & 6.2$\pm$1.08 & 80.01$\pm$3.52 & 246.11$\pm$10.29 & 2.35$\pm$0.29 & 5.38$\pm$0.81 & 7.38$\pm$1.91 & 41.08$\pm$1.24 & 91.91$\pm$2.9 & 22.81$\pm$0.95 & 18.45$\pm$0.9 \\
2132-53493-0224 & - & - & - & - & - & - & - & - & - & - & - & - \\
2170-53875-0347 & - & 4.93$\pm$0.56 & - & - & 1.21$\pm$0.26 & - & - & - & 0.72$\pm$0.08 & 2.03$\pm$0.07 & 0.54$\pm$0.17 & 0.3$\pm$0.09 \\
2227-53820-0018 & - & - & - & - & - & - & - & - & - & - & - & - \\
2234-53823-0120 & 11.92$\pm$2.26 & 11.87$\pm$2.12 & - & - & - & 0.42$\pm$0.19 & - & 0.55$\pm$0.15 & 1.3$\pm$0.04 & 4.05$\pm$0.12 & 0.88$\pm$0.4 & - \\
2268-53682-0424 & 1.0$\pm$1.02 & 1.12$\pm$1.13 & - & - & 0.81$\pm$0.62 & - & - & - & 0.6$\pm$0.34 & 1.7$\pm$0.93 & 0.44$\pm$0.31 & 0.27$\pm$0.15 \\
2273-53709-0027 & - & - & 1.23$\pm$0.2 & - & - & - & 0.97$\pm$0.2 & 1.13$\pm$0.17 & 0.75$\pm$0.02 & 2.11$\pm$0.03 & 1.75$\pm$0.1 & 1.19$\pm$0.05 \\
2277-53705-0378 & - & - & 0.24$\pm$0.04 & - & 0.18$\pm$0.04 & - & 0.27$\pm$0.04 & 0.12$\pm$0.03 & 0.18$\pm$0.03 & 0.45$\pm$0.03 & 0.33$\pm$0.02 & 0.18$\pm$0.01 \\
2350-53765-0539 & - & - & 1.13$\pm$0.26 & - & 3.42$\pm$1.31 & 0.8$\pm$0.06 & 2.88$\pm$0.45 & 1.85$\pm$0.17 & 5.65$\pm$0.13 & 15.95$\pm$0.09 & 5.9$\pm$0.12 & 4.06$\pm$0.06 \\
2352-53770-0006 & - & - & - & - & - & - & - & - & - & - & - & - \\
2426-53795-0583 & 9.97$\pm$1.54 & 17.03$\pm$2.65 & - & - & 1.4$\pm$0.54 & - & 1.7$\pm$0.44 & 1.04$\pm$0.15 & 2.57$\pm$0.09 & 7.56$\pm$0.1 & 3.75$\pm$0.14 & 2.6$\pm$0.1 \\
2496-54178-0330 & - & - & - & - & - & - & - & - & - & - & - & - \\
2515-54180-0046 & - & - & 0.1$\pm$0.04 & 1.84$\pm$0.01 & 5.67$\pm$0.08 & - & 0.34$\pm$0.09 & 0.25$\pm$0.04 & 0.14$\pm$0.01 & 0.45$\pm$0.03 & 0.64$\pm$0.01 & 0.41$\pm$0.0 \\
2518-54243-0066 & 6.86$\pm$1.93 & 100.9$\pm$3.87 & 2.58$\pm$0.58 & 15.01$\pm$0.31 & 48.99$\pm$1.49 & - & 3.04$\pm$0.96 & 4.53$\pm$1.18 & 21.79$\pm$0.3 & 62.17$\pm$1.01 & 17.29$\pm$0.2 & 15.22$\pm$0.27 \\
2523-54572-0428 & - & 22.8$\pm$1.08 & 2.17$\pm$0.6 & - & 2.42$\pm$0.69 & - & 1.42$\pm$0.28 & - & 7.2$\pm$0.07 & 21.17$\pm$0.24 & 6.14$\pm$0.05 & 5.29$\pm$0.05 \\
2580-54092-0232 & - & - & 0.4$\pm$0.17 & - & 0.68$\pm$0.19 & - & 0.23$\pm$0.06 & 0.27$\pm$0.1 & 0.36$\pm$0.08 & 0.89$\pm$0.17 & 0.61$\pm$0.18 & 0.2$\pm$0.08 \\
2599-54234-0571 & - & 4.54$\pm$0.48 & - & - & - & - & 0.33$\pm$0.06 & 0.24$\pm$0.07 & 0.28$\pm$0.01 & 1.05$\pm$0.03 & 0.58$\pm$0.03 & 0.37$\pm$0.04 \\
2613-54481-0304 & 1.51$\pm$0.75 & 8.44$\pm$0.94 & 0.28$\pm$0.11 & 1.4$\pm$0.07 & 3.75$\pm$0.2 & 0.57$\pm$0.1 & 0.22$\pm$0.11 & 0.47$\pm$0.11 & 1.01$\pm$0.04 & 3.1$\pm$0.12 & 1.23$\pm$0.04 & 0.84$\pm$0.06 \\
2746-54232-0165 & 109.85$\pm$28.71 & 189.8$\pm$25.56 & - & - & 7.09$\pm$2.82 & 2.36$\pm$0.76 & - & 3.14$\pm$0.53 & 4.2$\pm$0.31 & 11.68$\pm$0.22 & 4.78$\pm$0.86 & 2.49$\pm$1.05 \\
2764-54535-0355 & - & 5.0$\pm$0.33 & - & - & 1.2$\pm$0.33 & - & 0.44$\pm$0.07 & - & 1.57$\pm$0.05 & 4.61$\pm$0.15 & 1.31$\pm$0.22 & 1.34$\pm$0.18 \\
2765-54535-0129 & - & 53.71$\pm$3.05 & 2.74$\pm$0.76 & - & 3.57$\pm$1.15 & - & 2.56$\pm$0.43 & 1.51$\pm$0.21 & 6.59$\pm$0.18 & 19.46$\pm$0.13 & 6.32$\pm$0.19 & 4.77$\pm$0.08 \\
2774-54534-0448 & - & 22.51$\pm$0.96 & - & 2.48$\pm$0.22 & 6.7$\pm$0.38 & 0.52$\pm$0.08 & - & 1.49$\pm$0.44 & 4.28$\pm$0.29 & 13.81$\pm$0.49 & 3.76$\pm$0.49 & 2.07$\pm$0.21 \\
2776-54554-0098 & - & 208.23$\pm$237.22 & - & - & 5.07$\pm$4.8 & 1.65$\pm$1.42 & 3.22$\pm$2.35 & - & 4.75$\pm$2.9 & 10.36$\pm$6.25 & 3.65$\pm$2.37 & - \\
2782-54592-0240 & 0.25$\pm$0.13 & 1.73$\pm$0.18 & - & - & - & 0.15$\pm$0.06 & 0.29$\pm$0.04 & 0.21$\pm$0.06 & 0.21$\pm$0.02 & 0.59$\pm$0.02 & 0.34$\pm$0.01 & 0.26$\pm$0.01 \\
1693-53446-0629 & - & 1.54$\pm$0.19 & - & 0.1$\pm$0.03 & 0.37$\pm$0.11 & - & 0.39$\pm$0.14 & - & 0.5$\pm$0.05 & 1.49$\pm$0.03 & 0.49$\pm$0.03 & 0.32$\pm$0.05 \\
\end{tabular}
\tablefoot{\footnotesize (1) Control sample galaxies Plate-MJd-Fiberid information from SDSS DR9 and matched on Table~\ref{tab:radial_vel}.}
\end{sidewaystable*}

\begin{table*}[ht]
\caption{Extinction corrected permitted emission lines measurements in the CTR sample.}
\label{tab:CTRHflux}
\centering
\small
\begin{tabular}{cccccccc} 
\hline\hline
Object & Av & H$\delta$ & H$\gamma$ & H$\beta$ & H$\alpha$ & EW H$\alpha$\\
(1) & [mag] & flux x $\rm 10^{-15}$ & flux x $\rm 10^{-15}$ & flux x $\rm 10^{-15}$ & flux x $\rm 10^{-15}$& [$\mathring{A}$]
\\
& & $[\ erg\ s^{-1}\ cm^{-2}]$ & $[\ erg\ s^{-1}\ cm^{-2}]$ & $[\ erg\ s^{-1}\ cm^{-2}]$ & $[\ erg\ s^{-1}\ cm^{-2}]$ &
\\
\hline \hline
0275-51910-0320 & 1.3$\pm$0.1 & 2.58$\pm$0.11 & 3.86$\pm$0.24 & 6.2$\pm$0.09 & 17.79$\pm$0.16 & 38.8$\pm$0.3 \\
0420-51871-0285 & 1.2$\pm$0.1 & 0.19$\pm$0.02 & 0.23$\pm$0.03 & 0.43$\pm$0.01 & 1.22$\pm$0.03 & 12.2$\pm$0.3 \\
0420-51871-0517 & 1.3$\pm$0.1 & 1.02$\pm$0.05 & 1.8$\pm$0.05 & 4.09$\pm$0.1 & 11.73$\pm$0.14 & 25.7$\pm$0.3 \\
0421-51821-0003 & 2.3$\pm$1.0 & - & 2.1$\pm$1.03 & 1.35$\pm$0.6 & 3.88$\pm$1.23 & 2.4$\pm$0.5 \\
0731-52460-0345 & 1.5$\pm$0.2 & 1.48$\pm$0.32 & 1.29$\pm$0.25 & 2.06$\pm$0.14 & 5.92$\pm$0.09 & 4.5$\pm$0.1 \\
0860-52319-0120 & 1.2$\pm$0.1 & 1.02$\pm$0.16 & 1.48$\pm$0.06 & 2.87$\pm$0.03 & 8.24$\pm$0.15 & 21.3$\pm$0.4 \\
1266-52709-0142 & 2.2$\pm$1.1 & - & 4.59$\pm$2.42 & 1.11$\pm$0.62 & 3.18$\pm$1.07 & 1.5$\pm$0.2 \\
1588-52965-0534 & 0.8$\pm$0.3 & 0.22$\pm$0.03 & 0.39$\pm$0.08 & 0.56$\pm$0.05 & 1.6$\pm$0.03 & 6.4$\pm$0.1 \\
1601-53115-0280 & 1.0$\pm$1.2 & 0.75$\pm$0.52 & 0.77$\pm$0.47 & 0.45$\pm$0.28 & 1.3$\pm$0.5 & 2.1$\pm$0.3 \\
1607-53083-0452 & 1.7$\pm$0.1 & 12.13$\pm$0.57 & 18.93$\pm$0.95 & 31.87$\pm$1.21 & 91.46$\pm$1.16 & 38.7$\pm$0.5 \\
1652-53555-0578 & 0.1$\pm$0.1 & 0.07$\pm$0.04 & 0.11$\pm$0.01 & 0.37$\pm$0.0 & 1.05$\pm$0.01 & 19.3$\pm$0.2 \\
1693-53446-0629 & 1.2$\pm$0.1 & 0.43$\pm$0.12 & 0.7$\pm$0.09 & 1.29$\pm$0.06 & 3.7$\pm$0.04 & 9.1$\pm$0.1 \\
1771-53498-0483 & 2.1$\pm$0.3 & 7.06$\pm$1.43 & 7.32$\pm$1.15 & 15.23$\pm$1.42 & 43.71$\pm$3.27 & 15.9$\pm$1.1 \\
1776-53858-0098 & 1.2$\pm$0.2 & 0.94$\pm$0.12 & 0.65$\pm$0.11 & 1.01$\pm$0.06 & 2.91$\pm$0.04 & 7.9$\pm$0.1 \\
1852-53534-0298 & 2.5$\pm$0.1 & 24.06$\pm$2.7 & 28.45$\pm$1.76 & 37.19$\pm$0.78 & 106.75$\pm$1.28 & 21.1$\pm$0.3 \\
1929-53349-0372 & 1.6$\pm$0.3 & 1.63$\pm$0.54 & 1.43$\pm$0.14 & 1.83$\pm$0.15 & 5.26$\pm$0.1 & 5.9$\pm$0.1 \\
1937-53388-0453 & 3.2$\pm$0.3 & 11.58$\pm$1.86 & 13.86$\pm$1.81 & 12.56$\pm$1.1 & 36.04$\pm$0.83 & 5.9$\pm$0.1 \\
2094-53851-0261 & 3.4$\pm$0.2 & - & 20.15$\pm$3.63 & 31.01$\pm$2.21 & 89.0$\pm$3.35 & 17.4$\pm$0.6 \\
2124-53770-0272 & - & - & - & - & - & 0.4$\pm$0.1 \\
2126-53794-0075 & 1.4$\pm$0.1 & - & - & 1.21$\pm$0.03 & 3.46$\pm$0.04 & 2.1$\pm$0.0 \\
2130-53881-0553 & 2.0$\pm$0.2 & 16.02$\pm$0.76 & 23.83$\pm$0.79 & 36.83$\pm$2.26 & 105.7$\pm$2.22 & 19.9$\pm$0.4 \\
2132-53493-0224 & - & - & - & - & - & 2.3$\pm$0.8 \\
2170-53875-0347 & 1.2$\pm$0.3 & 0.55$\pm$0.14 & 0.23$\pm$0.13 & 0.96$\pm$0.09 & 2.76$\pm$0.08 & 2.9$\pm$0.1 \\
2227-53820-0018 & - & - & - & - & - & 0.3$\pm$0.1 \\
2234-53823-0120 & 1.6$\pm$0.3 & 3.05$\pm$0.53 & 2.22$\pm$0.21 & 2.79$\pm$0.28 & 8.01$\pm$0.2 & 4.5$\pm$0.1 \\
2268-53682-0424 & 1.7$\pm$1.5 & 0.86$\pm$0.88 & 0.87$\pm$0.8 & 0.68$\pm$0.57 & 1.95$\pm$1.1 & 2.6$\pm$0.6 \\
2273-53709-0027 & 1.0$\pm$0.1 & 2.12$\pm$0.3 & 1.69$\pm$0.02 & 2.21$\pm$0.03 & 6.36$\pm$0.11 & 11.0$\pm$0.2 \\
2277-53705-0378 & 0.8$\pm$0.1 & 0.18$\pm$0.09 & 0.25$\pm$0.06 & 0.43$\pm$0.01 & 1.24$\pm$0.03 & 5.5$\pm$0.1 \\
2350-53765-0539 & 1.5$\pm$0.1 & 8.18$\pm$1.22 & 8.68$\pm$0.56 & 12.66$\pm$0.4 & 36.35$\pm$0.5 & 12.5$\pm$0.2 \\
2352-53770-0006 & - & - & - & - & - & 0.8$\pm$0.2 \\
2426-53795-0583 & 1.9$\pm$0.1 & 2.69$\pm$0.35 & 4.66$\pm$0.08 & 7.44$\pm$0.23 & 21.36$\pm$0.19 & 21.0$\pm$0.2 \\
2496-54178-0330 & - & - & - & - & - & 0.4$\pm$0.0 \\
2515-54180-0046 & 0.0$\pm$0.1 & 0.56$\pm$0.06 & 0.93$\pm$0.01 & 1.77$\pm$0.02 & 5.09$\pm$0.07 & 48.9$\pm$0.7 \\
2518-54243-0066 & 2.1$\pm$0.1 & 10.64$\pm$0.28 & 17.24$\pm$0.56 & 32.11$\pm$0.4 & 92.15$\pm$0.99 & 40.0$\pm$0.4 \\
2523-54572-0428 & 2.1$\pm$0.1 & 5.97$\pm$0.39 & 9.63$\pm$0.4 & 17.57$\pm$0.37 & 50.42$\pm$0.45 & 30.8$\pm$0.3 \\
2580-54092-0232 & 1.0$\pm$0.9 & 1.02$\pm$0.39 & 0.9$\pm$0.4 & 0.57$\pm$0.21 & 1.64$\pm$0.33 & 3.2$\pm$0.3 \\
2599-54234-0571 & 0.9$\pm$0.2 & 0.31$\pm$0.04 & 0.47$\pm$0.04 & 0.8$\pm$0.04 & 2.29$\pm$0.03 & 5.6$\pm$0.1 \\
2613-54481-0304 & 1.6$\pm$0.3 & 0.46$\pm$0.26 & 0.61$\pm$0.07 & 0.81$\pm$0.09 & 2.33$\pm$0.09 & 2.7$\pm$0.1 \\
2746-54232-0165 & 3.4$\pm$0.3 & 10.48$\pm$3.74 & 7.61$\pm$1.2 & 4.16$\pm$0.38 & 11.95$\pm$0.22 & 2.0$\pm$0.0 \\
2764-54535-0355 & 1.8$\pm$0.3 & 1.18$\pm$0.14 & 2.09$\pm$0.16 & 2.59$\pm$0.29 & 7.44$\pm$0.2 & 7.8$\pm$0.1 \\
2765-54535-0129 & 2.6$\pm$0.1 & 8.41$\pm$1.55 & 10.2$\pm$1.5 & 13.67$\pm$0.49 & 39.23$\pm$0.24 & 16.7$\pm$0.1 \\
2774-54534-0448 & 1.4$\pm$0.3 & 1.05$\pm$0.45 & 1.71$\pm$0.18 & 1.88$\pm$0.15 & 5.41$\pm$0.16 & 2.0$\pm$0.0 \\
2776-54554-0098 & 3.5$\pm$1.6 & 12.29$\pm$13.48 & 5.47$\pm$6.5 & 4.82$\pm$4.79 & 13.83$\pm$8.44 & 2.3$\pm$0.2 \\
2782-54592-0240 & 0.6$\pm$0.3 & 0.24$\pm$0.03 & 0.26$\pm$0.03 & 0.58$\pm$0.05 & 1.66$\pm$0.05 & 7.1$\pm$0.2 \\
1693-53446-0629 & 1.2$\pm$0.1 & 0.43$\pm$0.12 & 0.7$\pm$0.09 & 1.29$\pm$0.06 & 3.7$\pm$0.04 & 9.1$\pm$0.1 \\ \hline
\end{tabular}
\tablefoot{\footnotesize (1) Control sample galaxies Plate-MJd-Fiberid information from SDSS DR9 and matched on Table~\ref{tab:radial_vel}.}
\end{table*}

\begin{table*}
\centering
\small
\caption{Classification of each galaxy from the INT sample using BPTs and WHAN diagrams, and the one obtained when we required the same classification in at least four out of the five diagrams.}\label{tab:INT_bptclass}
\begin{tabular}{cccccccc}
\hline
Object & [NII] BPT & [SII] BPT & [OI] BPT & [OII] BPT & WHAN & 4/5 DGs \\
(1)&&&&&&\\
\hline
AM0012-235 A & - & - & - & - & RG & INC \\
AM0012-235 B & - & - & - & - & RG & INC \\
AM0103-520 A & Composite & SF & SF & SF & sAGN & INC \\
AM0103-520 B & - & - & - & - & RG & INC \\
AM0302-611 G1 & LINER & - & - & - & wAGN & INC \\
AM0302-611 G2 & SF & SF & SF & SF & SF & SF \\
AM0302-611 G3 & - & - & - & - & RG & INC \\
AM0302-611 G4 & Seyfert & LINER & - & - & RG & INC \\
AM0338-320 A & Composite & SF & SF & SF & sAGN & INC \\
AM0338-320 B & Seyfert & Seyfert & SF & SF & sAGN & INC \\
AM0348-502 G1 & LINER & Seyfert & Seyfert & Seyfert & sAGN & AGN \\
AM0348-502 G2 & Composite & SF & SF & SF & sAGN & INC \\
AM0348-502 G3 & SF & SF & - & - & SF & INC \\
AM0737-764 A & LINER & LINER & LINER & SF & wAGN & AGN \\
AM0737-764 B & - & - & - & - & RG & INC \\
AM0830-235 A & Composite & SF & SF & SF & sAGN & INC \\
AM0830-235 B & Seyfert & - & Seyfert & LINER & RG & INC \\
AM1058-243 A & SF & SF & LINER & LINER & sAGN & INC \\
AM1058-243 B & Composite & SF & SF & SF & sAGN & INC \\
AM1125-374 A & Composite & SF & Seyfert & SF & wAGN & INC \\
AM1125-374 B & Composite & SF & SF & SF & wAGN & INC \\
AM1132-450 A & Composite & SF & SF & SF & sAGN & INC \\
AM1132-450 B & SF & SF & SF & SF & SF & SF \\
AM1228-260 A & SF & SF & SF & SF & sAGN & SF \\
AM1228-260 B & SF & SF & SF & SF & SF & SF \\
AM1238-340 A & Composite & SF & SF & SF & sAGN & INC \\
AM1238-340 B & Composite & LINER & LINER & SF & wAGN & AGN \\
AM1411-434 A & - & - & - & - & RG & INC \\
AM1411-434 B & - & - & - & - & RG & INC \\
AM1445-274  G1 & Seyfert & LINER & LINER & LINER & sAGN & AGN \\
AM1445-274  G2 & - & - & - & - & RG & INC \\
AM1722-595 A & SF & SF & SF & SF & sAGN & SF \\
AM1722-595 B & SF & SF & SF & SF & SF & SF \\
AM1933-422 A & SF & SF & SF & SF & sAGN & SF \\
AM1933-422 B & Seyfert & Seyfert & Seyfert & SF & sAGN & AGN \\
AM1933-422 C & SF & SF & SF & SF & SF & SF \\
AM1941-564 A & Composite & SF & - & - & sAGN & INC \\
AM1941-564 B & Seyfert & LINER & - & - & RG & INC \\
AM2004-662  G1 & SF & SF & SF & SF & sAGN & SF \\
AM2004-662  G3 & LINER & LINER & - & - & RG & INC \\
AM2036-510  G1 & Composite & LINER & Seyfert & LINER & wAGN & AGN \\
AM2036-510  G2 & SF & SF & SF & SF & SF & SF \\
AM2036-510  G3 & SF & SF & SF & SF & SF & SF \\
AM2106-624  G2 & Composite & SF & SF & SF & sAGN & INC \\
AM2108-452 G1 & SF & SF & SF & SF & sAGN & SF \\
AM2108-452 G2 & - & - & - & - & RG & INC \\
AM2108-452 G3 & SF & SF & SF & SF & SF & SF \\
AM2117-515  G1 & LINER & SF & Seyfert & SF & wAGN & INC \\
AM2117-515  G2 & SF & SF & SF & SF & SF & SF \\
AM2117-515  G4 & SF & SF & SF & SF & SF & SF \\
AM2117-515  G5 & SF & Seyfert & SF & SF & SF & SF \\
AM2117-515  G6 & SF & SF & SF & SF & SF & SF \\
AM2117-515  KNOTS & SF & SF & SF & SF & SF & SF \\
AM2132-664  G1 & Composite & SF & SF & SF & sAGN & INC \\
AM2132-664  G3 & SF & SF & SF & SF & SF & SF \\
AM2217-490  G1 & Composite & SF & SF & SF & sAGN & INC \\
AM2217-490  G2 & SF & SF & SF & SF & SF & SF \\
AM2220-493  G1 & Seyfert & Seyfert & Seyfert & Seyfert & sAGN & AGN \\
AM2220-493  G2 & Seyfert & Seyfert & Seyfert & Seyfert & sAGN & AGN \\
\hline
\end{tabular}
\tablefoot{\footnotesize (1) System name and component according to \cite{ArpMadore+87} catalogue and Table~\ref{tab:Sample}.}
\end{table*}

\begin{table*}
\centering
\caption{Classification of each galaxy from the CTR sample using BPTs and WHAN diagrams, and the one obtained when we required the same classification in at least four out of the five diagrams.}
\label{tab:CTR_bptclass}
\begin{tabular}{ccccccc}
\hline
{\small Plate-MJd-Fiberid} & [NII] BPT & [SII] BPT & [OI] BPT & [OII] BPT & WHAN & 4/5 DGs \\
(1) &&&&&&\\
\hline
0275-51910-0320 & SF & SF & SF & - & SF & SF \\
0420-51871-0285 & SF & LINER & - & - & SF & INC \\
0420-51871-0517 & SF & SF & - & - & SF & INC \\
0421-51821-0003 & Composite & SF & - & - & RG & INC \\
0731-52460-0345 & Seyfert & LINER & Seyfert & SF & wAGN & AGN \\
0860-52319-0120 & SF & SF & SF & SF & sAGN & SF \\
1266-52709-0142 & Seyfert & LINER & - & - & RG & INC \\
1588-52965-0534 & SF & SF & SF & SF & SF & SF \\
1601-53115-0280 & Composite & - & LINER & SF & RG & INC \\
1607-53083-0452 & Composite & SF & SF & SF & sAGN & INC \\
1652-53555-0578 & SF & SF & Seyfert & SF & SF & SF \\
1693-53446-0629 & SF & SF & - & - & sAGN & INC \\
1771-53498-0483 & Composite & SF & SF & SF & sAGN & INC \\
1776-53858-0098 & SF & SF & SF & SF & SF & SF \\
1852-53534-0298 & SF & SF & SF & - & sAGN & INC \\
1929-53349-0372 & - & - & - & - & SF & INC \\
1937-53388-0453 & Composite & SF & SF & SF & wAGN & INC \\
2094-53851-0261 & Composite & SF & SF & SF & sAGN & INC \\
2124-53770-0272 & - & - & - & - & RG & INC \\
2126-53794-0075 & Seyfert & LINER & - & - & RG & INC \\
2130-53881-0553 & Seyfert & Seyfert & Seyfert & Seyfert & sAGN & AGN \\
2132-53493-0224 & - & - & - & - & RG & INC \\
2170-53875-0347 & Composite & SF & - & - & RG & INC \\
2227-53820-0018 & - & - & - & - & RG & INC \\
2234-53823-0120 & - & - & - & - & wAGN & INC \\
2268-53682-0424 & LINER & SF & - & - & RG & INC \\
2273-53709-0027 & - & - & - & - & SF & INC \\
2277-53705-0378 & SF & SF & LINER & - & SF & INC \\
2350-53765-0539 & SF & SF & SF & - & sAGN & INC \\
2352-53770-0006 & - & - & - & - & RG & INC \\
2426-53795-0583 & SF & SF & SF & SF & SF & SF \\
2496-54178-0330 & - & - & - & - & RG & INC \\
2515-54180-0046 & SF & SF & Seyfert & - & SF & INC \\
2518-54243-0066 & Composite & SF & SF & SF & sAGN & INC \\
2523-54572-0428 & SF & SF & - & - & sAGN & INC \\
2580-54092-0232 & Composite & SF & LINER & - & wAGN & INC \\
2599-54234-0571 & - & - & - & - & wAGN & INC \\
2613-54481-0304 & LINER & Seyfert & Seyfert & LINER & RG & AGN \\
2746-54232-0165 & LINER & LINER & SF & SF & RG & INC \\
2764-54535-0355 & Composite & SF & - & - & sAGN & INC \\
2765-54535-0129 & SF & SF & SF & SF & sAGN & SF \\
2774-54534-0448 & LINER & LINER & SF & LINER & RG & RG \\
2776-54554-0098 & Composite & - & - & - & RG & INC \\
2782-54592-0240 & - & - & - & - & SF & INC \\
1693-53446-0629 & SF & SF & - & - & sAGN & INC \\
\hline
\end{tabular}
\tablefoot{\footnotesize (1) Control sample galaxies Plate-MJd-Fiberid information from SDSS DR9 and matched on Table~\ref{tab:radial_vel}.}
\end{table*}

\FloatBarrier

\section{Excluded sources additional material\label{AP_B}}

\begin{figure*}[ht]
    \centering
    \includegraphics[width=0.64\columnwidth]{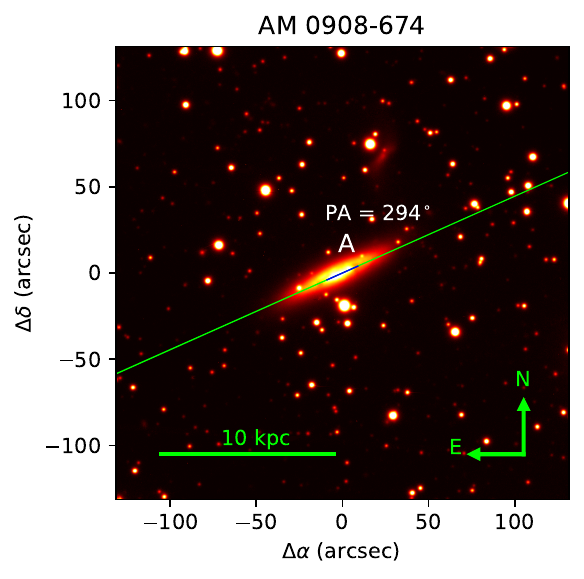}
    \includegraphics[width=0.64\columnwidth]{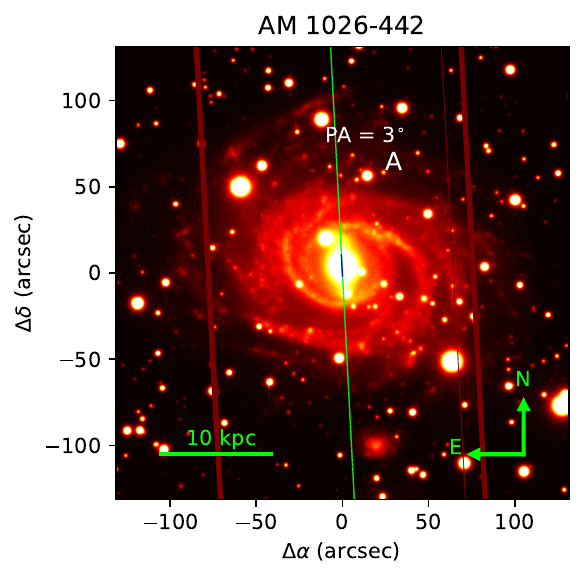}
    \includegraphics[width=0.64\columnwidth]{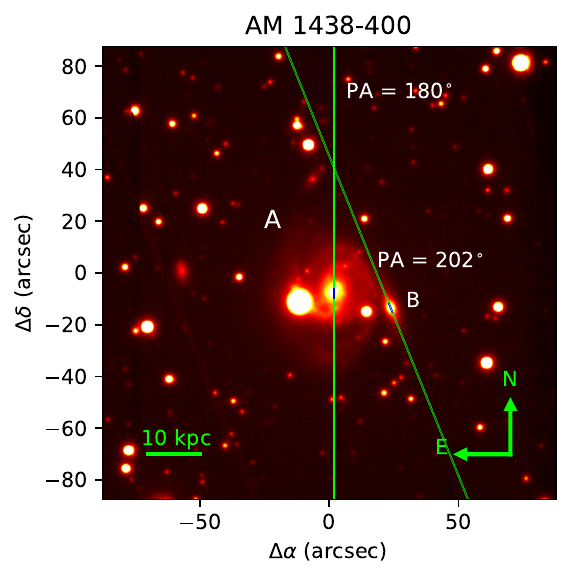}
    \includegraphics[width=0.64\columnwidth]{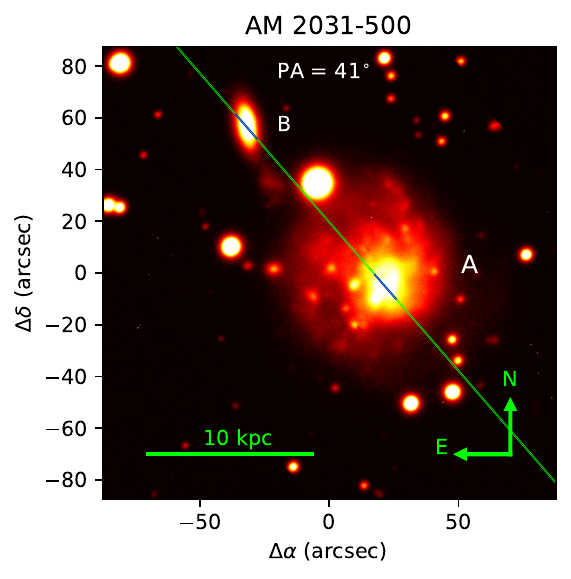}
    \includegraphics[width=0.64\columnwidth]{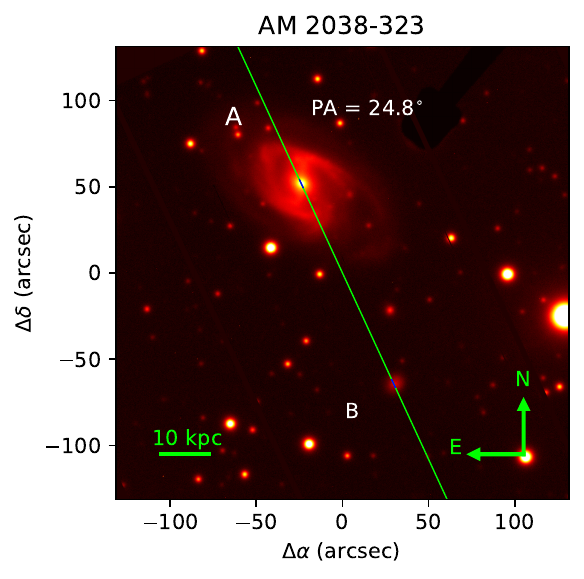}
    \includegraphics[width=0.64\columnwidth]{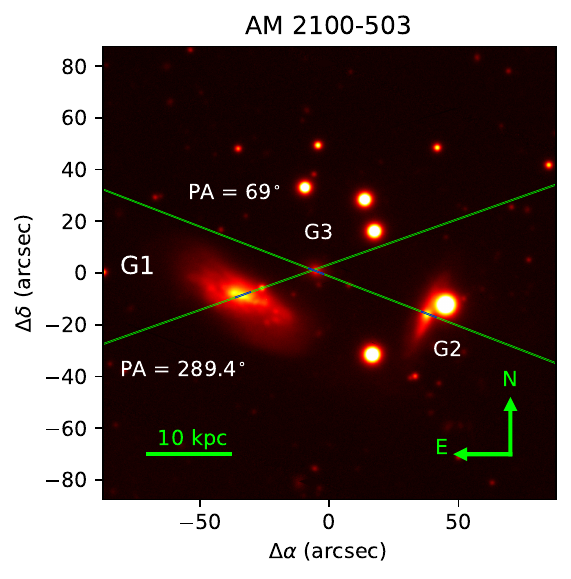}
    \includegraphics[width=0.64\columnwidth]{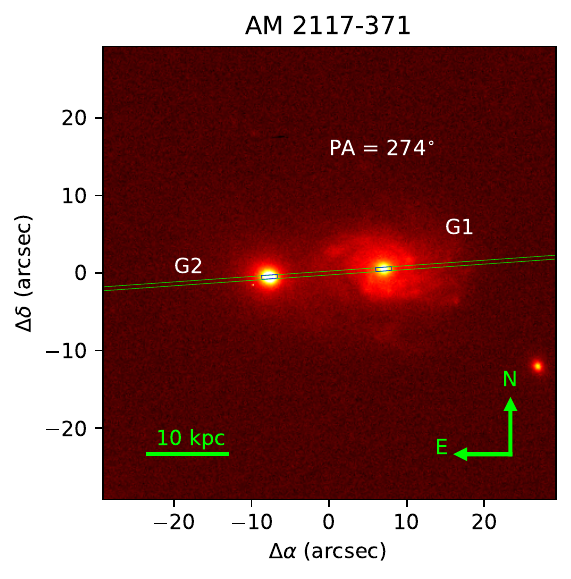}
    \includegraphics[width=0.64\columnwidth]{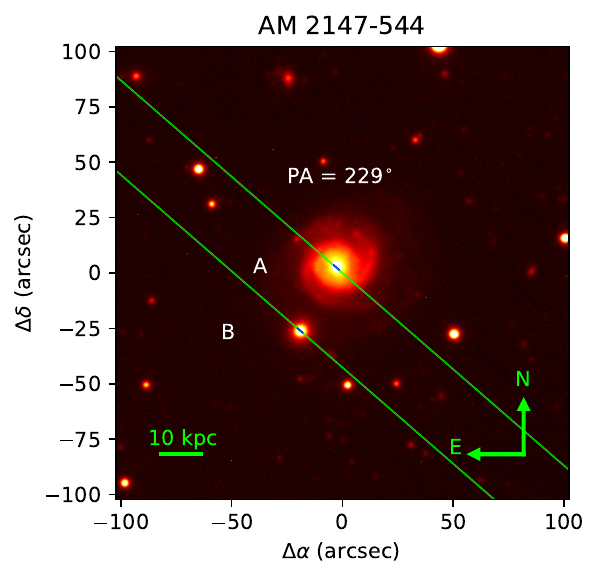}
    \includegraphics[width=0.64\columnwidth]{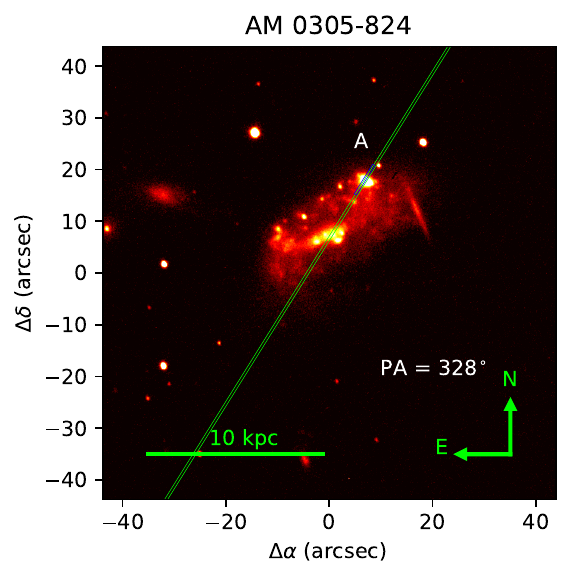}
    \caption{Same as in Fig.~\ref{fig:pairsimages}, but for the galaxies discarded as real pairs/groups. }
    \label{fig:excludedimages}
\end{figure*}

This appendix presents the information on the excluded sources from the initial interacting candidates sample of Table~\ref{tab:Sample}. They were excluded for not being part of an interacting system or if there was insufficient information to confirm or discard their nature. The images and spectra of the galaxies are in Figs~\ref{fig:excludedimages} and ~\ref{fig:exclspectra}. Table~\ref{tab:radial_vel_nonpairs} presents their radial velocities and the reason why they were excluded. Finally, their forbidden and permitted emission line fluxes are shown in Tables~\ref{tab:EXCforbflux} and \ref{tab:EXCHflux}, respectively.

\setcounter{figure}{0}
    \begin{figure*}[ht]
    \centering
    \includegraphics[width=0.64\columnwidth]{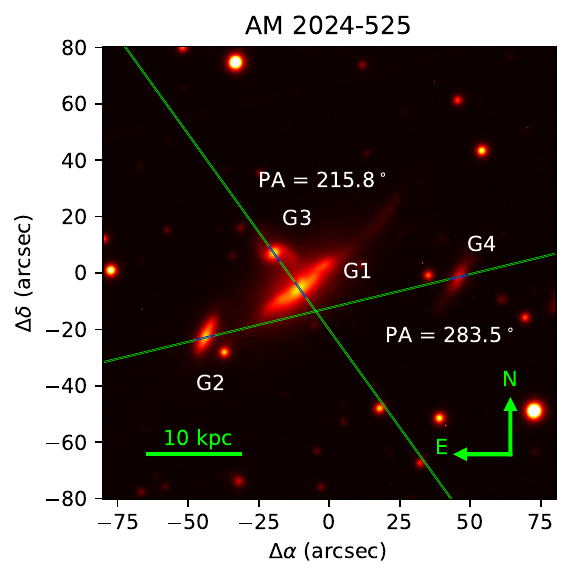}
    \includegraphics[width=0.64\columnwidth]{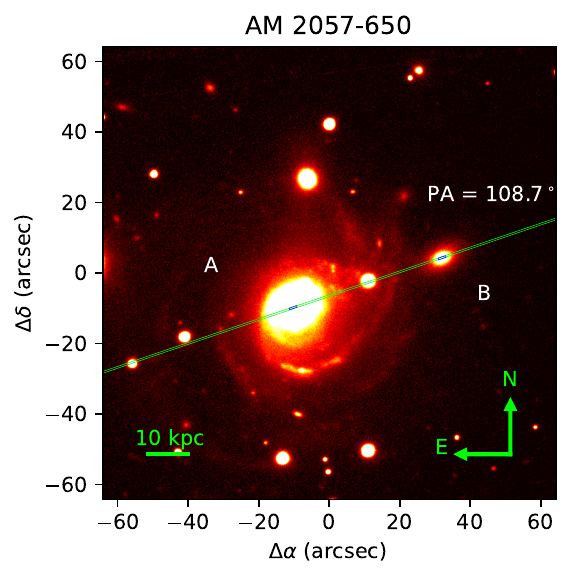}
    \includegraphics[width=0.64\columnwidth]{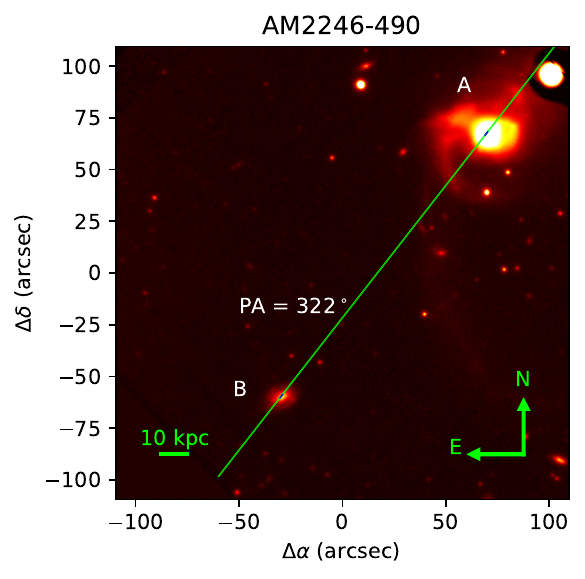}
    \includegraphics[width=0.64\columnwidth]{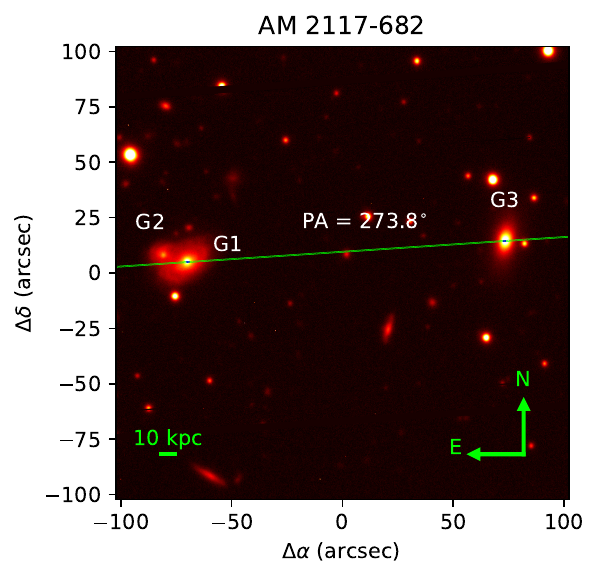}
    \includegraphics[width=0.64\columnwidth]{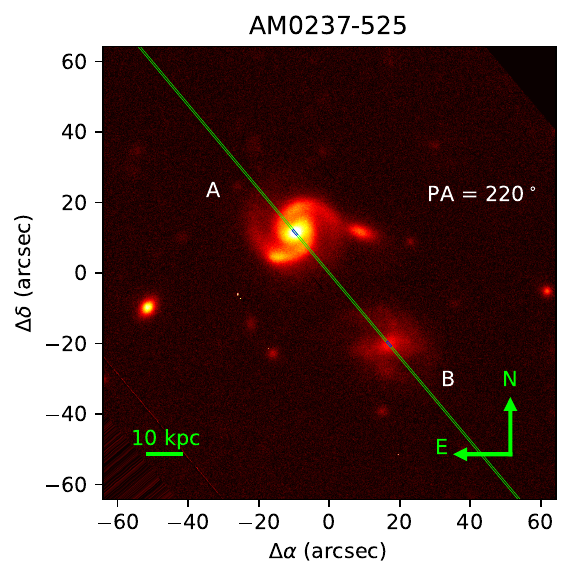}
    \includegraphics[width=0.64\columnwidth]{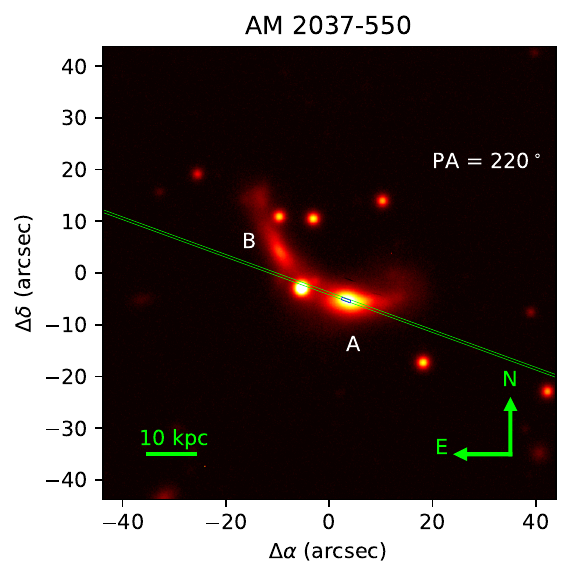}
    \includegraphics[width=0.64\columnwidth]{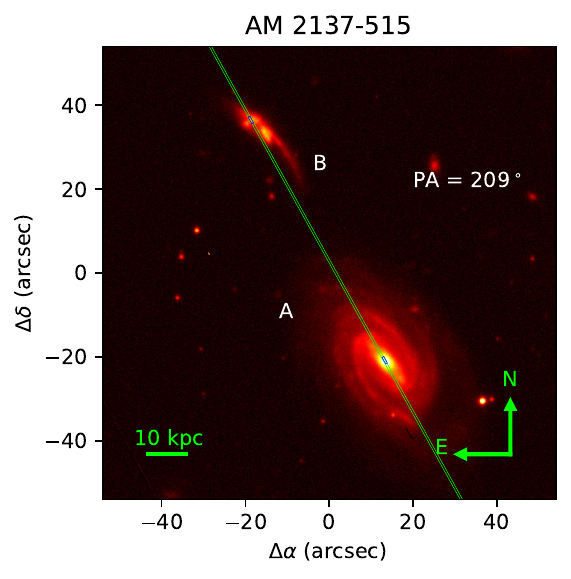}
    \includegraphics[width=0.64\columnwidth]{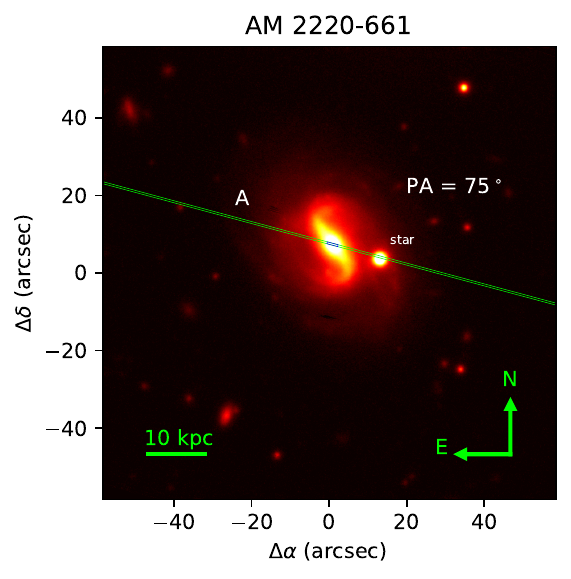}
    \caption{continue.}
\end{figure*}

\begin{figure*}
    \centering
    \includegraphics[width=0.97\columnwidth]{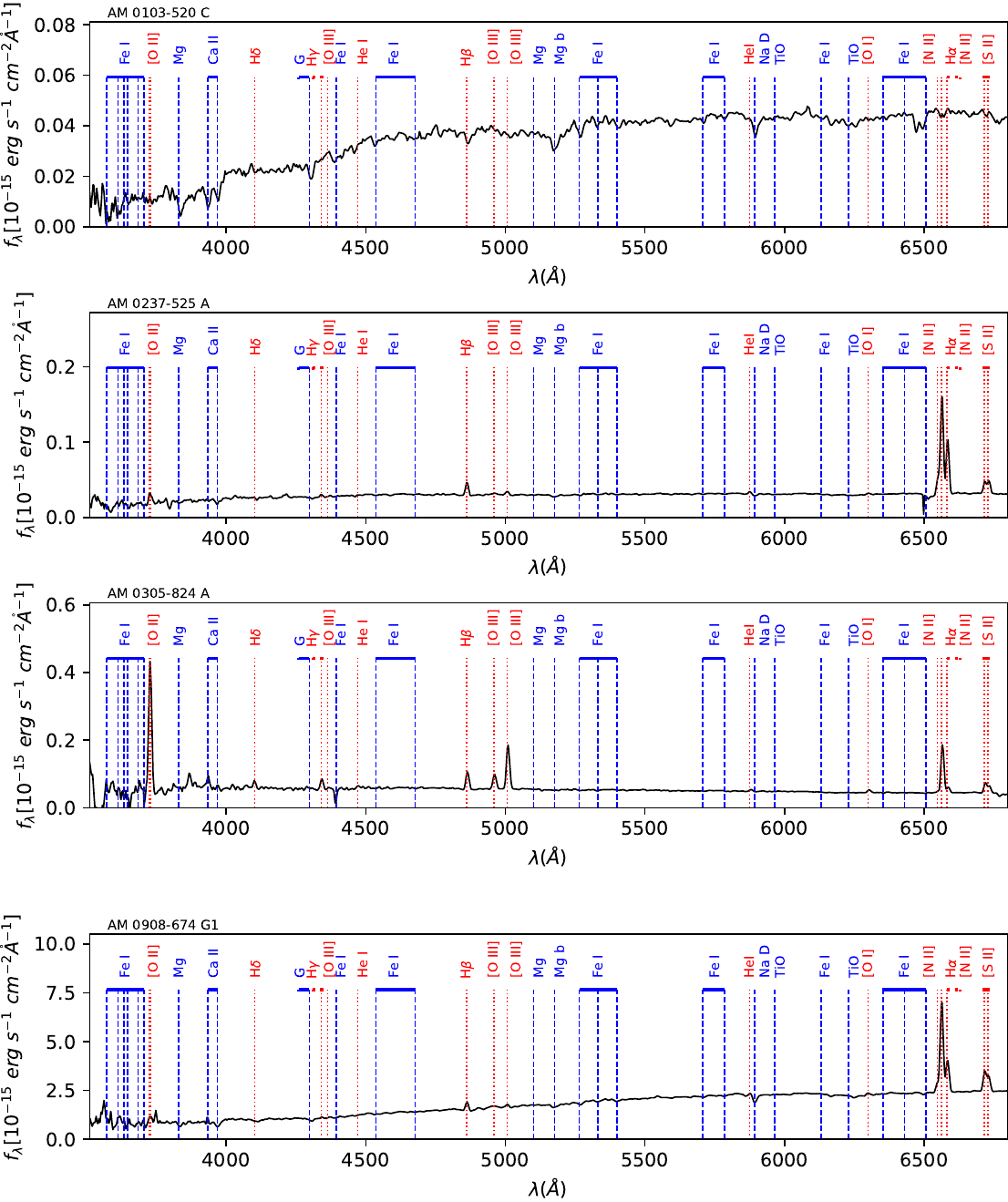}
    \includegraphics[width=0.97\columnwidth]{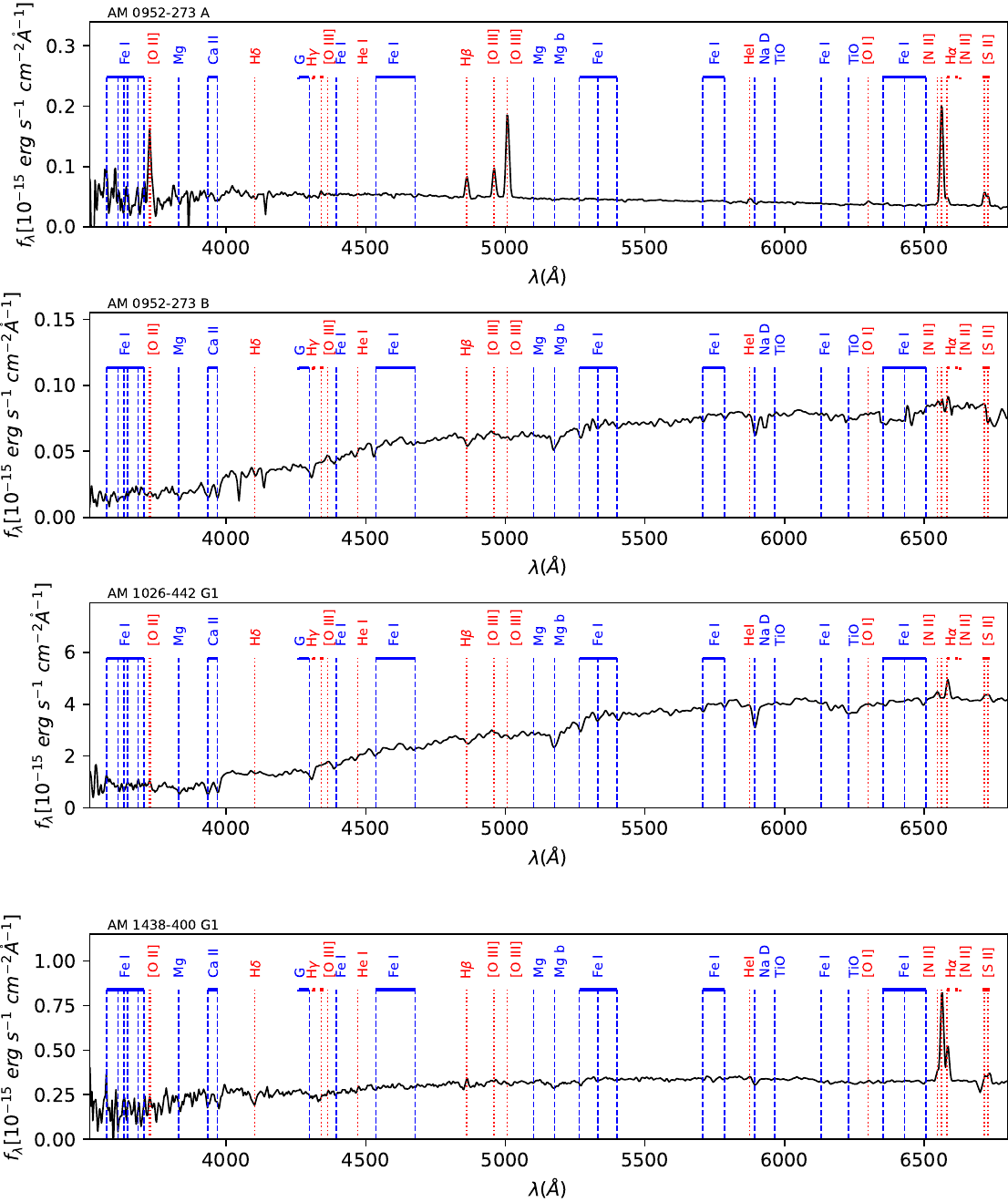}   
    \includegraphics[width=0.97\columnwidth]{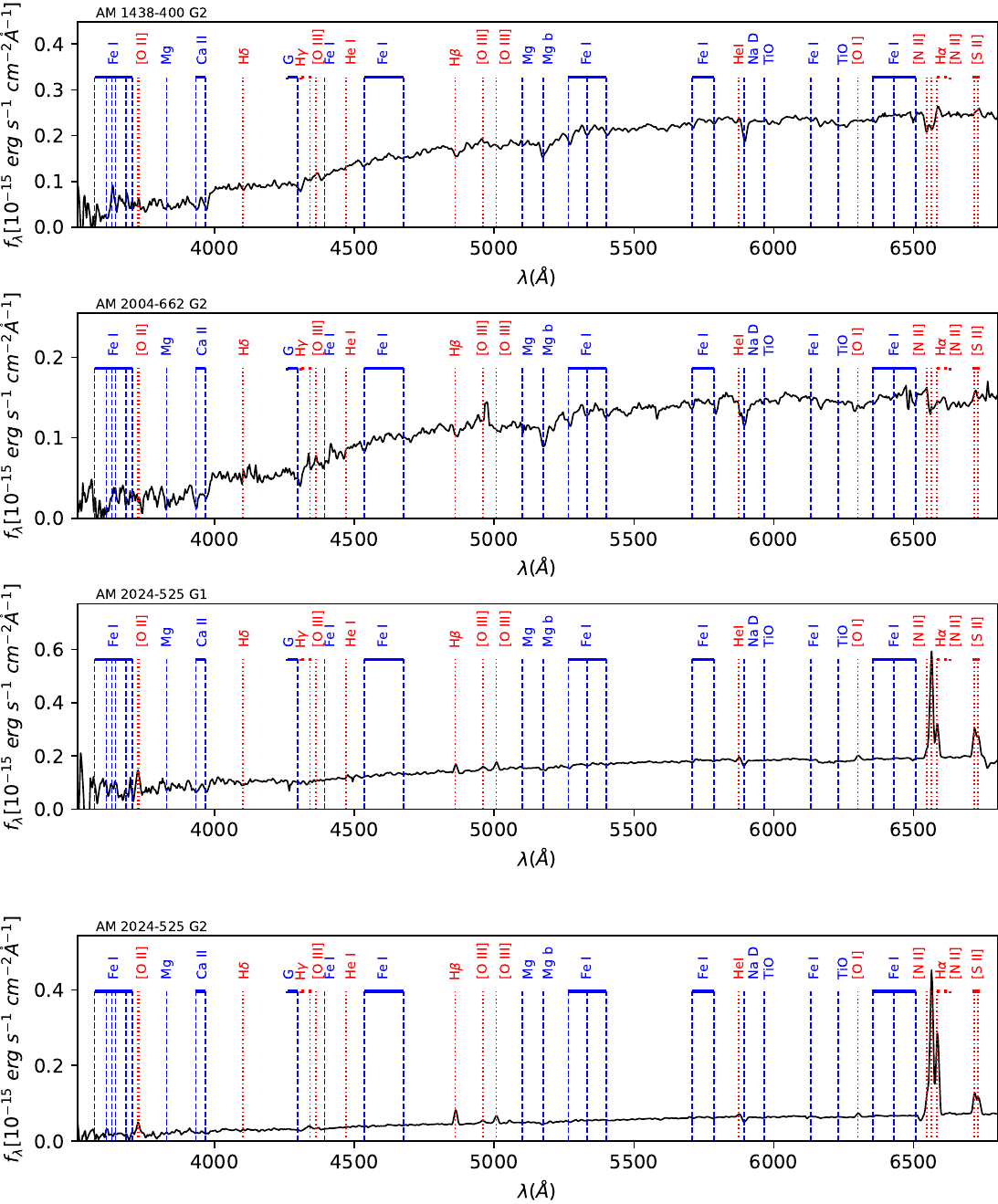}
    \includegraphics[width=0.97\columnwidth]{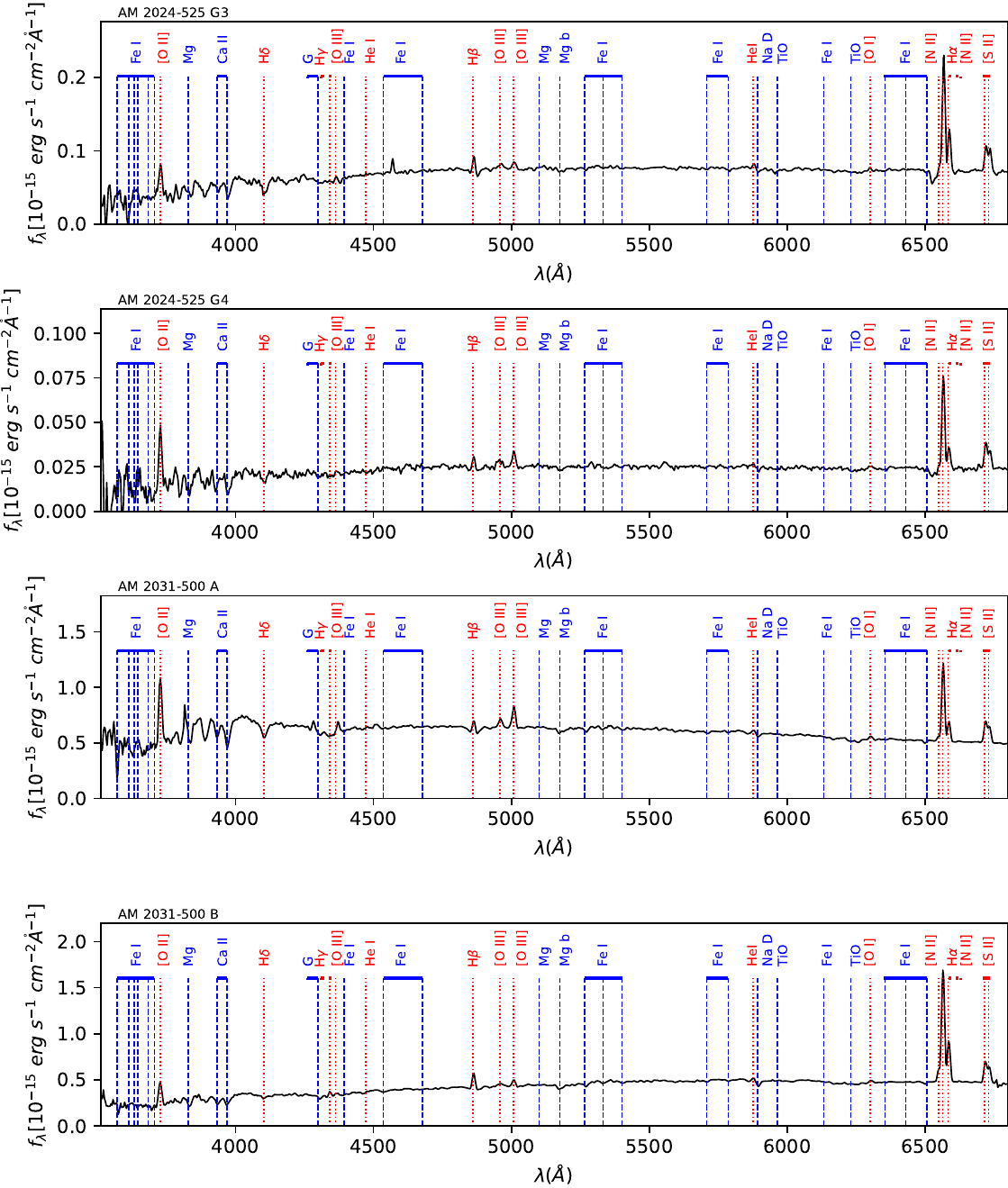}
    \caption{Same as in Fig.~\ref{fig:pairsspectra}, but for the galaxies discarded as real pairs/groups.}
    \label{fig:exclspectra}
\end{figure*}

\setcounter{figure}{1}
\begin{figure*}
    \centering
    \includegraphics[width=0.97\columnwidth]{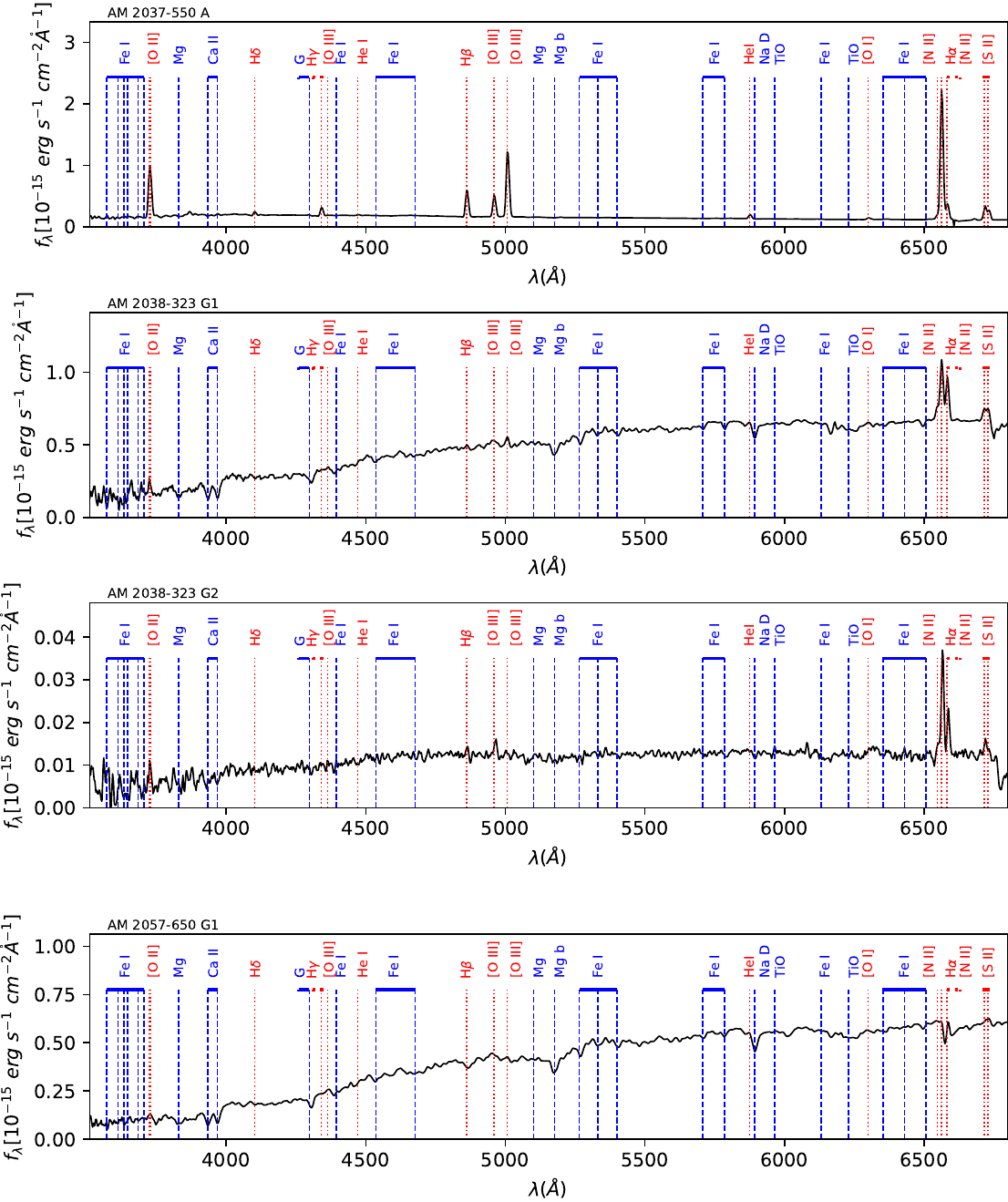}
    \includegraphics[width=0.97\columnwidth]{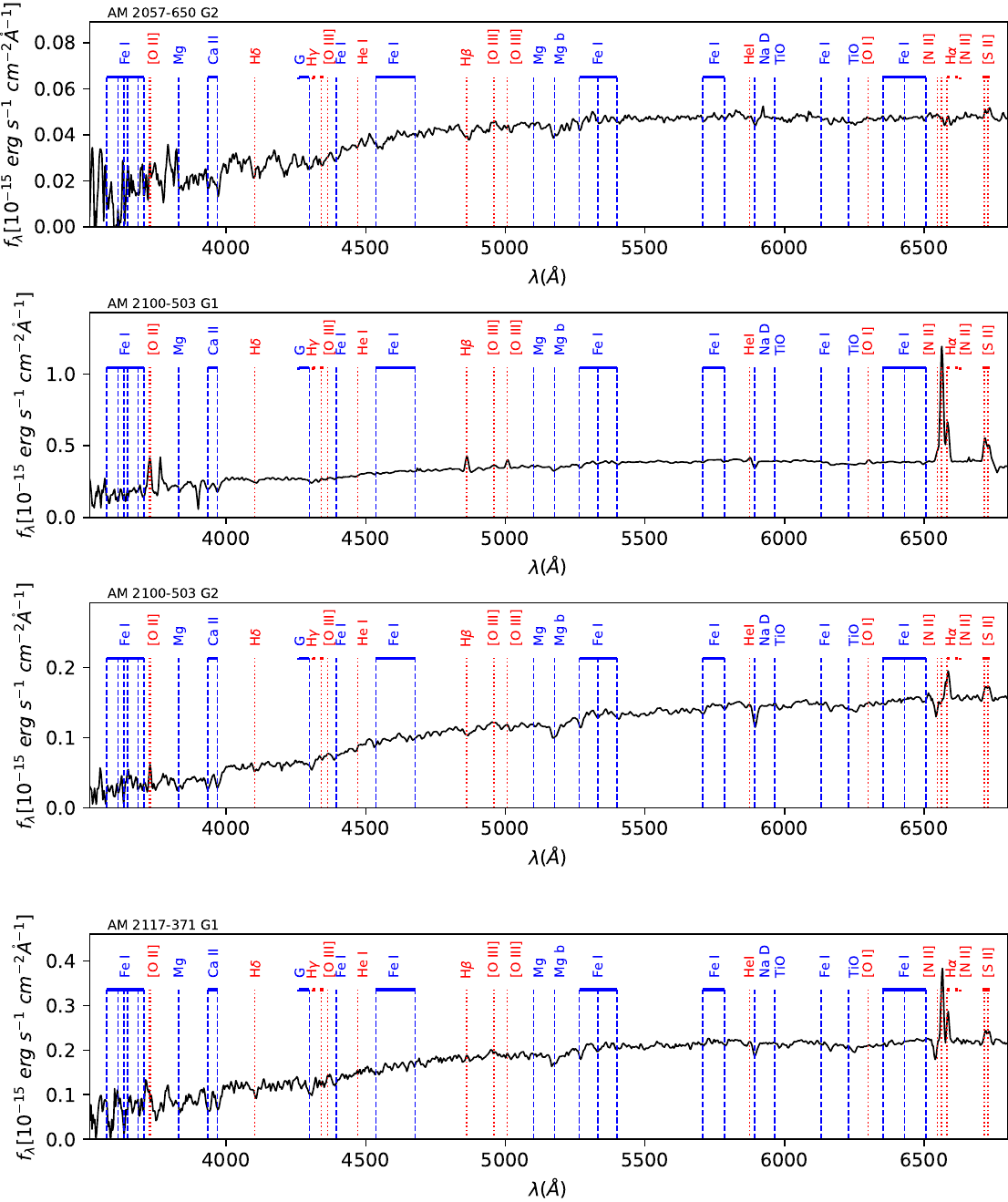}  
    \includegraphics[width=0.97\columnwidth]{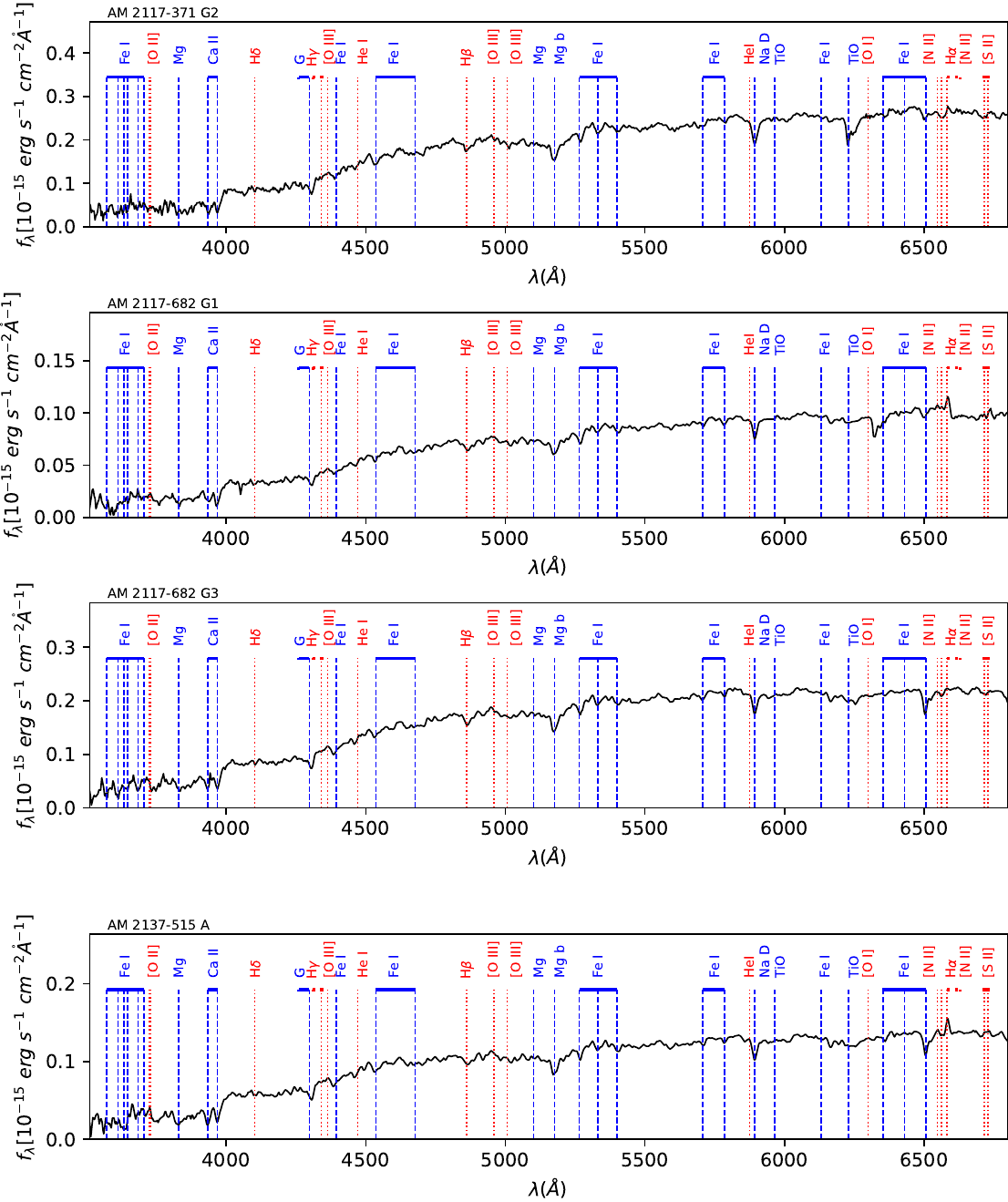}
    \includegraphics[width=0.97\columnwidth]{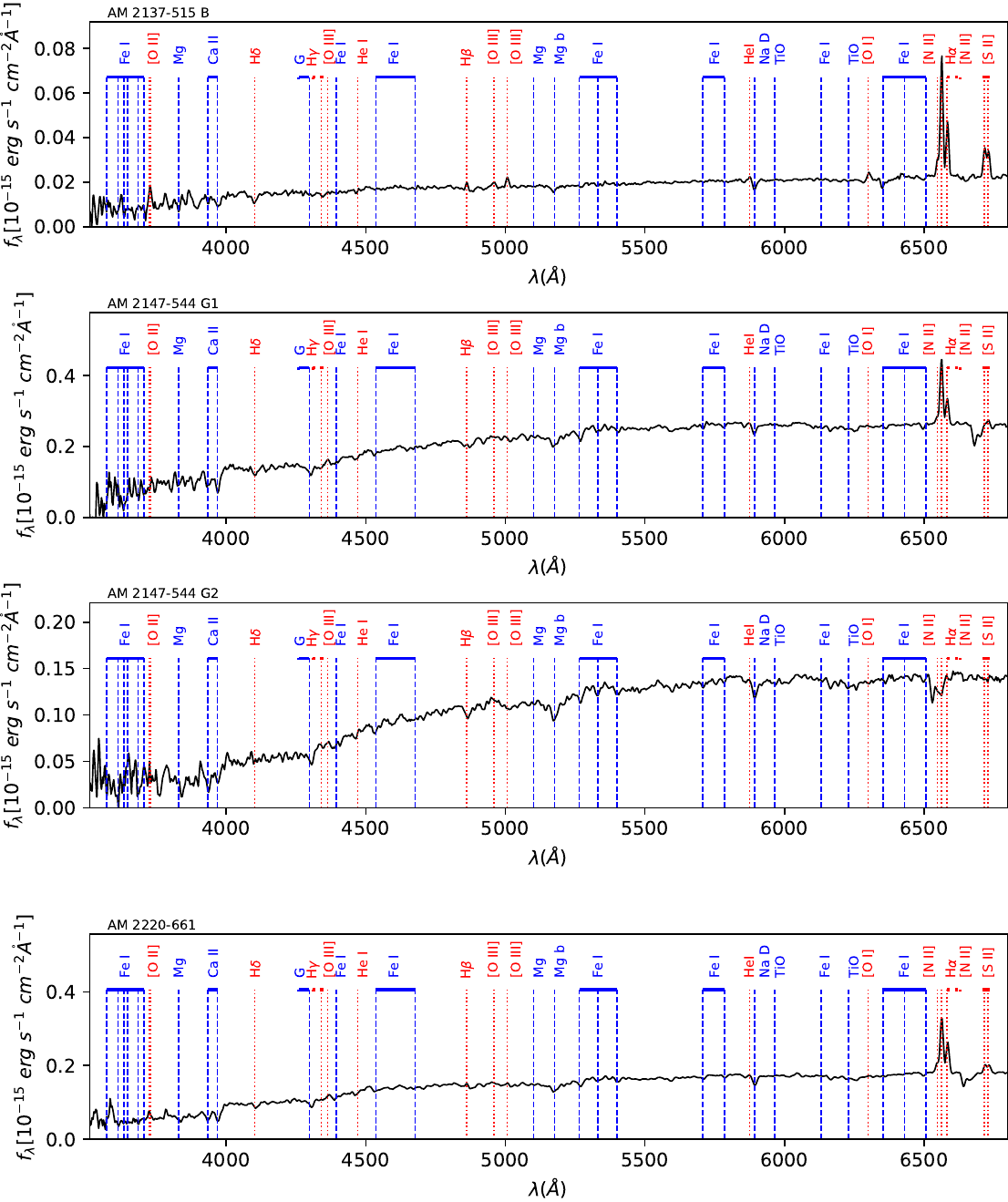} 
    \caption{continued.}
\end{figure*}

\setcounter{figure}{1}
\begin{figure*}
    \centering
    \includegraphics[width=0.97\columnwidth]{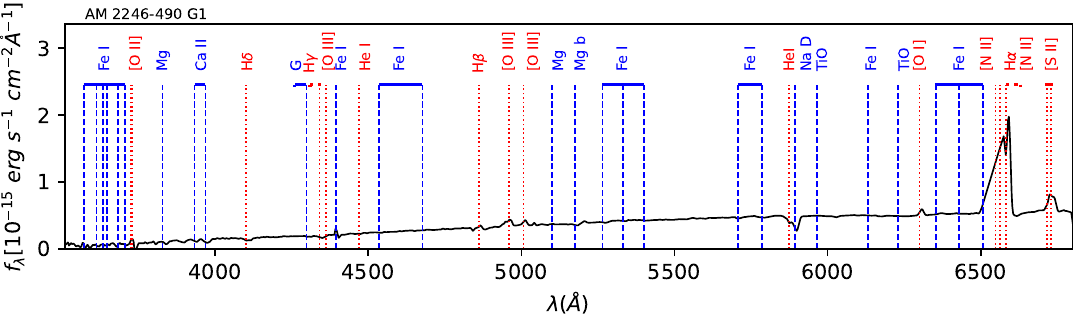}
    \caption{continued.}
\end{figure*}


\begin{table*}
\caption{List of non-physical, uncertain or not observed systems. The second column presents galaxies radial velocities when available.}
\label{tab:radial_vel_nonpairs}
\centering
\tiny
\begin{tabular}{ccc} 
System name & Radial velocity & Comments\\ (1) & [km/s] (2) &\\
\hline\hline
AM 0103-520~C& 18576 $\pm$  58       & Not make part of the system.\\ \hline
AM 0237-525~A& 16460 $\pm$  57               &\\
\quad ~B& & Without a good S/N spectrum.\\ \hline
AM 0305-824~A  & 4514 $\pm$  105      & Only this galaxy observed.\\ \hline
AM 0908-674~G1~(A)    & 1546 $\pm$ 126                & Possible companion (not observed) \\&& towards NE.\\
~B && It is a star towards the south-east of A.\\ \hline
AM 0952-273~A & 4352 $\pm$  86                &It is not a pair. \\
~B                & 19336 $\pm$  88               & \\ \hline
AM 1026-442~G1   & 2427 $\pm$  75                & Primary is NGC 3261.\\&& Two objects to the south were not observed.\\ \hline
AM 1438-400~G1    & 7684 $\pm$  91                &It is not a pair.\\
~G2        & 14924 $\pm$  86               &\\ \hline
AM 2004-662~G2 & 27853 $\pm$   84  & It is a foreground galaxy.\\
~G4 & 27977 $\pm$  102               & It is a foreground galaxy.\\ \hline
AM 2024-525~G1     & 4769 $\pm$  82                &Not make part of the system.\\
~G2                  & 15958 $\pm$  106 & Despite the redshits,\\ 
&& there are no signs of interaction\\ 
&& and the galaxies are far apart.\\
~G3  & 15924 $\pm$  101 \\                    
~G4                  & 16061 $\pm$  188 & -                   \\ \hline
AM 2031-500~A     & 2433 $\pm$  55                & It is not a pair.\\
&& A looks like a double or triple nucleus.\\ 
~B     & 4488 $\pm$  71                &\\ \hline
AM 2037-550~A     & 11366 $\pm$  68               &\\
B && We do not have a spectrum. \\ \hline
AM 2038-323~G1~(A)& 5566 $\pm$  96                & It is not a pair.\\
~G2~(B)                & 36821 $\pm$ 123           &\\ \hline
AM 2057-650~G1~(A)  & 13517 $\pm$   94 &\\
~G2~(B)                  & 13869 $\pm$   97 & B has faint continuum and no emission lines.\\
&& Redshift is uncertain.\\
&& The possible companion is at 36 arcsec north of A.\\ \hline      
AM 2100-503~G1~(A)     & 4901 $\pm$ 107                &It is not a group.\\
~G2~(B)                & 15037 $\pm$  87               &\\
~G3~(C)                & 59459 $\pm$  -                &\\ \hline
AM 2117-371~G1~(A)     & 15185 $\pm$  83               & It is not a pair.\\
~G2~(B)                & 31066 $\pm$  83               &\\ \hline
AM 2117-515~G3         &&The slit does not go through the nucleus.\\\hline
AM 2117-682~G1     & 25960 $\pm$  81               & It is not a group with G3.\\
~G2                 &  -                             & Close to G1. Not observed. \\ 
~G3                 & 16922 $\pm$  94&\\ \hline
AM 2132-664~G2 &   & Star towards the south of G1.\\\hline
AM 2137-515~A     & 16899 $\pm$  90   & Not part of the pair.\\
~B                & 24729 $\pm$   95   & B is likely a pair, no spectrum available.\\ \hline
AM 2147-544~G1~(A)     & 8614 $\pm$  86                & It is not a pair.\\
G2 - B                 & 15747 $\pm$  89               &\\ \hline
AM 2220-661 A     & 10513 $\pm$  109              & It is not a pair.\\ && A spiral + star.\\\hline
AM 2246-490~G1~(A)      & 12684 $\pm$   -                & Redshift is uncertain. The Ha+NII is inside\\        
&& the CCD gap in both spectra.\\
~G2~(B)                  & 12894 $\pm$   -                & - \\       \hline  
\hline
\end{tabular}
\tablefoot{\footnotesize (1) System name and component according to \cite{ArpMadore+87} catalogue and Table~\ref{tab:Sample}. (2) Systemic velocity obtained as described in Sec.~\ref{section:obs}}
\end{table*}

\begin{sidewaystable*}
\caption{Extinction corrected emission lines measurements for forbidden and helium transitions of the excluded galaxies.}
\label{tab:EXCforbflux}
\centering
\renewcommand{\tabcolsep}{.5mm}
\tiny
\begin{tabular}{ccccccccccccc}
\hline\hline
Object & [O\,{\sc ii}]$\lambda$3726 & [O\,{\sc ii}]$\lambda$3729 & [O\,{\sc iii}]$\lambda$4363 & [O\,{\sc iii}]$\lambda$4959 & [O\,{\sc iii}]$\lambda$5007 & [N\,{\sc ii}]$\lambda$5755 & He\,{\sc i}$\lambda$5876 & [O\,{\sc i}]$\lambda$6300 & [N\,{\sc ii}]$\lambda$6548 & [N\,{\sc ii}]$\lambda$6583 & [S\,{\sc ii}]$\lambda$6716 & [S\,{\sc ii}]$\lambda$6731 \\
& flux x $\rm 10^{-15}$ & flux x $\rm 10^{-15}$ & flux x $\rm 10^{-15}$ & flux x $\rm 10^{-15}$ & flux x $\rm 10^{-15}$ & flux x $\rm 10^{-15}$ & flux x $\rm 10^{-15}$ & flux x $\rm 10^{-15}$ & flux x $\rm 10^{-15}$ & flux x $\rm 10^{-15}$ & flux x $\rm 10^{-15}$ & flux x $\rm 10^{-15}$
\\
& $[\ erg\ s^{-1}\ cm^{-2}]$ & $[\ erg\ s^{-1}\ cm^{-2}]$ & $[\ erg\ s^{-1}\ cm^{-2}]$ & $[\ erg\ s^{-1}\ cm^{-2}]$ & $[\ erg\ s^{-1}\ cm^{-2}]$ & $[\ erg\ s^{-1}\ cm^{-2}]$ & $[\ erg\ s^{-1}\ cm^{-2}]$ & $[\ erg\ s^{-1}\ cm^{-2}]$ & $[\ erg\ s^{-1}\ cm^{-2}]$ & $[\ erg\ s^{-1}\ cm^{-2}]$ & $[\ erg\ s^{-1}\ cm^{-2}]$ & $[\ erg\ s^{-1}\ cm^{-2}]$
\\
\hline
AM0103-520 C & 32.96$\pm$19.95 & 13.66$\pm$8.41 & - & - & 2.85$\pm$1.27 & - & - & - & 1.84$\pm$0.6 & 2.57$\pm$0.83 & 1.7$\pm$0.57 & 1.6$\pm$0.51 \\
AM0237-525 A & - & 14.21$\pm$0.82 & 0.04$\pm$0.02 & 0.36$\pm$0.03 & 1.31$\pm$0.07 & - & 0.52$\pm$0.02 & 0.21$\pm$0.01 & 2.46$\pm$0.02 & 8.9$\pm$0.11 & 2.98$\pm$0.14 & 1.83$\pm$0.18 \\
AM0305-824 A & 2.52$\pm$0.04 & 2.73$\pm$0.05 & - & 0.61$\pm$0.01 & 1.88$\pm$0.02 & 0.01$\pm$0.0 & 0.05$\pm$0.01 & 0.08$\pm$0.01 & - & 0.21$\pm$0.01 & 0.48$\pm$0.03 & 0.28$\pm$0.02 \\
AM0908-674  G1 & 770.12$\pm$184.0 & 973.48$\pm$179.4 & - & 26.57$\pm$2.2 & 149.64$\pm$9.19 & 5.3$\pm$1.49 & 38.5$\pm$4.92 & - & 172.56$\pm$9.56 & 637.24$\pm$4.38 & 395.49$\pm$5.64 & 315.31$\pm$6.18 \\
AM0952-273 A & 9.84$\pm$0.26 & 3.6$\pm$0.28 & 0.24$\pm$0.04 & 2.98$\pm$0.03 & 9.14$\pm$0.09 & 0.04$\pm$0.0 & 0.24$\pm$0.02 & 0.2$\pm$0.01 & - & 0.52$\pm$0.01 & 0.87$\pm$0.02 & 0.54$\pm$0.05 \\
AM0952-273 B & 367.42$\pm$85.53 & 417.4$\pm$89.53 & 51.04$\pm$25.12 & 2.37$\pm$0.84 & 41.07$\pm$3.25 & 6.03$\pm$0.88 & 11.25$\pm$1.21 & 7.51$\pm$0.82 & 35.15$\pm$1.77 & 47.41$\pm$4.09 & 30.83$\pm$1.86 & - \\
AM1026-442  G1 & 83.18$\pm$11.03 & 25.86$\pm$9.24 & - & 6.28$\pm$0.82 & 23.07$\pm$2.03 & 5.12$\pm$0.8 & - & - & 61.56$\pm$3.11 & 135.85$\pm$6.75 & 46.03$\pm$2.1 & 48.52$\pm$2.68 \\
AM1438-400  G1 & 17.9$\pm$1.45 & 5.8$\pm$1.16 & - & 0.76$\pm$0.12 & 0.83$\pm$0.12 & - & 1.6$\pm$0.11 & - & 6.79$\pm$0.34 & 20.85$\pm$0.18 & 2.93$\pm$0.34 & 5.0$\pm$0.19 \\
AM1438-400  G2 & - & - & - & - & - & - & - & - & - & - & - & - \\
AM2004-662  G2 & - & - & - & - & - & - & - & - & - & - & - & - \\
AM2024-525  G1 & 208.77$\pm$7.22 & 116.19$\pm$9.8 & - & 5.96$\pm$0.14 & 22.43$\pm$0.98 & 1.2$\pm$0.14 & 4.48$\pm$0.3 & 4.28$\pm$0.22 & 10.55$\pm$0.48 & 37.12$\pm$0.25 & 31.64$\pm$0.39 & 20.79$\pm$0.29 \\
AM2024-525  G2 & 90.25$\pm$1.98 & 84.82$\pm$2.21 & - & 6.45$\pm$0.19 & 21.71$\pm$0.22 & 0.82$\pm$0.2 & 4.33$\pm$0.2 & 2.46$\pm$0.14 & 18.0$\pm$0.31 & 66.98$\pm$0.58 & 17.14$\pm$0.2 & 12.97$\pm$0.21 \\
AM2024-525  G3 & 5.56$\pm$0.21 & 6.73$\pm$0.26 & - & 0.69$\pm$0.04 & 1.48$\pm$0.06 & - & 0.34$\pm$0.05 & 0.18$\pm$0.0 & 0.58$\pm$0.07 & 3.45$\pm$0.07 & 2.17$\pm$0.04 & 1.86$\pm$0.03 \\
AM2024-525  G4 & 6.0$\pm$0.16 & 6.21$\pm$0.16 & - & 0.48$\pm$0.07 & 1.44$\pm$0.04 & 0.05$\pm$0.01 & 0.14$\pm$0.02 & 0.12$\pm$0.01 & 0.14$\pm$0.03 & 0.89$\pm$0.02 & 1.06$\pm$0.02 & 0.73$\pm$0.02 \\
AM2031-500  A & 36.54$\pm$1.34 & 43.32$\pm$1.18 & - & 5.04$\pm$0.07 & 15.43$\pm$0.23 & 0.12$\pm$0.02 & 1.51$\pm$0.12 & 0.61$\pm$0.07 & 3.01$\pm$0.25 & 8.94$\pm$0.16 & 9.06$\pm$0.19 & 7.12$\pm$0.19 \\
AM2031-500  B & 76.2$\pm$1.96 & 59.68$\pm$1.62 & - & 2.89$\pm$0.13 & 13.75$\pm$0.64 & 1.98$\pm$0.12 & 3.37$\pm$0.26 & 0.72$\pm$0.22 & 13.38$\pm$0.71 & 44.21$\pm$0.49 & 22.29$\pm$0.41 & 14.59$\pm$0.49 \\
AM2037-550 A & 105.86$\pm$2.02 & 81.93$\pm$2.04 & - & 37.02$\pm$0.35 & 112.38$\pm$1.05 & 0.2$\pm$0.01 & 4.42$\pm$0.16 & 1.38$\pm$0.07 & 2.92$\pm$0.81 & 15.27$\pm$0.13 & 12.15$\pm$0.07 & 8.45$\pm$0.06 \\
AM2038-323  G1 & 30.18$\pm$3.76 & - & - & 4.01$\pm$0.26 & 14.12$\pm$0.99 & 3.02$\pm$0.23 & 1.6$\pm$0.11 & 0.64$\pm$0.04 & 12.97$\pm$0.23 & 34.33$\pm$0.32 & 11.28$\pm$0.3 & 9.66$\pm$0.3 \\
AM2038-323  G2 & 1.62$\pm$0.19 & 0.93$\pm$0.09 & - & 0.69$\pm$0.07 & 0.27$\pm$0.04 & - & 0.05$\pm$0.01 & 0.2$\pm$0.02 & 0.3$\pm$0.01 & 0.95$\pm$0.02 & 0.58$\pm$0.03 & 0.08$\pm$0.02 \\
AM2057-650  G1 & 2.88$\pm$0.64 & 4.04$\pm$0.69 & - & 1.73$\pm$0.18 & 3.16$\pm$0.25 & - & - & 0.06$\pm$0.01 & 2.11$\pm$0.17 & - & 2.55$\pm$0.25 & 3.18$\pm$0.2 \\
AM2057-650  G2 & 0.09$\pm$0.02 & 0.06$\pm$0.01 & - & 0.02$\pm$0.0 & 0.03$\pm$0.0 & - & - & 0.0$\pm$0.0 & 0.03$\pm$0.0 & 0.02$\pm$0.0 & 0.06$\pm$0.01 & 0.08$\pm$0.01 \\
AM2100-503  G1 & 121.63$\pm$2.3 & 19.63$\pm$2.22 & - & 1.8$\pm$0.12 & 11.51$\pm$0.49 & 1.3$\pm$0.1 & 2.74$\pm$0.15 & 0.98$\pm$0.1 & 7.75$\pm$0.6 & 29.37$\pm$0.23 & 17.7$\pm$0.24 & 12.28$\pm$0.28 \\
AM2100-503  G2 & 0.15$\pm$0.01 & 0.12$\pm$0.01 & - & 0.06$\pm$0.0 & 0.09$\pm$0.01 & 0.02$\pm$0.0 & 0.02$\pm$0.0 & 0.03$\pm$0.0 & - & 0.82$\pm$0.04 & 0.29$\pm$0.01 & 0.36$\pm$0.01 \\
AM2117-371  G1 & 38.13$\pm$3.47 & - & - & 1.66$\pm$0.21 & 2.75$\pm$0.09 & 0.7$\pm$0.11 & 0.34$\pm$0.06 & - & - & 6.57$\pm$0.21 & 3.57$\pm$0.2 & 3.85$\pm$0.13 \\
AM2117-371  G2 & - & - & - & - & - & - & - & - & - & - & - & - \\
AM2117-682  G1 & 160.69$\pm$35.26 & 202.01$\pm$35.47 & - & 18.23$\pm$1.49 & 21.15$\pm$2.13 & - & - & 0.54$\pm$0.1 & 9.52$\pm$0.71 & 18.69$\pm$1.35 & 1.15$\pm$0.32 & 3.69$\pm$0.71 \\
AM2117-682  G3 & - & - & - & - & - & - & - & - & - & - & - & - \\
AM2137-515 A & 3.66$\pm$0.44 & 1.34$\pm$0.4 & - & 0.4$\pm$0.05 & 0.65$\pm$0.06 & - & 0.1$\pm$0.01 & 0.25$\pm$0.02 & 0.69$\pm$0.08 & 1.89$\pm$0.13 & 0.73$\pm$0.06 & 0.81$\pm$0.07 \\
AM2137-515 B & 33.5$\pm$1.05 & 42.03$\pm$0.72 & 0.31$\pm$0.12 & 1.93$\pm$0.07 & 4.83$\pm$0.16 & - & 1.01$\pm$0.03 & 1.35$\pm$0.06 & 3.21$\pm$0.06 & 8.92$\pm$0.07 & 4.43$\pm$0.06 & 3.65$\pm$0.05 \\
AM2147-544  G1 & - & 21.57$\pm$5.76 & - & 0.91$\pm$0.09 & 4.61$\pm$0.45 & 0.25$\pm$0.07 & 0.92$\pm$0.08 & - & 4.19$\pm$0.26 & 13.22$\pm$0.37 & 1.01$\pm$0.25 & 2.77$\pm$0.08 \\
AM2147-544  G2 & - & - & - & - & - & - & - & - & - & - & - & - \\
AM2220-661 & 11.75$\pm$1.45 & 3.3$\pm$1.56 & - & 0.15$\pm$0.02 & 1.79$\pm$0.09 & 0.25$\pm$0.05 & 0.52$\pm$0.03 & - & 4.49$\pm$0.15 & 11.56$\pm$0.08 & 2.86$\pm$0.15 & 3.05$\pm$0.15 \\
AM2246-490  G1 & 34.18$\pm$6.93 & 57.21$\pm$6.9 & - & 37.27$\pm$2.15 & 32.39$\pm$2.08 & 4.26$\pm$0.49 & - & 7.26$\pm$1.75 & 312.25$\pm$12.41 & 424.09$\pm$21.63 & 16.92$\pm$1.29 & 80.68$\pm$2.46 \\
\hline
\end{tabular}
\tablefoot{\footnotesize (1) System name and component according to \cite{ArpMadore+87} catalogue and Table~\ref{tab:Sample}.}
\end{sidewaystable*}

\begin{table*}[ht]
\caption{Extinction corrected permitted emission lines measurements of the excluded galaxies.}
\label{tab:EXCHflux}
\centering
\small
\begin{tabular}{cccccccc} 
\hline\hline
Object & Av & H$\delta$ & H$\gamma$ & H$\beta$ & H$\alpha$ & EW H$\alpha$\\
(1) & [mag] & flux x $\rm 10^{-15}$ & flux x $\rm 10^{-15}$ & flux x $\rm 10^{-15}$ & flux x $\rm 10^{-15}$& [$\mathring{A}$]
\\
& & $[\ erg\ s^{-1}\ cm^{-2}]$ & $[\ erg\ s^{-1}\ cm^{-2}]$ & $[\ erg\ s^{-1}\ cm^{-2}]$ & $[\ erg\ s^{-1}\ cm^{-2}]$ &
\\
\hline \hline
AM0103-520 C & 5.5$\pm$1.2 & 10.13$\pm$5.75 & 4.62$\pm$2.61 & 0.88$\pm$0.5 & 2.52$\pm$0.85 & 0.9$\pm$0.1 \\
AM0237-525 A & 3.0$\pm$0.1 & 2.73$\pm$0.31 & 3.55$\pm$0.27 & 6.09$\pm$0.27 & 17.47$\pm$0.12 & 58.1$\pm$0.3 \\
AM0305-824 A & 0.0$\pm$0.1 & 0.33$\pm$0.03 & 0.44$\pm$0.02 & 0.74$\pm$0.02 & 2.11$\pm$0.03 & 48.0$\pm$0.6 \\
AM0908-674  G1 & 4.4$\pm$0.1 & 156.15$\pm$3.4 & 244.24$\pm$24.14 & 640.7$\pm$21.13 & 1838.8$\pm$10.71 & 28.1$\pm$0.1 \\
AM0952-273 A & 1.5$\pm$0.1 & 0.55$\pm$0.06 & 1.25$\pm$0.05 & 2.53$\pm$0.1 & 7.25$\pm$0.05 & 68.3$\pm$0.4 \\
AM0952-273 B & 7.4$\pm$0.4 & - & 203.76$\pm$26.33 & 20.38$\pm$2.92 & 58.5$\pm$4.27 & 2.9$\pm$0.2 \\
AM1026-442  G1 & 2.8$\pm$0.4 & 44.01$\pm$3.7 & - & 12.67$\pm$1.76 & 36.35$\pm$2.47 & 1.1$\pm$0.1 \\
AM1438-400  G1 & 2.7$\pm$0.1 & 4.59$\pm$1.03 & 9.54$\pm$1.01 & 19.83$\pm$0.78 & 56.92$\pm$0.34 & 24.0$\pm$0.1 \\
AM1438-400  G2 & - & - & - & - & - & 0.0$\pm$0.0 \\
AM2004-662  G2 & - & - & - & - & - & 0.0$\pm$0.0 \\
AM2024-525  G1 & 4.0$\pm$0.1 & 22.09$\pm$0.64 & 21.02$\pm$0.47 & 43.13$\pm$1.39 & 123.79$\pm$0.89 & 31.1$\pm$0.2 \\
AM2024-525  G2 & 4.2$\pm$0.1 & 15.07$\pm$0.39 & 20.0$\pm$1.47 & 42.59$\pm$2.03 & 122.24$\pm$0.9 & 76.1$\pm$0.4 \\
AM2024-525  G3 & 2.0$\pm$0.2 & - & 1.4$\pm$0.18 & 3.58$\pm$0.29 & 10.26$\pm$0.16 & 31.3$\pm$0.2 \\
AM2024-525  G4 & 2.3$\pm$0.2 & 0.23$\pm$0.05 & 0.53$\pm$0.07 & 1.38$\pm$0.1 & 3.95$\pm$0.05 & 30.9$\pm$0.2 \\
AM2031-500  A & 1.6$\pm$0.2 & 4.61$\pm$0.62 & 5.2$\pm$0.21 & 12.24$\pm$0.61 & 35.11$\pm$0.25 & 20.6$\pm$0.1 \\
AM2031-500  B & 2.6$\pm$0.1 & 12.21$\pm$0.79 & 21.99$\pm$0.67 & 43.19$\pm$1.84 & 123.94$\pm$0.99 & 36.2$\pm$0.2 \\
AM2037-550 A & 2.0$\pm$0.1 & 12.1$\pm$0.8 & 19.25$\pm$0.8 & 43.45$\pm$1.45 & 124.69$\pm$0.75 & 240.5$\pm$1.3 \\
AM2038-323  G1 & 2.7$\pm$0.1 & 8.45$\pm$0.8 & 7.04$\pm$0.24 & 18.31$\pm$0.65 & 52.54$\pm$0.48 & 10.4$\pm$0.1 \\
AM2038-323  G2 & 2.7$\pm$0.3 & 0.46$\pm$0.11 & 0.53$\pm$0.07 & 0.81$\pm$0.07 & 2.33$\pm$0.04 & 26.9$\pm$0.2 \\
AM2057-650  G1 & 2.0$\pm$0.3 & - & - & 1.18$\pm$0.1 & 3.38$\pm$0.18 & 1.3$\pm$0.1 \\
AM2057-650  G2 & 0.2$\pm$0.5 & - & - & 0.02$\pm$0.0 & 0.06$\pm$0.01 & 1.1$\pm$0.1 \\
AM2100-503  G1 & 2.7$\pm$0.1 & 11.7$\pm$0.4 & 11.85$\pm$0.52 & 30.8$\pm$1.15 & 88.41$\pm$0.53 & 30.5$\pm$0.2 \\
AM2100-503  G2 & 0.0$\pm$0.3 & - & - & 0.04$\pm$0.0 & 0.11$\pm$0.01 & 0.7$\pm$0.1 \\
AM2117-371  G1 & 2.9$\pm$0.2 & - & - & 6.87$\pm$0.51 & 19.71$\pm$0.28 & 10.6$\pm$0.1 \\
AM2117-371  G2 & - & - & - & - & - & 0.3$\pm$0.0 \\
AM2117-682  G1 & 6.0$\pm$0.4 & - & - & 5.32$\pm$0.6 & 15.27$\pm$1.02 & 1.7$\pm$0.1 \\
AM2117-682  G3 & - & - & - & - & - & 0.4$\pm$0.1 \\
AM2137-515 A & 2.6$\pm$0.4 & 0.37$\pm$0.06 & 0.58$\pm$0.07 & 0.34$\pm$0.04 & 0.98$\pm$0.07 & 1.1$\pm$0.1 \\
AM2137-515 B & 4.5$\pm$0.1 & 1.47$\pm$0.56 & 6.29$\pm$0.59 & 6.82$\pm$0.25 & 19.58$\pm$0.14 & 30.6$\pm$0.2 \\
AM2147-544  G1 & 3.5$\pm$0.1 & 5.35$\pm$0.17 & 1.77$\pm$0.88 & 14.63$\pm$0.57 & 41.99$\pm$0.25 & 12.0$\pm$0.1 \\
AM2147-544  G2 & - & - & - & - & - & 0.0$\pm$0.0 \\
AM2220-661 & 2.9$\pm$0.1 & 1.96$\pm$0.35 & 4.06$\pm$0.12 & 7.2$\pm$0.32 & 20.67$\pm$0.23 & 13.2$\pm$0.1 \\
AM2246-490  G1 & 3.4$\pm$0.2 & - & 7.7$\pm$1.07 & 48.26$\pm$3.33 & 138.5$\pm$4.44 & 20.0$\pm$0.6\\  
\hline
\end{tabular}
\tablefoot{\footnotesize (1) System name and component according to \cite{ArpMadore+87} catalogue and Table~\ref{tab:Sample}.}
\end{table*}

\end{appendix} 
\end{document}